# A Comprehensive Survey on Fog Computing: State-of-the-art and Research Challenges

Carla Mouradian, Diala Naboulsi, Sami Yangui, Roch H. Glitho, Monique J. Morrow, and Paul A. Polakos

*Abstract* — Cloud computing with its three key facets (i.e., IaaS, PaaS, and SaaS) and its inherent advantages (e.g., elasticity and scalability) still faces several challenges. The distance between the cloud and the end devices might be an issue for latency-sensitive applications such as disaster management and content delivery applications. Service Level Agreements (SLAs) may also impose processing at locations where the cloud provider does not have data centers. Fog computing is a novel paradigm to address such issues. It enables provisioning resources and services outside the cloud, at the edge of the network, closer to end devices or eventually, at locations stipulated by SLAs. Fog computing is not a substitute for cloud computing but a powerful complement. It enables processing at the edge while still offering the possibility to interact with the cloud. This article presents a comprehensive survey on fog computing. It critically reviews the state of the art in the light of a concise set of evaluation criteria. We cover both the architectures and the algorithms that make fog systems. Challenges and research directions are also introduced. In addition, the lessons learned are reviewed and the prospects are discussed in terms of the key role fog is likely to play in emerging technologies such as Tactile Internet.

## I. Introduction

OVER the years, computing paradigms have evolved from distributed, parallel, and grid to cloud computing. Cloud computing [1][2] comes with several inherent capabilities such as scalability, on-demand resource allocation, reduced management efforts, flexible pricing model (pay-as-you-go), and easy applications and services provisioning. It comprises three key service models: Infrastructure-as-a-Service (IaaS), Platform-as-a-Service (PaaS), and Software-as-a-Service (SaaS). IaaS provides the virtualized resources, such as compute, storage, and networking. The PaaS provides software environments for the development, deployment, and management of applications. The SaaS provides software applications and composite services to end-users and other applications.

Nowadays, cloud computing is widely used. However, it still has some limitations. The fundamental limitation is the connectivity between the cloud and the end devices. Such connectivity is set over the Internet, not suitable for a large set of cloud-based applications such as the latency-sensitive ones [3]. Well-known examples include connected vehicles [4], fire detection and firefighting [5], smart grid [4], and content delivery [6]. Furthermore, cloud-based applications are often distributed and made up of multiple components [7]. Consequently, it is not uncommon to sometimes deploy application components separately over multiple clouds (e.g., [8] and [9]). This may worsen the latency due to the overhead induced by inter-cloud communications. Yet, as another limitation, the regulations may prescribe processing at locations where the cloud provider may have no data center [10].

Fog computing [11] is a computing paradigm introduced to tackle these challenges. It is now being promoted by the OpenFog Consortium which has recently published a few white papers (e.g., [12]). Fog is "cloud closer to ground". It is a novel architecture that extends the traditional cloud computing architecture to the edge of the network. With fog, the processing of some application components (e.g., latency-sensitive ones) can take place at the edge of the network, while others (e.g., delay-tolerant and computational intensive components) can happen in the cloud. Compute, storage, and networking services are the building blocks of the cloud and the fog that extends it. However, the fog provides additional advantages, such as low-latency, by allowing processing to take place at the network edge, near the end devices, by the so-called fog nodes and the ability to enable processing at specific locations. It also offers densely-distributed points for gathering data generated by the end devices. This is done through proxies, access points, and routers positioned at the network edge, near the sources. In the literature (e.g., [11][13]) it is widely acknowledged that cloud computing is not viable for most of Internet of Things (IoT) applications and fog could be used as an alternative. However, it is important to note that the applicability of fog goes beyond IoT and includes areas such as content delivery as shown later in this paper.

Several surveys and tutorials related to fog computing have been published over the past years. The next subsection outlines how our survey differs from them. The following subsection

C. Mouradian and D. Naboulsi are with Concordia University, Montreal, QC H3G 1M8, Canada (e-mail: {ca_moura, d_naboul}@encs.concordia.ca).

S. Yangui was with Concordia University, Montreal, QC H3G 1M8, Canada. He is now with LAAS-CNRS, Université de Toulouse, INSA, Toulouse, France (e-mail: syangui@laas.fr).

R. Glitho is with Concordia University, Montreal, QC H3G 1M8, Canada, and also with the University of Western Cape, Bellville 7535, South Africa (e-mail: glitho@ece.concordia.ca).

Monique J. Morrow was with CISCO Systems, Zurich, Switzerland (e-mail: mmorrow@cisco.com).

Paul A. Polakos was with CISCO Systems, New York, NY, USA (e-mail: ppolakos@cisco.com).

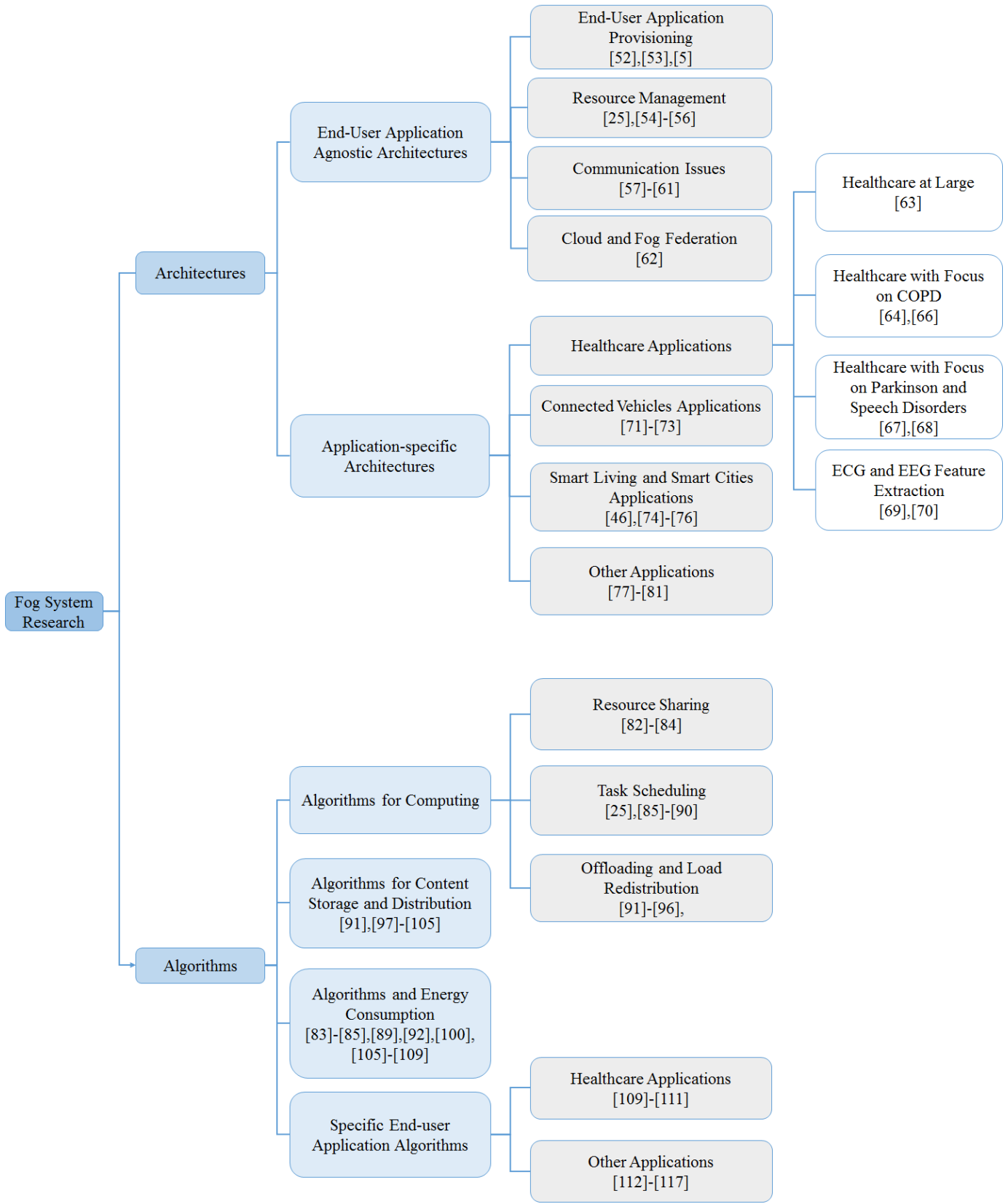

Fig. 1. Overview of the Surveyed Research Works Classified According to the Proposed Classification

presents the classification scheme that we propose to review the literature. We end the introduction by presenting the survey's organization and a reading map. This paper targets several categories of readers: readers interested in detailed architectural aspects, readers interested in detailed algorithmic aspects, and readers interested in a general overview. Our proposed reading map enables a reading a la carte.

*A. Existing Surveys and Tutorials on Fog Computing*

Several tutorials have been published, to formally define fog computing and the related challenges. LM Vaquero *et al.* [14] provide an overview of the concept of fog computing in terms of enabling technologies and emerging trends in usage patterns. They also briefly discuss the challenges ahead. Yi *et al.* [15] discuss the definition of fog computing and closely related concepts. They introduce application scenarios and discuss as well the challenges ahead. Yannuzzi *et al.* [16] discuss some of the challenges in IoT scenarios and demonstrate that fog computing is a promising enabler for IoT applications. Dasterji *et al.* [17] provide an overview of fog computing along with its characteristics. They introduce various applications that benefit from fog and present several challenges. More recently, Chiang *et al.* [18] provide a tutorial on fog computing. They discuss at a very high level the differences between fog computing, edge computing, and cloud computing. They also present the advantages of fog computing and discuss the research challenges. Several surveys have also been published on fog computing at large [4][19][20] and also in the context of specific application domains, i.e., vehicular Ad-hoc NETworks (VANETs) [21], Radio Access Networks (RAN) [22] [23], and Internet of Things [24].

Nevertheless, these tutorials and surveys do not provide a critical evaluation of each reviewed contribution in the light of well-defined and well-motivated criteria, as it is done in this survey. Neither do they present an exhaustive literature review. In particular, algorithmic aspects are not considered in any of these papers despite the critical role algorithms play in fog computing. In contrast, this paper comprehensively reviews the work done so far in the field from both architectural and algorithmic perspectives. In addition, the discussions of research directions in the existing surveys are sketchy, while in our case, they are comprehensive. We cover unaddressed issues, remaining challenges for the reviewed problems, and propose hints to deal with them. Moreover, we derive several lessons learned relevant to fog systems based on our review of the existing literature. Furthermore, the prospects of fog computing in terms of emerging technologies are sketched with a focus on Tactile Internet.

*B. Literature Classification*

In this survey, we present a structured classification of the related literature. The literature on fog computing is very diverse; structuring the relevant works in a systematic way is not a trivial task.

The outline of the proposed classification scheme is shown in Fig. 1. At the left side, we identify two main categories; proposed architectures for fog systems and proposed algorithms for fog systems. The review of the literature from these two perspectives separately was a natural choice, because most researchers in the area tackle the issues from either perspective. It allows grouping the reviewed papers under common umbrellas. However, there is one work (i.e., [25]) that tackles both architectural and algorithmic aspects. It is reviewed in both sections. Within the first category; architectures for fog systems, two subcategories have been proposed: application agnostic architectures and application-specific architectures. In the second category, algorithms for fog systems, four subcategories have been identified: algorithms for computing, algorithms for content storage and distribution, algorithms and energy consumption, and application-specific algorithms. By having this classification, we simplify the reader's access to references tackling a specific issue.

We note that we exclude from our survey efforts on security, confidentiality, and data protection, as commonly done nowadays in the community. We do acknowledge the importance of these aspects and many research activities currently tackle them (e.g., [26], [27], [28]). However, these issues are generally better covered in dedicated surveys. In IoT for instance, reference [29] surveys the technologies, protocols, and applications, while reference [30] reviews the security aspects.

This leaves us with a set of sixty-two papers published between 2013 and 2016. In addition, more papers have been published in 2017 including 6 papers in special issues of IEEE publications (i.e., [31] [32]). This brings the total number of the published and reviewed papers as part of this survey to sixty-eight for the 2013-2017 period. In fact, the very first publication related to fog computing is published in 2012. It introduces the concept of fog computing and discusses its role in the Internet of Things. More elaborate contributions in the field started from 2013. One can easily notice the fast growth of the interest for fog computing over the years. Starting with four papers in 2013 and three papers in 2014, the number of publications jumped to thirty-three in 2015 and twenty-two in 2016. Interestingly, almost half of these papers (i.e., thirty-two) are dedicated to fog-enabled architectures while the other half is dedicated for algorithms for fog systems. For fog-enabled architectures works, application-specific architectures for fog systems have gained more attention (19 papers, 59.3%) compared to end-user application agnostic architectures (13 papers, 40.6%). More specifically, most of the application-specific architectures were designed in the context of healthcare (7 papers, 36.8%) and smart environments (4 papers, 21.05%). There are also proposed architectures for connected vehicles (3 papers, 15.7%) and the rest are related to various kinds of applications (5 papers, 26.3%). As for algorithms for fog systems, computing aspects have received significant attention (16 papers, 34.7%), compared to content storage and distribution (10 paper, 21.7%) and energy efficiency (10 paper, 21.7%). The specific end-user applications are the least considered (9 papers, 19.5%). We remark that some algorithmic papers fall in different subcategories and the percentage presented are derived accordingly.

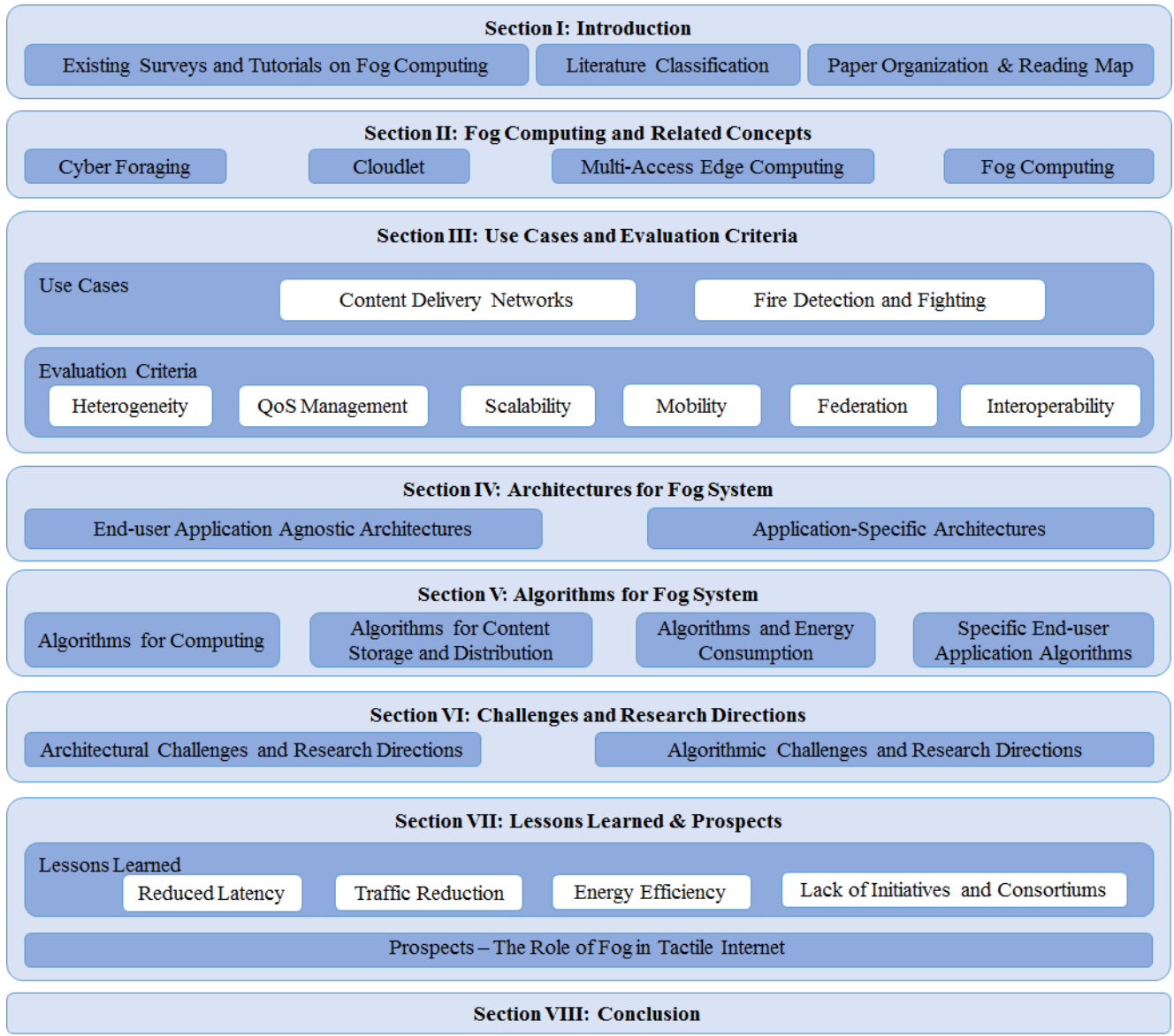

Fig. 2. The Structure of the Survey

*C. Paper Organization and Reading Map*

The structure of the survey is shown in Fig. 2. It is organized as follows: Section II introduces fog computing and related concepts, such as cyber foraging, cloudlets, and Multi-access Edge Computing (MEC). Section III discusses fog application use cases and derives the evaluation criteria for fog systems. In this survey, we define a fog system as a system of architectural modules/interfaces with the accompanying algorithms that enable fog applications. Section IV reviews the fog systems proposed so far from the architectural modules/interfaces perspective while section V reviews them from an algorithmic perspective. We complement sections IV and V with Fig. 1, showing the classification of the papers reviewed in the two sections, to simplify the user's access to references in a particular category. Section VI focuses on the challenges and the research directions. Section VII presents the lessons learned and discusses the prospects. In the prospects portion of the section, we discuss the vital role fog computing is expected to play in emerging technologies such as Tactile Internet. Finally, Section VIII concludes the survey.

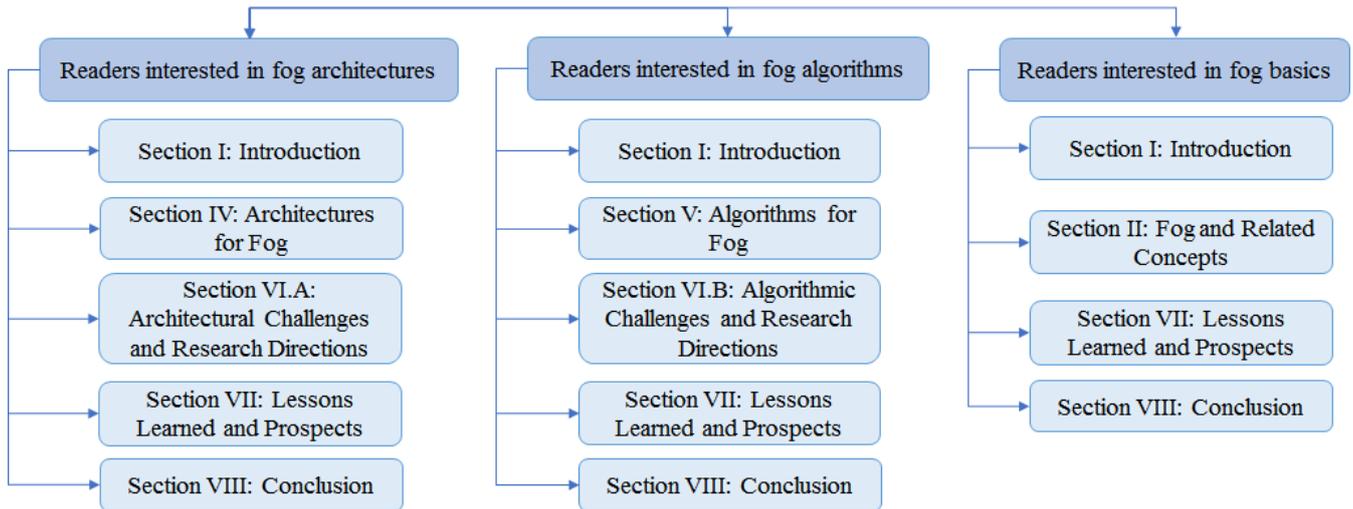

Fig. 3. A Reading Map

An "a la carte" approach can be followed to read this survey. Fig. 3 provides a reading map. Readers with main interests in the detailed architectural aspects of fog computing can focus their reading on Sections I, IV, VI.A, VII, and VIII. When it comes to those mainly interested in the detailed algorithmic aspects, they can read only Sections I, V, VI.B, VII, and VIII. Finally, we recommend Sections I, II, VII, and VIII to the readers interested in getting a very high-level overview of fog computing including the similarities and differences with closely related concepts such as cyber-foraging, cloudlet, and MEC.

## II. FOG COMPUTING AND RELATED CONCEPTS

Provisioning resources at the edge of networks (closer to end-devices) brings several benefits such as low latency and enables provisioning of new applications such as mobile data offloading. This section discusses and contrasts the key concepts that enable application provisioning at the edge. The next subsections introduce cyber foraging, cloudlet, MEC, and fog. The last subsection, illustrated by a Venn diagram, discusses the similarities and differences between these concepts.

It is important to note that we do not discuss in this section the industry initiatives which focus solely on implementations. An example is the Open Edge computing initiative which aims at reference implementations, live demonstrations, and a real-world testbed based on OpenStack technology [33]. Yet another example is the Open Cord initiative which aims at providing a reference implementation of central offices re-architected as data centers, using technologies such as OpenStack and ONOS [34]. Let us also stress that a few surveys that discuss mobile edge computing at large have been published in the very recent past (e.g., [35]).

### A. Cyber Foraging

Cyber foraging is among the first concepts for edge computing but has now been superseded by more recent concepts such a cloudlet, MEC, and fog. We discuss it here because of its seminal aspects. It was introduced by Satyanarayanan [36] in 2001 and was further refined by Balan *et al.* [37] in 2002. In cyber foraging, resource-limited mobile devices exploit the capabilities of nearby servers, connected to the Internet through high-bandwidth networks. These servers are called surrogates and perform computing and data staging. Data staging is the process of prefetching distant data to nearby surrogates. For instance, when a mobile device has to process a request of a compute-intensive component such as face recognition, which needs to access a large volume of data for face matching process, for example, it captures raw images and offloads the complex processing to a surrogate. The surrogate performs the face detection and matching processes using a database. This surrogate may stage the database on its local disk and perform the whole or some part of the processing on behalf of the mobile device. It then delivers the result to the mobile device with low latency, since it is close to the device. If the mobile device does not find a surrogate server nearby, it may provide a degraded service to the end-user due to its limited capabilities.

### B. Cloudlet

The concept of cloudlet was proposed next by Satyanarayanan [38] in 2009. It is referred to as cloudlet-based cyber foraging in [39] and [40]. Cloudlets reuse modern cloud computing techniques such as virtual machine (VM) based - virtualization. They are resource-rich servers or clusters of servers located in a single-hop proximity of mobile devices. They run one or more VMs in which mobile devices can offload components for expensive computation. Back to the face recognition application, using cloudlets, the face detection and matching processes will be performed on VMs instead of real machines. Thanks to the VM technology, cloudlets can expand and shrink dynamically, leading eventually to scalability with respect to the mobile users' service requests. Moreover, the VM separates the guest software environment from the host software environment of the cloudlet, which in turn increases the chance of mobile users finding a compatible cloudlet to offload their computation-intensive requests anywhere in the world.

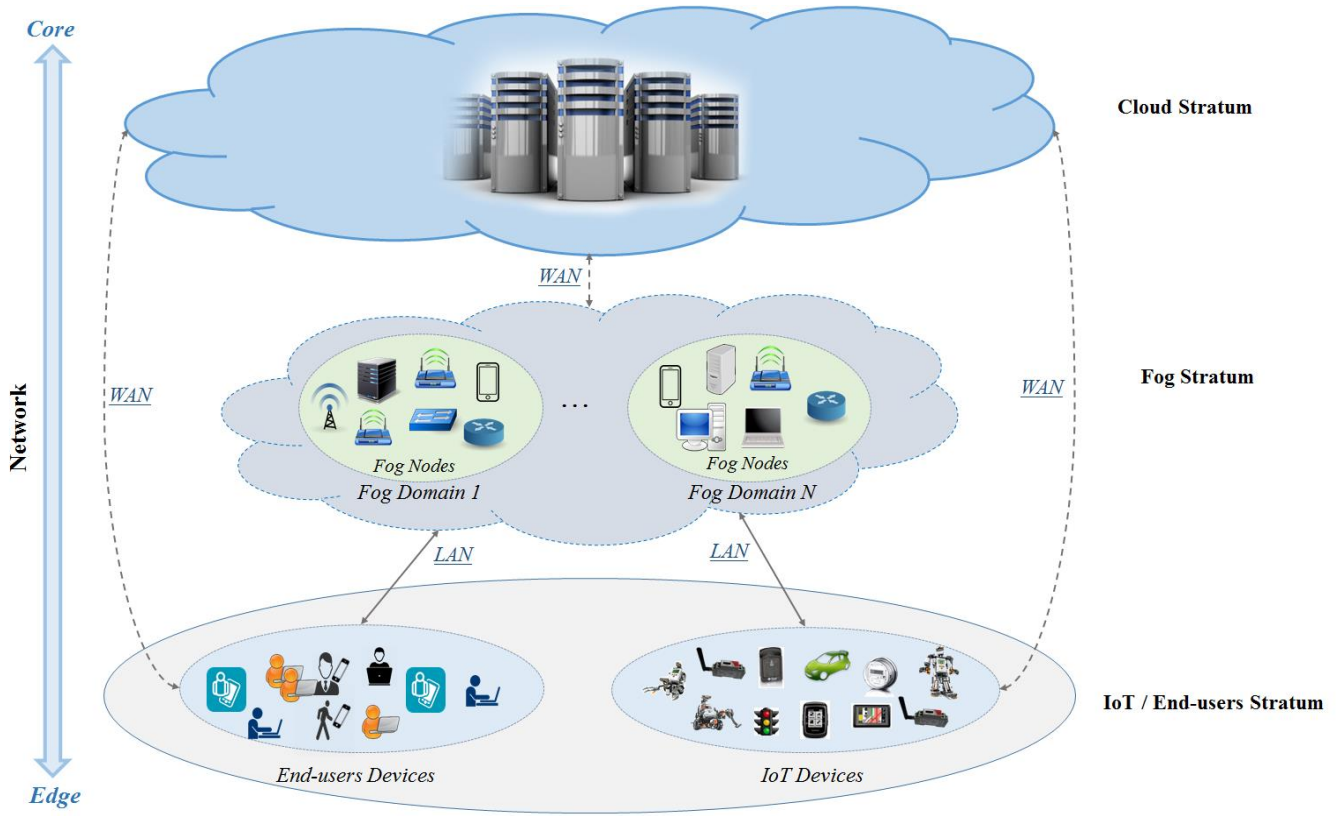

Fig. 4. The Fog System

Using cloudlets, resource-poor mobile devices offload their intensive computations (e.g., face recognition) to the cloudlets they use, thereby guaranteeing real-time interactive responses. If the mobile device moves away from the cloudlet, it may connect to a distant cloud, hence providing a degraded service. Although cloudlets represent the middle tier of a three-tier hierarchy (i.e., mobile device – cloudlet – cloud), in the current definition of cloudlets, there is no particular focus on the interactions with the cloud. Cloudlets can also act as a full cloud on the edge. Even when totally isolated from the cloud, they can exist as a standalone environment, since VM provisioning of the cloudlets is done without cloud intervention.

### C. Multi-Access Edge Computing (MEC)

Multi-Access Edge Computing is an industry initiative under the auspices of the European Telecommunication Standards Institute (ETSI). It was initiated in 2014 under the name of Mobile Edge Computing (MEC), with a focus on mobile networks and VM as virtualization technology. However, its scope has been expanded in March 2017 to encompass non-mobile network requirements (thus the replacement of "Mobile" by "Multi-Access" in the name), as well as virtualization technologies other than VM.

Prior to the scope expansion, the concept (envisioned as a key technology towards 5G) [34] aimed at providing cloud computing capabilities at the edge of mobile networks, and within the Radio Access Network (RAN). These capabilities are provided by mobile edge computing servers which can be deployed at LTE macro base stations (eNodeB) sites, 3G Radio Network Controller (RNC) sites, and at multi-Radio Access Technology (RAT) sites. The envisioned applications include augmented reality, intelligent video acceleration, and connected cars. The edges of non-mobile networks and related applications will certainly now be considered due to the new scope.

The functional entities [41] (before the scope expansion) include the mobile edge platform and the mobile edge host. The mobile edge platform provides the functionality required to provision mobile edge applications on a specific virtualization infrastructure while the mobile edge host anchors the platform and the virtualization infrastructure. Other functional entities

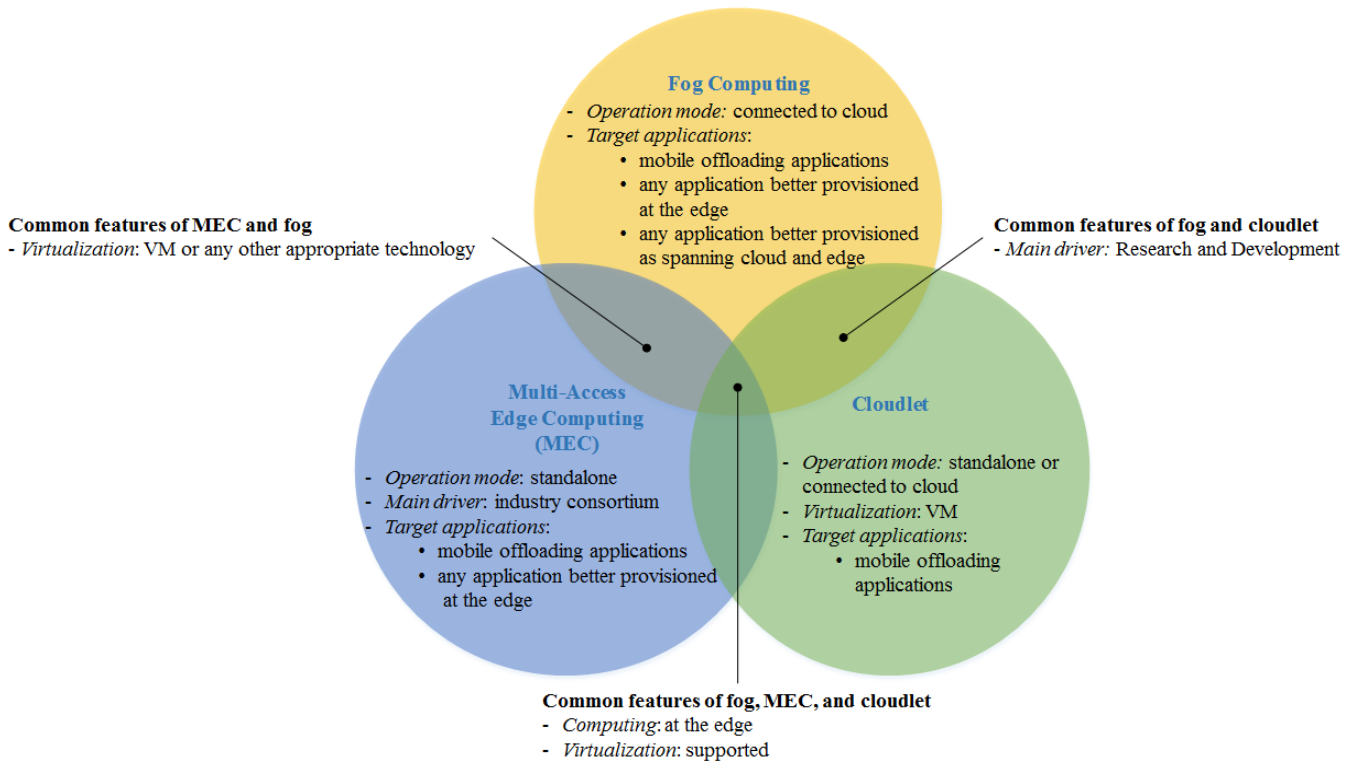

Fig. 5. A Venn diagram for the Relationship between Fog Computing, Cloudlet, and MEC

might now be added to the architecture as a result of the scope expansion.

A fundamental goal assigned to the ETSI initiative is to standardize the APIs between the mobile edge platform and the applications in order to foster innovation in an open environment. Several APIs have already been standardized (e.g., Mobile application enablement API [42], radio network API [43], Location API [44]). Reference [45] provides a survey of MEC.

### D. Fog Computing

Fog computing, a concept introduced by CISCO in 2012, is an extension of cloud computing paradigm from the core to the edge of the network. It enables computing at the edge of the network, closer to IoT and/or the end-user devices. It also supports virtualization. However, unlike cloudlet and MEC, fog is tightly linked to the existence of a cloud, i.e., it cannot operate in a standalone mode. This has driven a particular attention on the interactions between the fog and the cloud [11]. Moreover, fog has an n-tier architecture, offering more flexibility to the system [13][16].

Fig. 4 shows a fog system with a three-tier architecture. It has three strata: The cloud stratum, the fog stratum, and the IoT/end-users stratum. The fog stratum can be formed by one or more fog domains, controlled by the same or different providers. Each of these fog domains is formed by the fog nodes that can include edge routers, switches, gateways, access points, PCs, smartphones, set-top boxes, etc. As for the IoT/end-users stratum, it is formed in turn by two domains, the first including end-user devices and the second including IoT devices. It should be noted that one of these two domains may be absent in the stratum. It is, for instance, the case of fog systems based – content delivery. There is no IoT domain. The communication between the IoT/end-users stratum and the fog stratum is done through Local Area Network (LAN). Instead, the communication between the IoT/end-users stratum and the cloud stratum requires connection over the Wide Area Network (WAN), through the fog or not. There are several visual presentations for fog in the literature Most of them, however, focus on IoT only (e.g., [4][16][46]) and exclude end-users from the bottom stratum. This precludes non-IoT applications such as Content Delivery Networks (CDNs). The possibility of having several fog domains is not presented in any of them, either.

### E. Discussion of The Similarities and The Differences

This discussion focuses on cloudlets, MEC, and fog, seeing that cyber foraging has now been superseded. Although cloudlets, MEC, and fogs aim at computing at the edge and rely on virtualization, there are few subtle differences that need to be pinpointed.

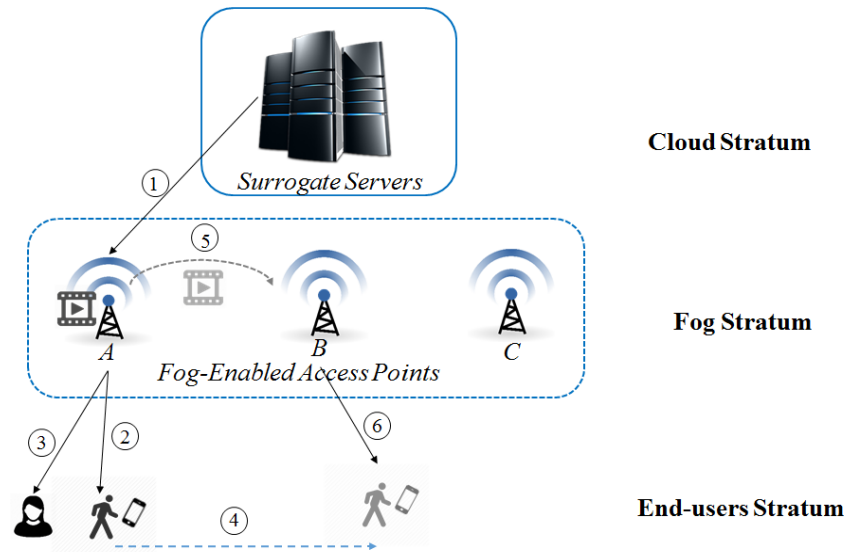

Fig. 6. A Fog System Use Case for CDN

A first difference is that it is only MEC which is mainly driven by an industry consortium, ETSI. Cloudlet and fog are more driven by R&D. The OpenFog consortium, for instance, does aim at producing standard specifications for fog computing. However, it is still at an early stage, has not yet gathered full speed as evidenced by the very few specifications it has produced so far. Yet another difference is that cloudlet relies solely on VM technology for virtualization, while MEC and fog do consider virtualization technologies other than VM. A third difference is that MEC functions only in stand-alone mode. There has been no work done so far on how it could interact with a distant cloud. Cloudlets can function in either stand-alone mode or connected to a cloud, although there is almost no work on how they can interact with the cloud. On the other hand, fog is designed as an extension to the cloud.

A fourth difference is the application(s) targeted by these concepts. Cloudlets focus solely on mobile offloading application, while MEC aims at any application that is better provisioned at (mobile and non-mobile) edges (including mobile offloading). Fog computing goes far beyond this, it enables, in addition, applications that can span cloud and edge. Actually, with fog, applications can be fully provisioned in either cloud or fog and can eventually span over both, as illustrated by the firefighting application presented in reference [5]. Fig. 5 provides a Venn diagram that summarizes the similarities and differences between cloudlets, MEC, and fog computing.

### III. ILLUSTRATIVE USE CASES AND FOG SYSTEM EVALUATION CRITERIA

This section introduces illustrative use cases that highlight the benefits of fog computing. This is followed by the listing of the evaluation criteria derived from the use cases.

#### A. Illustrative Use Cases

Fog computing holds promising capabilities for a variety of applications. This potential has been unveiled through several use cases [11][4][47][15] in the context of cyber-physical systems, smart grid, micro-grid decentralized smart grid, smart traffic light and connected vehicle, and decentralized smart building control. This paper discusses two use cases in details. The first is on CDN and the second is on fire detection and fighting application. The CDN use case is built on the use case presented in reference [6].

It should be noted that all fog use cases (including the ones presented in this paper and more generally in the literature) are so far hypothetical although some of them have been prototyped. Fog computing is an emerging paradigm at a very early stage of deployment. To the best of our knowledge, no description of real-world use case/implementation is available in the public domain. However, there are commercial fog devices such as the CISCO IOx[1].

##### 1) A Fog System Use Case for CDN

A CDN aims at delivering content (e.g., video) to end-users in a cost-efficient manner and with the required quality of experience (QoE). A CDN [48] consists of origin servers, surrogate servers (also known as replica servers), and a controller. The origin servers store the original content and this content is replicated on the surrogate server(s). For each end-user request, the controller selects the most appropriate surrogate server using certain criteria such as physical distance, network conditions, content availability, and delivery costs and then redirects the requester to the selected server. It should be noted that the content is usually served to end-users via access points such as cellular base stations.

---

[1] http://cisco.com/c/en/us/products/cloud-systems-management/iox/

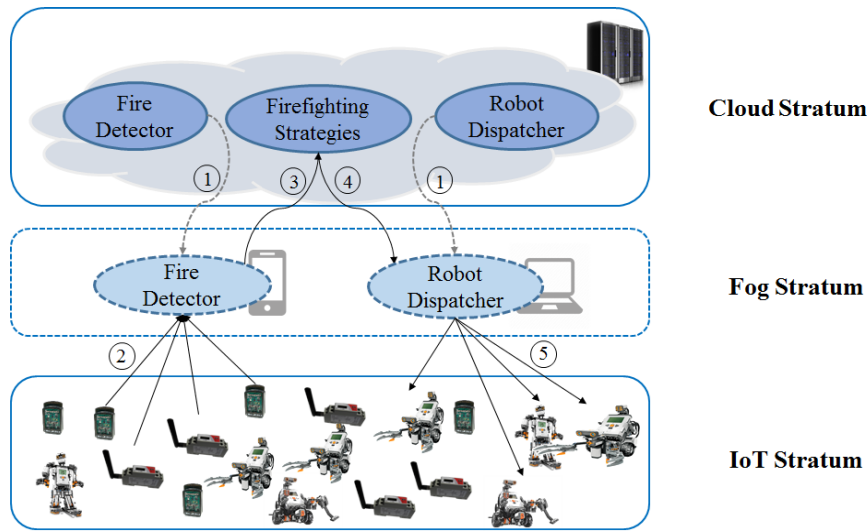

Fig. 7. A Fog System Use Case for Fire Detection and Fighting

Cloud-based CDNs (Cloud-CDN, CCDN for short) [49] have recently emerged to bring to the CDN world the benefits inherent to cloud computing (e.g., scalability and elasticity). Some examples of CCDNs are Amazon CloudFront[2], CloudFlare[3], and Rackspace[4]. In CCDNs, the CDN components are located in the cloud. The cloud-based surrogate server that serves a given end-user might, therefore, be located too far from the end-user. In the case of video streaming, for instance, this may degrade the QoE through initial delay and stall. A potential solution is to have fog stratum between the cloud and the end-user. Fig. 3 shows a fog system for CDN in which the fog stratum could be made of fog-enabled access points, and popular videos could be cached on these access points in a pro-active and/or reactive mode.

In this use case, the video John wishes to access is cached in a proactive mode, meaning that the surrogate server pushes the video to access point A (Fig. 6, action 1). John then accesses it from the access point (Fig. 6, action 2). The initial delay will certainly be shorter and the feedback information sent to the fog by John's device is less likely to be stale compared to the case where the video is streamed from the cloud. Mary who subsequently connects to the same access point and sees the same video will benefit from the same improved QoE (Fig. 6, action 3). Let us now assume that John moves to access point B, considered to be closer to it than to the cloud (Fig. 6, action 4). Access point A will replicate the video on access point B in a reactive mode, just in case the video has not yet been placed there in a pro-active mode by the surrogate server (Fig. 6, action 5). This reduces the delay compared to the case where it is the surrogate server that replicates the video on access point B in a reactive mode. John will now be served by access point B (Fig. 6, action 6).

*2) A Fog System Use Case for Fire Detection and Fighting*

A fire detection and fighting application is introduced in reference [5]. It monitors geographic areas (e.g., city, forest) and dispatches fleets of robots when fire is detected. The monitoring is done through the gathering of information such as wind speed, moisture, and temperature. When fire is detected, the application evaluates its intensity and contour and dispatches the most appropriate robots to extinguish it.

The application is made up of three components: *Fire Detector*, *Firefighting Strategies*, and *Robots Dispatcher*. These three components could be located in a cloud that might be geographically far from IoT devices (i.e., the sensors and the robots). This may cause transmission delays between the IoT devices and the application components. Chances to detect and fight fire in a timely manner might be compromised.

Fig. 7 depicts a potential fog system approach to the problem. The *Fire Detector* and/or the *Robot Dispatcher* can be moved from the cloud stratum to the fog stratum (Fig. 7, action 1). Moving either of both of these components to the fog stratum helps to achieve tight end-to-end latency constraint. For example, locating the *Fire Detector* at the fog stratum allows it to collect real-time data from the sensors and detect if there is fire (Fig. 7, action 2). This detection can be done in a timely manner, seeing that the fog stratum is closer to the IoT devices.

---

[2] https://aws.amazon.com/cloudfront/
[3] https://www.cloudflare.com/cdn/
[4] https://rackspace.com/cloud/cdn-content-delivery-network

TABLE I
Summary of the Evaluation Criteria

| Criteria | Definition | Architectural Dimension | Algorithmic Dimension |
|---|---|---|---|
| **Heterogeneity $C_1$** | Nodes in fog stratum and nodes in cloud stratum are very heterogeneous in terms of computational and storage capabilities. The fog system should be able to cope with this heterogeneity. | The need to take the heterogeneity into account when deciding which application component(s) should be deployed and where. | The need to factor the limitations of specific nodes in the models and operations of the algorithms. |
| **QoS Management $C_2$** | Fog system is a promising enabler for real-time applications due to the proximity of fog nodes to IoT/end-user devices. However, the latency varies greatly depending on where the application components are located. The QoS should be managed. | The need for architectural modules for QoS management such as migration engine. | The need for QoS management algorithms that ensure QoS of each application is met. |
| **Scalability $C_3$** | Fog systems are expected to cover millions of IoT/end-user devices. Moreover, they may encompass a large number of applications, fog domains, and fog nodes. Some applications may also have a large number of components. Fog systems need to be operational at such large scales. They should scale down and up in an elastic manner. | The need for architectural modules to ensure this scalability such as elasticity engine. | The need for algorithms to make the actual scaling decisions (scale up, down, in, or out). The need for algorithms to remain operational over large scales. |
| **Mobility $C_4$** | IoT/end-user devices and fog nodes can be mobile. Fog system should be able to handle this mobility. | The need for architectural modules such as a mobility engine to ensure the continuity of a service for the end-user. | The need for algorithms to manage this mobility. |
| **Federation $C_5$** | Fog has geographically distributed deployment in a wide scale, where each fog domain may be owned by a different provider. Moreover, cloud can be operated by a different provider. Provisioning applications requires the federation of these different providers, which might host the different components making the applications. | The need for cooperation among different providers in order to ensure the proper coordination of the necessary interactions between application components. | The need for cooperation algorithms at each provider's side responsible for outsourcing and insourcing decisions. |
| **Interoperability $C_6$** | As part of a federated system, an application can be executed with its components spread over different providers. Fog system should be interoperable at the level of providers and architectural modules. | The need for appropriate signaling and control interfaces, and appropriate data interfaces to enable this interoperability. | ---- |

In case of fire, the *Fire Detector* notifies the *Firefighting Strategies* (Fig. 7, action 3) located in the cloud stratum to implement the firefighting procedures. The *Firefighting Strategies* as a component communicates with the *Robots Dispatcher* (Fig. 7, action 4) that dispatches the robots in order to extinguish the fire (Fig. 7, action 5). As shown in reference [5], based on concrete measurements, the fog system approach reduces end-to-end delays in a significant manner.

*B. Evaluation Criteria*

We propose here a set of criteria to evaluate the work done so far on fog systems (i.e., architectural modules/interfaces and algorithms). The pertinence of these criteria is illustrated in the previously discussed fog system use cases. It should be noted that similar evaluation criteria are sketched in the literature (e.g., [50] and [11]) but not discussed in depth. All proposed criteria except the very last have both architectural and algorithmic dimensions. As the algorithmic dimension is concerned, we consider that a given system will meet a given criterion if the algorithm either has the criterion as the main objective or the criterion is part of the set of constraints the algorithm should satisfy. We further discuss these aspects as we present each criterion.

The first criterion ($C_1$) is the need to support heterogeneity in resources. Nodes in fog stratum and nodes in cloud stratum are very heterogeneous in terms of computational and storage capabilities. Fog nodes are likely to have limited capabilities compared to their counterparts in the cloud. There might also be significant differences between nodes in different fog domains and even between nodes in the same fog domain. In the CDN use case, for instance, the fog-enabled access points are certainly less powerful than the surrogate server in the cloud. It is critical for the fog system to be able to cope with this heterogeneity. In the architecture, this heterogeneity needs to be taken into account when deciding which application component(s) should be deployed and where. Algorithms also need to take this heterogeneity into account. The limitations of specific nodes need to be factored in the models and operations of the algorithms. In the very same CDN use case, a caching algorithm should consider the storage limitations of the potential caches.

The second criterion ($C_2$) is the need to meet the QoS required for each application deployed in the fog system. Fog system is envisioned as a promising enabler for real-time applications due to the proximity of fog nodes to IoT/end-user devices. This proximity reduces latency. However, this latency varies greatly depending on where the application components are located as shown in reference [5] for fire detection and fighting. Architectural modules for QoS management are therefore needed. An example of such modules is the migration engine that will move the content from the cloud stratum to the fog stratum, and also from a fog-enabled access point to another fog-enabled access point to ensure QoS in the CDN use case. In the fire detection and fighting use case, it is also the same module that will move the *Firefighting Strategy* module from the cloud stratum to the fog stratum or vice versa depending on the QoS requirements on robot dispatching. QoS management algorithms are needed as well. They need to enable decisions in the system, while still ensuring the QoS of each application is met. Their decisions will be executed by architectural modules. In the CDN use case, a migration algorithm triggering migration decisions of the migration engine is required. Another example is that of an application component placement algorithm in the case of the fire detection and fighting use case: When making the decision for placing the *Fire Detector* component, the algorithm should make sure the maximum latency threshold is not exceeded.

The third criterion ($C_3$) is the need for elastic scalability. Fog systems are expected to cover millions of IoT/end-user devices. Moreover, they may encompass a large number of applications, fog domains, and fog nodes. Some applications may also have a large number of components. In the fire detection and fighting use case, there may be thousands of sensors and thousands of robots. When it comes to the CDN use case, there may be millions of end-users and videos and hundreds or even thousands of fog-enabled access points. Accordingly, fog systems need to be operational at such large scales. They should scale down and up in an elastic manner. Architectural modules are needed to ensure this scalability. An example is an elasticity engine that allocates resources in terms of virtual machines for instance as the number of end-users' devices and applications grows or shrinks. Algorithms are also needed to manage this scalability. They will make the actual decisions (scale up, down, in, or out) carried out by the elasticity manager. Furthermore, all architectural modules and algorithms in the system should scale, i.e., remain operational over large scales. Back to the CDN use case, a content placement algorithm should remain efficient even in the presence of a flash-crowd event, i.e., the presence of a very large number of end-users requesting a video content. When it comes to the *Fire Detector* component, it should remain operational when the number of sensors increases significantly.

The fourth criterion ($C_4$) is the need to support mobility. IoT/end-user devices and fog nodes can be mobile. Accordingly, the system should be able to handle this mobility. Back to the CDN use case, the end-user watching a video content may roam from a source fog-enabled access point to a destination fog-enabled access point. The system should be able to provide the end-user with the same video from where it was left without interrupting the service. Architectural modules such as a mobility engine are needed to ensure this. The mobility engine would execute decisions by a mobility handling algorithm. If there are several end-users watching the same video, the mobility handling algorithm might require the mobility engine to duplicate the video and push a copy to the destination fog- enabled access point. In the case of a fog node's mobility, resource displacement takes place, with implications on resource management algorithms. In the fire detection and fighting use case, a *Fire Detector* component may be running in a target fog domain, over a specific mobile device. In case the latter moves, a resource management algorithm, e.g., a task scheduling algorithm, may decide to reschedule the component over another device in the same target fog domain.

The fifth criterion ($C_5$) is the need for federation. The fog stratum has geographically distributed deployment on a wide scale, where each fog domain may be owned by a different provider. Moreover, the cloud stratum can be operated by a different provider. Accordingly, provisioning applications requires the federation of these different providers, which might host the different components making the applications. This implies cooperation among these providers in order to ensure the proper coordination of the necessary interactions between application components. Back to the fire detection and fighting use case, from an architectural perspective, this federation allows the provisioning of the *Fire Detector* and the *Robot Dispatcher* components in different fog domains owned by different providers. It also allows for the *Firefighting Strategy* in the cloud stratum owned by a third provider, while ensuring a seamless provision of the application. Indeed, the federation among providers grants a higher degree of flexibility in the system. However, it also implies challenges for individual fog providers inside a single federation. For instance, each fog provider needs to make outsourcing decisions, towards other fog providers in the same federation, i.e., decisions to whether execute its own computations over a different fog provider's resources or not. The same holds for insourcing decisions, i.e., to enable a fog provider to execute computations from another fog provider in the same federation. Thus, cooperation algorithms at each provider's side are needed, responsible for such decisions, according to the availability of resources and the corresponding costs. Back to the fire detection and fighting use case, a fog provider may consider outsourcing decisions for running the *Fire Detector* and the *Robot Dispatcher* components over the resources of another fog provider, in case this allows it to reduce its overall costs.

The sixth criterion ($C_6$) is the need for interoperability. As part of a federated system, an application can be executed with its components spread over different providers. From an architectural perspective, this implies the need for appropriate signaling and control interfaces, as well as appropriate data interfaces to enable interoperability at the level of providers and architectural modules. More precisely, control interfaces are needed to enable interactions between the different involved domains to support the application's lifecycle. Meanwhile, data interfaces are needed between different components of the same application deployed in the cloud and the fog strata. Back to the fire detection and fighting use case, control interfaces will enable the deployment of *Fire Detector* component in the fog stratum, for instance, and the *Firefighting Strategies* in the cloud stratum. In this case, the data interfaces will be used by the *Fire Detector* component in the fog stratum to send the sensed data to the *Firefighting Strategies* component deployed in the cloud stratum.

### IV. ARCHITECTURES FOR THE FOG SYSTEM

Similar to other large-scale distributed computing systems, the architectures proposed so far for fog systems are either application agnostic or application specific. This section is structured accordingly. The applications agnostic architectures focus on specific architectural facets although they target end-user applications at large. The following facets have attracted the interest of the researchers: end-user application provisioning, resource management, communication issues, and cloud and fog federation. On the other hand, most of the application-specific architectures focus on healthcare. In addition, there are architectures proposed for connected vehicles and smart living. A few other application areas have also been considered. Table II and Table III provide a summary of the main features of the papers reviewed in this section. For each paper, we outline its scope, the approach that is followed, the evaluation methodology, its major contribution, as well as the criteria it meets. The first subsection presents application-agnostic architectures and the second subsection is devoted to the application-specific architectures.

#### A. End-User Application Agnostic Architectures for The Fog systems

##### 1) End-User Application Provisioning Architectures

Two programming architectures are proposed. In addition, a more general architecture is proposed. The concept of application lifecycle is used in this paper to review these proposals. The lifecycle of the end-user applications is made up of three main phases: Development, deployment, and management (including execution) [51].

The first programming architecture is called mobile-fog and is proposed by Hong *et al.* [52]. It allows developers to write programs for specific nodes in the IoT/end-users, the fog, and the cloud strata. The deployment can be done in any of the strata and the same code can be deployed on several different nodes that belong to any of the strata. Once the code is written, the developer compiles it and generates the mobile-fog process image that can be deployed on these nodes with an associated unique identifier. Mobile-fog allows the management of applications distributed over the IoT/end-users, the fog, and the cloud strata. During the execution, the application retrieves information about the underlying node resources, capabilities, location, and the stratum it belongs to. Then, based on the retrieved information, it decides the set of instructions that should be run. As an illustration, the *query_capability (type t)* instruction retrieves a specific set of capabilities such as sensing and actuation functionality. A descriptor of the capability set is returned if the node has those capabilities and the appropriate sensing or actuation instruction is executed. Mobile-fog also provides a set of control interfaces that allows managing the applications using the identifiers of the generated mobile-fog process. It also offers communication APIs that enable the interaction between the distributed application components. Moreover, during the management phase, it creates on-demand instances when a computing instance is overloaded.

The heterogeneity criterion ($C_1$) is obviously met since mobile-fog offers an API to query the underlying capabilities of nodes and react accordingly. The performed simulation has demonstrated that the use of mobile-fog reduces the end-to-end latency of end-user applications significantly, compared to the pure cloud-based approach. However, the related latency may fluctuate due to the absence of the QoS management module. The QoS ($C_2$) criterion is therefore not met. Mobile-fog can handle dynamic workloads and scale. So, the scalability criterion ($C_3$) is met. In addition, mobile-fog includes prediction models for the future locations of mobile end-users. Accordingly, it can move the application or the application component(s) to the new location and start an early processing of events, such that the required service is available once the end-user arrives at the future location. Hence, the work meets the mobility criterion ($C_4$). However, coordinating the execution between the distributed nodes is not discussed, leaving the federation criterion ($C_5$) unmet. Finally, mobile-fog provides control and data interfaces as discussed in the management phase. Consequently, the interoperability criterion ($C_6$) is met.

The second programming architecture deals with the distributed data flow (DDF) and is proposed by Giang *et al.* [53]. The essence is that the application topology is expressed as a directed graph (flow) consisting of nodes with each node corresponding to an application component. DDF provides a flexible way to develop the end-user applications that span the cloud and the fog strata. Two types of developers are considered: Component developers and IoT application developers. Component developers are responsible for developing components that ensure the communication with things. IoT application developers are responsible for the flow of components and creating scripting components. Scripting components are responsible for implementing new protocols or functionalities that do not exist in the current system. Components can be deployed in both the cloud and the fog strata. In the fog stratum, the authors have classified nodes into three types: Edge, IO, and Compute nodes, each with different computational capabilities. The components are deployed on the nodes with the appropriate capabilities. For instance, some components are constrained to run only on compute nodes. Moreover, the application components in the flow are deployed on more than one node in order to handle the movement of devices. For the management phase, DDF allows managing the applications with the components distributed on both the cloud and the fog strata. The authors have considered that each cloud or fog node may execute one or more components in the flow. These nodes include modules responsible for executing the application according to the topology in the flow's design. They are also responsible for allowing the communication with other participating nodes. These nodes also include modules responsible for deciding when to scale up.

The proposed programming model allows developing and deploying an application for heterogeneous nodes based on the classification of the hosting nodes according to their capabilities. Consequently, it meets the heterogeneity criterion ($C_1$). The proposed model takes into consideration the application topology and the latency requirement when deploying application components on the cloud and the fog nodes. Accordingly, the QoS criterion ($C_2$) is met. The scalability criterion ($C_3$) of the proposed model in terms of the fog nodes is met thanks to the modules responsible for scaling. However, it should be noted that the scalability in terms of the IoT devices and/or the fog domains is not considered. Moreover, the mobility criterion ($C_4$) in terms of the IoT devices mobility is met. This is achieved by duplicating the components in the flow and deploying them in a possible location where the IoT devices might move. However, it should be noted that the fog nodes mobility is not considered. Also, coordinating the execution of the components is static as developers wire the flow of the components manually. There is no orchestration support when the application is executed in a distributed manner. So, the federation ($C_5$) criterion is not met. Finally, in the proposed model, the data and control interfaces are provided. The components in different strata can communicate and exchange data. If they do not support the same interface, new special-purpose components can be developed and then deployed to provide such interfaces. Accordingly, the interoperability criterion ($C_6$) is met.

Yangui *et al.* [5] propose a more general architecture compared to Hong *et al.* [52] and Giang *et al.* [53] that is based on two principles: Designing the architecture as an extension to the existing PaaS and using the REST paradigm for the interactions. The authors validate their architecture by using a fire detection and firefighting application. In the IoT/ end user stratum, the proposed architecture uses temperature sensors and robots as firefighters. In the cloud stratum, the authors propose a layer-based architecture consisting of four layers: Application development layer, application deployment layer, application hosting and execution layer, and application management layer. For development, there is an extended Integration Development Environment (*IDE*) module. It allows the composition of components and the communication with the IoT devices. For deployment, there is a *Deployer* module, responsible for placing the applications either in the cloud or the fog. The *Controller* sets the locations of the components. The *Fog Resources Repository* lists the descriptors that detail the capabilities and the specificities of all the involved nodes as part of the cloud and the fog strata. *Orchestrator* is the other module that is responsible for orchestrating the execution flow between the application components spanned across the cloud and the fog strata. In addition to these modules, other modules of the management phase are included. Examples are the *SLA Manager,* responsible for managing the application's QoS, the *Elasticity Engine,* responsible for scaling up/down the application components, and the *Migration Engine,* responsible for migrating the components from the cloud to the fog and vice versa or from one fog node to another. In addition, the architecture is composed of a set of appropriate REST-based interfaces that enable the communication between the cloud and the fog. Control, data, operation, and management interfaces are designed.

The heterogeneity criterion ($C_1$) is not met since the authors do not provide details about the description models in the *Fog Resources Repository* that specifies the capabilities of the involved cloud and fog nodes. The *SLA Manager* module allows meeting the QoS criterion ($C_2$). Moreover, the scalability criterion ($C_3$) is met as a result of the *Elasticity Engine*. The mobility criterion ($C_4$) is not met. The authors do not provide details about the mobile fog nodes. However, their architecture supports the mobile IoT devices by using appropriate gateways. For instance, they use mobile Lego robots as firefighters when implementing the validation use case. In addition, the *Orchestrator* module allows meeting the federation criterion ($C_5$). Finally, the application components deployed in different strata or different domains in the fog stratum can interact and exchange messages through the data and control the interfaces the authors have proposed. Consequently, the interoperability criterion ($C_6$) is met.

All the architectures reviewed in this subsection support the three phases of the lifecycle. They all provide appropriate signaling and control interfaces ($C_6$); however, only Yangui *et al.* [5] provide a module that enables the federation ($C_5$) between different cloud and fog providers. Moreover, the scalability is ensured in the three of them.

*2) Architectures for Resource Management in The Fog Systems*

When the cloud stratum, the fog stratum, and the IoT/end-users stratum are integrated into a fog system, resource management becomes a critical issue. Resource migration, allocation, and scheduling are the topics addressed by researchers so far.

Bittencourt *et al.* [54] have worked on resource migration, by focusing on VM migration between the fog nodes. The goal is to keep VM available when the user moves. They assume that VM contains the user's data and application component(s). The migration is done in a way that users do not notice any degradation in their application's performance. The authors propose a layer-based architecture. For the IoT/end-users stratum, the authors propose the mobile devices layer. It includes several modules. The *Application Migration* module supports the VM migration decision-making. The *Processing Location/Offloading* allows the partitioning of the application into components and deploying these components in one of the three strata. This module allows the application developer to decide where to deploy the components based on the QoS parameters, the node's processing and the storage capacity, or the network delays between the cloud and the fog. For the fog stratum, the authors propose the cloudlets layer. This layer also includes several modules. The *QoS Monitoring* is responsible for checking the QoS requirements by considering different metrics (e.g., bandwidth and latency). The *Mobility Behaviour and Handoff Analysis* is responsible for monitoring the user and deciding about the time and the location to perform VM migration. The *VM/Container Migration* is responsible for performing the migration according to the decisions of the *Mobility Behavior* module. And, the cloud stratum includes the cloud layer including modules responsible for functionalities such as load balancing.

The heterogeneity criterion ($C_1$) is met thanks to the *Processing Location/Offloading* module. The latter allows the developer to deploy components when matching between the component's requirement and the hosting node's capabilities. The QoS criterion ($C_2$) is met as a result of the *Processing Location/Offloading* module and the *QoS Monitoring* module. The scalability of the architecture in terms of the supported number of mobile devices, the fog nodes, and/or the fog domains is not discussed. So, the scalability criterion ($C_3$) is not met. The mobility criterion ($C_4$) is met thanks to the *Mobility Behaviour and Handoff Analysis* module that detects the user's movement and informs the *VM/Container Migration* module to perform the VM/container migration. Coordinating the execution between the distributed components is not discussed, the federation criterion ($C_5$) is then not met. Moreover, the authors do not discuss any common control or data interfaces needed for the interaction between the cloud and the fog strata. Consequently, the interoperability criterion ($C_6$) is not met.

Agarwal *et al.* [25] focus on resource allocation. The proposed architecture implements an algorithm that distributes the workload between the cloud and the fog strata. The proposed architecture is layer-based, consisting of three layers. In the IoT/end-users stratum, the authors propose the client layer, consisting of clients, mobile devices, and sensors. In the fog stratum, the authors propose the fog layer that involves a module called *fog server manager*. This module is responsible for checking whether enough computational resources are available to the host components. Based on this availability, it either executes all the components or executes some while postpones the execution of others, or even dispatch some of the components to the node(s) in the cloud stratum. The *fog server manager* is also responsible for the VMs' lifecycle management. For the cloud stratum, the authors propose the cloud layer that includes modules responsible for executing the components received from the *fog server manager*.

In this paper, the authors take the heterogeneity of the hosting nodes into consideration, with the *fog server manager* module responsible for matching between the node's capabilities and the components requirement. Accordingly, the heterogeneity criterion ($C_1$) is met. However, there is no architectural module in the proposed architecture for QoS management. Besides, the scalability of the proposed architecture in terms of the supported number of end-users, the fog nodes, and/or the fog domains is not discussed. So, the QoS ($C_2$) and the scalability ($C_3$) criteria are not met. In addition, the authors do not discuss the user's or the fog node's mobility. So, the mobility ($C_4$) criterion is not met. The cooperation between the cloud and the fog providers to ensure the proper execution of distributed components is not discussed. Finally, the common control and data interfaces needed to allow the communication between the cloud and the fog and to manage the application lifecycle are not discussed, which leaves the federation ($C_5$) and the interoperability ($C_6$) criteria unmet.

Cardellini *et al.* in [55] focus on task scheduling. They propose an extension to an existing architecture called Storm[5], in order to execute a distributed QoS-aware scheduler. The scheduler is responsible for deploying the application components on the pool of available resources. Storm is an open source Data Stream Processing (DSP) system. This extension allows Storm to operate in geographically distributed and highly dynamic environments. The authors have added new modules to the architecture that allows executing a distributed QoS-aware scheduler. They have also included self-adaptation capabilities to the architecture. The proposed scheduling strategy deploys Storm-based applications close to data sources and consumers. An application in a Storm is represented by a flow called topology. The topology includes several entities such as spout, bolt, and task. Some of these entities are described below. These entities provide an abstraction of the underlying hosting nodes' capabilities. The IoT/end-users stratum consists of data sources/sensors generating a flow of data. Those data sources are called *Spout* in Storm topology. In the fog and the cloud strata, the components are called *Bolt*. Both the fog and the cloud nodes can be a Storm node. Whether a Storm node is a fog or a cloud node, the corresponding stratum includes several modules. The *QoSMonitor* module estimates the network latency and monitors the QoS attributes for the nodes in the cloud or the fog stratum. The *AdaptiveScheduler* module executes the distributed placement policy and takes the mobility of the hosting nodes into consideration. The *BootstrapScheduler* module is responsible for defining the initial assignment of the application components, monitoring its execution, and rescheduling the application in case of failure. Moreover, Storm includes a module called *Nimbus,* a centralized component responsible for coordinating the topology execution.

With the provided abstraction, the development of the components (i.e., *Spout* or *Bolt*) can be independent of the hosting node's capabilities. By that, the heterogeneity criterion ($C_1$) is met. The QoS criterion ($C_2$) is met thanks to the

---

[5] https://storm.apache.org/

*QoSMonitor* module. The proposed architecture can scale in terms of the number of applications and nodes since it does not need a global knowledge of the whole DSP system. So, the scalability criterion ($C_3$) is met in terms of the number of applications and nodes (whether in the cloud or the fog). The mobility criterion ($C_4$) is met thanks to the *AdaptiveScheduler* module. The federation criterion ($C_5$) between the cloud and the fog providers has met thanks to the *Nimbus* module. The nodes (whether on the cloud or the fog) that host the application components are interconnected by an overlay network. The overlay provides the logical links between the nodes so that they can communicate. Accordingly, the interoperability criterion ($C_6$) is met.

Kapsalis *et al.* [56] presents an architecture for fog-enabled platform that is responsible for allocating and managing the computational resources needed to host application components. It utilizes a distributed communication method based on publication/subscription pattern and MQTT (Message Queueing Telemetry Transport) protocol. The proposed architecture has four layers. In the IoT/end-users stratum, the authors propose the device layer and the hub layer. The device layer consists of constrained physical devices, such as sensors and actuators, and the hub layer consists of gateways. The gateways do not perform any computation. Their role is to convert the communication protocol of the devices in the device layer and send it over MQTT (when needed) to the upper layers. They act as mediators. In the fog stratum, authors propose the fog layer consisting of stable edges and mobile edges. The stable edge includes an architectural module called fog broker which is responsible for enriching the messages received from the lower layers. They are also responsible for task management and allocation. And the mobile edge is responsible for communicating with the devices in the lower layer. Finally, for the cloud stratum, authors propose the cloud layer which is responsible for long-term analysis and storage.

The fog broker performs workload balancing. It sorts the available hosts and chooses the most suitable one for each component based on its current utilization, latency, battery, and the required resources such that the heterogeneity of the hosting nodes are taken into consideration and the QoS is maintained. Accordingly, the heterogeneity ($C_1$) and the QoS criteria ($C_2$) are met. Large-scale experiments with thousands of components and hosting nodes are conducted. Accordingly, the scalability criterion ($C_3$) is met in terms of number of IoT/end-users devices and fog nodes. However, the mobility of the devices and/or the fog nodes is not taken into consideration. So, the mobility criterion ($C_4$) is not met. The subscribed nodes in the fog system use a specific message proposed by the authors to communicate, called fog message. It is based on MQTT and extends it further. It provides a unified model for the data and the resources. Hence, the federation criterion ($C_5$) is met. Finally, by using a standard M2M publish/subscribe protocol (i.e., MQTT) for the data interfaces, the authors provide data interfaces. They also extend the MQTT message to include some metadata. However, the control interfaces between the different strata for handling the application's lifecycle management are not discussed. So, the interoperability criterion ($C_6$) is not met.

All the architectures reviewed in this subsection take the nodes' heterogeneity ($C_1$) into consideration when managing the resources. However, none fully meets the scalability criterion ($C_3$). The architecture by Cardellini *et al.* [55] partly meets this criterion, i.e., in terms of the number of the cloud and the fog nodes but not in terms of IoT devices.

*3) Communication Architectures*

Intra-stratum and inter-stratum communications are required in a fog system. Research has already been carried out on both inter-stratum and intra-stratum communications. Shi *et al.* [57] deal with the inter-stratum communication between the IoT/end-users stratum and the fog stratum. Specifically, they use Constrained Application Protocol (CoAP) for communication. CoAP is a specialized web transfer protocol used for constrained nodes in IoT. The main focus of their work is on enabling the nodes in the fog stratum and devices in the IoT/end-users stratum to share services and capabilities in a seamless way by using REST and IoT CoAP protocol. For the IoT/end-users stratum, the authors have considered both low-energy and high-end sensors that belong to this stratum. These devices act as CoAP clients. However, for experimentation, the authors have used a machine with Mac OS to send requests to the CoAP servers. In the fog stratum, CoAP servers act as the fog nodes. This stratum includes application components that are responsible for performing basic pre-processing of the data generated by the sensors. The authors have developed their CoAP server using Erlang language and Raspberry Pi.

In this work, there is no discussion on the heterogeneity of nodes in the cloud and the fog strata. The authors have considered one type of fog node. So, the heterogeneity criterion ($C_1$) is not met. Moreover, the authors have evaluated the performance of their CoAP servers in terms of throughput and latency. The results demonstrate the efficiency of their approach. However, performing QoS management in terms of latency is not discussed. The scalability in terms of the supported number of the IoT devices and/or the fog nodes is not considered either. Accordingly, the QoS ($C_2$) and the scalability ($C_3$) criteria are not met. Moreover, the mobility of IoT devices that has a direct impact on the latency is not discussed. So, the mobility ($C_4$) criterion is not met. In this paper, the authors do not discuss the federation between the cloud and the fog providers. So, the federation criterion ($C_5$) is not met. The CoAP servers implementing the fog nodes expose a southbound REST-based API to the IoT/end-users stratum. The API operations cover both the control and data interfaces. However, the northbound side (i.e., cloud side) is not discussed. So, the interoperability criterion ($C_6$) is not met.

While Shi *et al.* [57] focus on the inter-stratum communication between the IoT/end-users stratum and the fog stratum, Slabicki *et al.* [58] focus on the intra-stratum communication between the devices in the IoT/end-users stratum. The authors consider three cases: Direct communication between devices, communication through the fog, and communication through the cloud. They analyze the transmission delay for data exchange between the devices for CoAP, SNMP, and NETCONF protocols in the mentioned three cases. In the IoT/end-users stratum, the authors have considered a simple data exchange between a sensor and an actuator. First, they implement a direct message over a CoAP, SNMP, or NETCONF protocol. Then, they transmit the data between the sensors and the actuator through the fog and the cloud strata respectively. The results demonstrate that the direct communication between the devices is the fastest, the transmission through the fog stratum is two times higher, and the transmission through the cloud stratum is more than two times higher than that of the previous one.

In this work, the authors do not discuss the heterogeneity of nodes in the cloud and the fog strata. For instance, the authors do not take into account the transmitter node's capabilities, such as required speed, processing delay, or network delay when messages are being routed. This is critical because the location and the capabilities of the transmitter nodes may influence the performances and/or the delays. So, the heterogeneity criterion ($C_1$) is not met. Moreover, the authors take into account QoS in terms of latency when the sensors and/or the actuators communicate through the cloud or the fog. Accordingly, the QoS criterion ($C_2$) is met. Furthermore, the scalability in terms of the supported number of the IoT devices, the fog nodes, and/or the fog domains is not discussed. In addition, the authors do not discuss the mobility of the devices in the IoT/end-users stratum and its effect on the obtained results (i.e., latency). So, the scalability ($C_3$) and the mobility ($C_4$) criteria are not met. The federation between the cloud and the fog is not discussed. The federation criterion ($C_5$) is then not met. Finally, authors do not provide any common interfaces between the cloud and the fog. By that, the interoperability ($C_6$) criterion is unmet.

Krishnan *et al.* [59] propose an architecture that allows the user device to decide whether a component should be executed in the cloud or the fog strata. They deal with inter-stratum communication between the fog and the cloud strata. In the IoT/end-users stratum, the packets are tagged before being sent. These tags include information about the destination (i.e., the cloud or the fog). In the fog stratum, when a fog node receives a packet, it decodes it and sends it to the respective node for processing. If the node is in the cloud, the fog node puts the source IP address as that of the data generator and the destination address as that of the cloud. It then forwards the packet to the cloud stratum for processing.

In this work, the authors do not discuss the heterogeneity of the hosting nodes. They do not provide any architectural module that manages QoS when deciding where to execute the application components, either. Consequently, the heterogeneity ($C_1$) and the QoS ($C_2$) criteria are not met. Besides, the scalability of the proposed architecture in terms of the supported number of users, the fog nodes, and/or the fog domains is not taken into consideration. The user's mobility is not considered as well. Accordingly, the scalability ($C_3$) and the mobility ($C_4$) criteria are not met. In addition, the mechanisms to coordinate the execution flow among the components hosted in the cloud and the fog strata are not provided. So, the federation criterion ($C_5$) is not met. Finally, the common control and data interfaces needed for the communication and connection between the cloud and the fog are not discussed. So, the interoperability ($C_6$) criterion is not met.

Aazam *et al.* [60] deal with inter-stratum communications. However, unlike Shi *et al.* [57] who focus on communications between the IoT/end-users stratum and the fog stratum, they focus on the fog stratum/the cloud stratum communications. They propose an architecture that attempts to reduce the number of packets sent to the cloud. This is done in order to lessen the burden on the cloud and to alleviate the communication overhead on the cloud. The authors propose a layer-based architecture consisting of six layers. For the IoT/end-users stratum, the authors have proposed the physical and virtualization layer. This stratum includes the physical nodes, WSN, virtual nodes, etc. In the fog stratum, smart gateways act as the fog nodes. For this stratum, the authors propose five layers that include several architectural modules and application components: The monitoring layer, the preprocessing layer, the temporary storage layer, the security layer, and the transport layer. The monitoring layer includes modules that monitor the activities of the nodes in the physical layer. The preprocessing layer is responsible for data management. It is basically responsible for reducing the number of packets sent to the cloud stratum. It first analyzes the collected data, then performs data filtering and trimming and, finally, it generates more meaningful and necessary data to send to the cloud. The temporary storage layer includes the components that are responsible for storing the data locally. The security layer includes a module that can encrypt/decrypt the data. Finally, the transport layer involves the modules that are responsible for uploading the ready-to-send data to the cloud. The cloud stratum includes the components that are responsible for storing the received data and provide it as a service to the users.

However, the heterogeneity of nodes in the cloud and the fog are not discussed in this paper. So, the heterogeneity criterion ($C_1$) is not met. Moreover, the authors do not discuss the QoS management in terms of latency when distributing the components across the cloud and the fog. So, the QoS criterion ($C_2$) is not met. The scalability of the architecture in terms of the IoT devices, the fog nodes, and/or the fog domains is not discussed. The authors do not discuss the mobility of the IoT devices and/or the fog nodes, either. Accordingly, the scalability ($C_3$) and the mobility ($C_4$) criteria are not met. In this work, the authors do not provide any model to enable the federation between the cloud and the fog providers. So, the federation criterion ($C_5$) is not met. The interoperability criterion ($C_6$) is met thanks to the transport layer.

Similar to Aazam *et al.* [60], Moreno-Vozmediano *et al.* [61] deal with inter-stratum communication between fog stratum/and cloud stratum. They proposed an architecture called Hybrid Fog and Cloud (HFC) Interconnection Framework to enable simple and efficient configuration of virtual networks to interconnect geographically distributed fog and cloud domains. The proposed architecture at the bottom includes mobile devices requesting applications. They constitute the IoT/end-users stratum. In the fog stratum, the fog nodes are implemented as micro-clouds and managed by fog management platform (similar to cloud management platform such as Openstack). The cloud stratum includes cloud nodes managed by cloud management platform. The proposed architecture includes an HFC manager through which tenants/end-users interact with the architecture. This manager provides abstraction and simplicity for tenants independently of the cloud and fog nodes. Each cloud and fog node includes an agent called HFC agent responsible for building HFC virtual network as an interconnection of different network segments deployed on different fog and cloud domains. The HFC virtual network is built through an L2 and L3 overlay network on top of the physical network.

In this work, the heterogeneity criterion ($C_1$) is met by the provided abstraction by the HFC manager. However, the authors do not take into account the application QoS when building the virtual network including cloud and fog nodes. Accordingly, the QoS criterion ($C_2$) is not met. The use of per-tenant HFC agents provides scalability in terms of fog and cloud nodes since each virtual network (that includes HFC agents as fog and cloud nodes) is managed independently from another tenant's network. So, the scalability criterion ($C_3$) is met. Moreover, the authors do not discuss the effect of the mobility of end-user devices on the proposed framework. Accordingly, the mobility criterion ($C_4$) is not met. The proposed HFC manager interacts with different cloud and fog management platforms and HFC agents. It is responsible for instantiating the different network segments on each fog and cloud node and coordinating and controlling the behavior of HFC agents. So, the federation criterion ($C_5$) is met. Besides, the nodes (whether on the cloud or the fog) are interconnected by an overlay network. The overlay provides the logical links between the nodes so that they can communicate. Accordingly, the interoperability criterion ($C_6$) is met.

Most of the papers reviewed in this subsection deal with inter-stratum communications except for Slabicki *et al.* [58] that focus on the intra-stratum communication between the devices in the IoT/end-users stratum. It can be concluded that all the architectures reviewed in this subsection do not meet most of the criteria.

*4) Architectures for The Cloud and The Fog Federation*

To the best of our knowledge, Zhankieev [62] is the only researcher who has so far tackled the federation issue by proposing an architecture called Cloud Visitation Platform (CVP). The proposed architecture enables the federation between the cloud and the fog. The participant cloud and fog providers need to install a CVP and register it with the federated cloud manager. This registration includes information about the nodes as part of the federation. The CVP can belong to the cloud or the fog stratum. It includes modules that are responsible for hiding the underlying hardware specificities. These modules create hardware awareness for VMs or container-based applications. The hardware awareness allows VMs to sense their local environment and adjust accordingly. This is done by APIs through which the cloud can provide details about its local hardware environment (e.g., RAM and storage). Moreover, it provides interfaces that are responsible for load balancing, queuing, etc.

In this work, the heterogeneity criterion ($C_1$) is met thanks to the modules creating hardware awareness. However, it should be noted that the QoS management when distributing application's components across the cloud and the fog is not discussed. Moreover, the scalability is discussed in terms of adding the cloud domains that belong to different providers and not in terms of the number of the fog nodes and/or the fog domains. Accordingly, the QoS ($C_2$) and the scalability ($C_3$) criteria are not met. Furthermore, the mobility of fog nodes is not taken into consideration. So, the mobility criterion ($C_4$) is not met. The goal of the proposed architecture is to provide the federation between the cloud and the fog providers. Providers can achieve this by installing the CVP platform. By that, the federation ($C_5$) criterion is met. Finally, the interoperability criterion ($C_6$) is met between the cloud and the fog nodes that installed CVP.

*B. Application-Specific Architectures for The Fog Systems*

*1) Architectures for Healthcare*

Applications for healthcare are latency-sensitive. They process patients' vital data (e.g., heart rate and glucose level) that are monitored by IoT devices (e.g., Body Area Network). Moreover, they send real-time notifications (e.g., heart attack alerts to family members). Consequently, researchers increasingly rely on the fog when designing such applications in order to address the latency drawback characteristics of the cloud. To the best of our knowledge, there has been only one architecture proposed for general healthcare. Meanwhile, there are several architectures proposed for healthcare applications but with a focus on specific health conditions. The specific health conditions considered so far are Chronic Obstructive Pulmonary Disease (COPD), Parkinson, speech disorders, and ECG and EEG feature extraction.

*a) Healthcare at Large*

The fog has major contributions in applications for home nursing services for elderly people. In this context, Stantchev *et al.* [63] propose an architecture. A process-oriented view of the healthcare application is provided as well. It is modeled by Business Process Modeling Notation (BPMN). The authors validated their architectural model by a use case for smart sensor-based healthcare infrastructure. The IoT/end-users stratum comprises of health sensors such as blood pressure gauge. The fog stratum comprises of gateways acting as the fog nodes. Application components in this stratum are responsible for short-term storage (e.g., informing the patient about his current glucose level). In the cloud stratum, components responsible for allowing permanent access and evaluation of the data are deployed. For instance, when the doctor receives the patient's data, he/she evaluates and decides whether medical intervention is necessary or not.

TABLE II
Main features and requirements for the works proposing architecture for Applications Agnostic Fog Systems. In the Evaluation column, P is prototype, S is simulation, and x means no evaluation is conducted. In the criteria columns, ✓ means the criterion is met and x means the criterion is not met.

| Scope | Paper | Evaluation | Performance metrics | Major Contribution | $C_1$ | $C_2$ | $C_3$ | $C_4$ | $C_5$ | $C_6$ |
|---|---|---|---|---|---|---|---|---|---|---|
| **Application Provisioning** | Hong et al. [52] | S | -Latency<br>-Network traffic<br>-Workload distribution | Propose a high-level programming model for IoT applications provisioning. | ✓ | x | ✓ | ✓ | x | ✓ |
| | Giang et al. [53] | P | n/a | Propose a DDF programming model for IoT applications provisioning, in which the application topology is expressed as directed graph consisting of nodes. | ✓ | ✓ | ✓ | ✓ | x | ✓ |
| | Yangui et al. [5] | P | -End to end delay variation | Propose a layer-based architecture for IoT application provisioning that spans cloud and fog. The architecture is based on two principles; extend the existing PaaS, use REST for interactions. | x | ✓ | ✓ | ✓ | ✓ | ✓ |
| **Resource Management** | Bittencourt et al. [54] | x | n/a | Propose a layer-based architecture that supports VM migration between fog nodes. | ✓ | ✓ | x | ✓ | x | x |
| | Agarwal et al. [25] | S | -Response time<br>-Data center Request Times and Processing Times<br>-Data center Loading and Total Cost | Propose an architecture for resource allocation. It includes an algorithm that distributes the workload between the cloud and fog. | ✓ | x | x | x | x | x |
| | Cardellini et al. [55] | P | -Average node utilization<br>-Application latency<br>-Inter-node traffic<br>-Average latency | Propose an extension to Storm in order to execute a distributed QoS-aware scheduler. | ✓ | ✓ | ✓ | ✓ | ✓ | ✓ |
| | Kapsalis et al. [56] | S | -Average execution time and delay<br>- Number of tasks<br>- number of failed tasks | Propose an architecture for resource management and load balancing between the cloud and the fog. | ✓ | ✓ | x | x | ✓ | x |
| **Communication Issues** | Shi et al. [57] | P | -Throughput<br>-Average round trip time<br>-Time out probability<br>-Average response time<br>-Relationship of number of packets and time | Study the inter-stratum communication between IoT/end-users stratum and fog stratum. | x | x | x | x | x | x |
| | Slabicki et al. [58] | P | Transmission delay for different communication protocols for:<br>-Direct communication<br>-Communication via fog<br>-Communication via cloud | Study the intra-stratum communication between devices in IoT/end-users stratum. | x | ✓ | x | x | x | x |
| | Krishnan et al. [59] | P | -Latency<br>-Internet Traffic | Propose an architecture composed of the fog and the cloud, and a method to move computation from the cloud to the fog. | x | x | x | x | x | x |
| | Aazam et al. [60] | P | -Data upload and synchronization delay in the cloud<br>-Jitter in the cloud | Propose an architecture for fog computing co-located within a gateway. | x | x | x | x | x | ✓ |
| | Moreno-Vozmediano et al. [61] | P | -Network Configuration Throughput | Propose an architecture HFC to interconnect geographically distributed cloud and fog domains. | ✓ | x | ✓ | x | ✓ | ✓ |
| **Cloud and Fog Federation** | Zhankieev et al [62] | x | n/a | Propose an architecture called CVP that enables federation between cloud and fog | ✓ | x | x | x | ✓ | ✓ |

In this paper, the authors discuss the heterogeneity of the IoT devices. However, the heterogeneity of the cloud and the fog nodes are not addressed, which affects the application components deployment. So, the heterogeneity criterion ($C_1$) is not met. Besides, the authors do not take into account the application QoS when the application components are distributed between the cloud and the fog. Accordingly, the QoS criterion ($C_2$) is not met. Moreover, the scalability in terms of the supported IoT devices, the fog nodes, and/or the fog domains is not taken into consideration and hence the scalability ($C_3$) criterion is not met. Although elderly people are equipped with wearable sensors to move, the authors do not provide any architectural module that handles this mobility. So, the mobility criterion ($C_4$) is not met. The authors do not provide any unified model between the cloud and the fog providers. The federation criterion ($C_5$) is then not met. Finally, they neither provide common data nor control interfaces that could enable the interoperability between the involved nodes. Therefore, the interoperability criterion ($C_6$) is not met.

*b) Healthcare with Focus on COPD*

Fratu *et al.* [64] present an architecture for an application that offers support for people affected by COPD and mild dementia. The architecture they propose relies on the fog in order to reduce the latency requirement, which is critical in case of emergency. It is based on the eWALL monitoring framework introduced in [65]. In the IoT/end-users stratum, different types of sensors in the patient's home are deployed, such as temperature sensors and infrared movement detectors. The fog stratum is responsible for real-time data processing and emergency case handling (e.g., when the patient's pulse or oxygen level is out of the normal range). This is done through appropriate application components deployed on the fog. In addition, it includes architectural modules that are responsible for monitoring the patient's mobility (i.e., *Mobility Management*) and the storage (i.e., *Local Data Manager*). The cloud stratum involves application components responsible for maintaining the patient's history for a long time and offering access to it for the caregivers. It also includes several modules such as the *Cloud Middleware* and the *SLA Management*. The *Cloud Middleware* implements a unified model that aggregates the strong diversity and heterogeneity of the cloud hosting nodes. The *SLA Management* is responsible for the application's QoS management.

It should be noted that the *Cloud Middleware* does not support the heterogeneity of the involved fog nodes. So, the heterogeneity criterion ($C_1$) is not met. The *SLA Management* handles the latency prospective variations when the application's components placement changes. Consequently, the QoS ($C_2$) criterion is met. The authors do not discuss the scalability of the architecture they propose in terms of the supported number of sensors, the fog nodes, and/or the fog domains. The scalability ($C_3$) criterion is then not met. The *Mobility Management* supports only indoor monitoring, i.e., when the patient is at home. Consequently, the mobility criterion ($C_4$) is not met. The data processing is distributed between the cloud and the fog strata. However, the authors do not provide any discussion about the federation between the cloud and the fog providers. The data and control interfaces are not discussed, either. So, the federation ($C_5$) and the interoperability ($C_6$) criteria are not met.

Masip-Bruin *et al.* [66] propose an architecture called Fog-to-Cloud computing (F2C), also aiming to support people affected by COPD. It consists of several components that span from the cloud to the fog. Those components ensure real-time monitoring of the patient's oxygen doses, collecting the data generated by the sensors attached to the patient and then deciding and tuning the oxygen doses. The IoT/end-users stratum consists of sensors attached to the patient. In the fog stratum, portable oxygen concentrator (POC) act as the fog nodes. These nodes are enriched with F2C capabilities. In this stratum, several components are deployed. An example of this component is *Patient Monitoring* responsible for monitoring the activity of the patient and positioning the patient geographically. *Context Data Processing* is responsible for processing the collected data from the sensors. *Smart Data Processing* is responsible for issuing the final decision about the oxygen doses and tuning it based on the data processed by the *Context Data Processing* module. For the cloud stratum, the authors do not discuss which application components can be deployed and what architectural modules it involves.

Although the authors have considered an application with components spanning the cloud and the fog, they did not address the heterogeneity of the targeted fog nodes, which affects this deployment. So, the heterogeneity criterion ($C_1$) is not met. Moreover, they do not discuss how to manage the QoS when the components are distributed across the cloud and the fog. Accordingly, the QoS criterion ($C_2$) is not met. In addition, the scalability of the architecture in terms of the supported number of sensors, the fog nodes, and/or the fog domains is not discussed. The scalability ($C_3$) criterion is then not met. It should be noted that, in contrast to Fratu *et al.* [64] who provide static monitoring of the patients, *Patient Monitoring* supports the dynamic monitoring of patients in outdoor activities (e.g., walking). So, the mobility criterion ($C_4$) is met. Finally, the authors do not provide any unified model for the handled data. The federation criterion ($C_5$) is then not met. Similarly, they neither provide common data nor control interfaces that could enable the interoperability between the involved nodes. Therefore, the interoperability criterion ($C_6$) is not met.

*c) Healthcare with Focus on Parkinson and Speech Disorders*

Monteiro *et al.* [67] propose a fog computing interface called FIT. It processes and analyzes the clinical speech data of patients with Parkinson's disease and speech disorders. In the IoT/end-users stratum, an Android smartwatch is used to acquire the clinical speech data of the patients. In the fog stratum, application components responsible for collecting and analyzing the speech data from the smartwatches are deployed. Clinical features are then extracted from this data and are forwarded to the cloud stratum for long-term analyses. In the cloud stratum, additional components store the data so that they can be accessed by clinicians to monitor the progress of their patients.

The proposed architecture does not take into account the heterogeneity of the involved cloud and fog nodes. So, the heterogeneity criterion ($C_1$) is not met. Moreover, the authors do not take into account the application QoS when the component placement between the cloud and the fog changes. Accordingly, the QoS criterion ($C_2$) is not met. Furthermore, authors do not consider the scalability of their architecture in terms of the supported number of smartwatches, the fog nodes, and/or the fog domains. So, the scalability ($C_3$) criterion is not met, either. Despite the fact that supporting mobility is critical for such applications and the authors consider wearable smartwatches for monitoring, the architecture they propose does not support the mobility of the patients. So, the mobility criterion ($C_4$) is not met. There is no discussion on the federation, needed to ensure the proper coordination of the interactions. The federation ($C_5$) criterion is then not met. Besides, there is no discussion of the interoperability between the cloud and the fog, although needed to enable the interactions between the cloud and the fog strata and the different application components deployed in them. So, the interoperability ($C_6$) is not met.

Dubey *et al.* [68] propose a service-oriented architecture for telehealth applications but with a use case for speech disorders. They apply filtering operations to speech data in the use case. A smart gateway acts as a fog node, connecting the patients equipped with wearable sensors to physicians to diagnose and treat them. The IoT/end-users stratum consists of smartwatches, wearable Electrocardiogram (ECG) sensors, and pulse glasses that can be used for health data acquisition. In the fog stratum, the authors propose a layer-based architecture. It consists of three layers: The hardware layer where the Intel Edison embedded processors are used, with the processors connected to the wearable telehealth sensors; the embedded operating system layer where an ubilinux operating system is installed; and the fog services layer that includes several application components such as signal processing, feature extraction, and onsite database. These components are responsible for processing the incoming data from the wearable sensors and producing medical logs that are sent to the cloud stratum. The cloud stratum includes components that are responsible for storing all the received data or features for comprehensive analyses.

In this work, the hosting nodes' capabilities (whether on the cloud or the fog) are taken into account when the application components are deployed on these nodes. For instance, considering that the fog nodes have limited capabilities compared to the cloud nodes, the authors deploy simple components with algorithms and methods that require simple computation on the fog nodes. By that, the heterogeneity criterion ($C_1$) is met. Moreover, the experiments in [68] show that the proposed architecture considerably reduces the ECG processing time. However, the related latency for such processing fluctuates due to the absence of a QoS manager. So, the QoS criterion ($C_2$) is not met. Furthermore, the scalability of the proposed architecture in terms of the supported number of sensors, the fog nodes, and/or the fog domains is not discussed. The scalability criterion ($C_3$) is then not met. Although the patients are equipped with wearable sensors, the authors do not provide any architectural modules to handle this mobility. Consequently, the mobility criterion ($C_4$) is not met. In addition, the federation between the cloud and the fog providers, to ensure a proper coordination of necessary interactions between application components running in the cloud and the fog strata, is not discussed. Finally, in the proposed architecture, the fog nodes transmit the necessary data to the cloud nodes after the preliminary filtering and analysis. However, the authors do not discuss any specification for the interfaces that enable the data transmissions. The control interfaces needed between the cloud and the fog strata are not provided, either. Consequently, the federation ($C_5$) and the interoperability criteria ($C_6$) are not met.

*d) ECG and EEG Feature Extraction*

Electrocardiogram or ECG is related to the heart while electroencephalogram or EEG is related to the brain. ECG and EEG are widely used as BANs in order to monitor people or patients in daily life. ECG sensors are used to measure the activities of the heart and EEG sensors are used to measure the electrical activities of the brain.

Gia *et al.* [69] propose an IoT-based health monitoring architecture. The proposed architecture exploits the fog and its advantages such as bandwidth, QoS assurance, and emergency notification. The ECG feature extraction at the edge of the network is used as a case study to help diagnose cardiac diseases. The IoT/end-users stratum consists of several physical devices including implantable and wearable sensors. These sensors generate different types of data such as temperature, ECG, and Electromyography (EMG). The sensed data is then sent to the fog stratum where smart gateways act as fog nodes. These nodes connect the sensors to the cloud nodes. The architecture proposed at the fog stratum consists of three layers: The hardware layer that acts as a middleware between the embedded operating system and all the physical components of the gateway; the embedded operating system where Linux is installed; and the fog computing service layer that includes modules such as *Heterogeneity and Interoperability* and *Location Awareness*. The *Heterogeneity and Interoperability* module aggregates the heterogeneity of the nodes in the fog stratum. This is important for deploying the application components across the cloud and the fog nodes. Moreover, it provides interoperability in terms of the ability to serve various nodes in the fog stratum with different manufactures, models, operating systems, and communication protocols. The *Location Awareness* module is used to provide the geographical location of the fog node. It is also used to locate the patient, an essential possibility in case of emergency. In the cloud stratum, the application components responsible for storing, processing, and broadcasting data are deployed.

In this paper, the heterogeneity ($C_1$) criterion is met thanks to the *Heterogeneity and Interoperability* module. The proposed architecture provides real-time notifications when it detects abnormal situations of the patient. However, it is not discussed how QoS is managed in terms of latency when the deployment of the applications' components between the cloud and the fog changes, hence leaving the QoS criterion ($C_2$) unmet. Moreover, the scalability of the architecture in terms of the supported number of sensors, the fog nodes, and/or the fog domains is not addressed. So, the scalability criterion ($C_3$) is not met. The mobility criterion ($C_4$) is met as a result of the *Location Awareness* module. The authors do not discuss the federation between the cloud and the fog providers, which is needed to ensure a proper coordination between the application components. Consequently, the federation criterion ($C_5$) is not met. Finally, the interoperability criterion ($C_6$) is met thanks to the *Heterogeneity and Interoperability* module.

Zao *et al.* [70] apply the fog concept in healthcare but in a different context. They have developed a BCI (Brain-Computer Interfaces) game called "EEG Tractor Beam". It is a brain-state monitoring game, running on a mobile app on the user's smartphone. Each player wears an EEG headset and is provided by a smartphone. Players are shown on a ring surrounding a target object. The goal of each player is to pull the target toward himself by concentrating. In the IoT/end-users stratum, EEG sensors are used to monitor the brain state of the individual players and generate raw data streams. These data are sent through their smartphones to the fog nodes located in the fog stratum. The fog nodes can be personal computers, televisions set-top-boxes, or gaming consoles. Application components deployed in this stratum perform continuous real-time brain state classifications and send the classification models to the cloud stratum for additional processing purposes.

In this work, the authors consider that the components may be deployed either on the cloud or the fog, however, they do not consider the heterogeneity of the fog nodes, which affects this deployment. Moreover, they do not discuss how to meet QoS when some components are executed on the cloud and some others are executed on the fog. So, the heterogeneity ($C_1$) and the QoS ($C_2$) criteria are not met. The scalability of the architecture in terms of the supported number of players, the fog nodes, and/or the fog domains is not discussed. The scalability criterion ($C_3$) is then not met. Furthermore, it is not discussed how the proposed architecture can provide the same service to the player when they move. Consequently, the mobility criterion ($C_4$) is not met. Authors do not provide appropriate mechanisms needed to run the application according to its execution chain, leaving by that the federation criterion ($C_5$) between the cloud and the fog providers unmet.

Finally, the authors use MQTT, for the data interfaces between the cloud and the fog strata. Thus, they provide data interfaces for interoperability. However, the control interfaces between these two strata for handling the application's lifecycle management are not discussed. So, the interoperability criterion ($C_6$) is not met.

It can be noticed that none of these architectures provides a solution to federate the cloud and the fog providers ($C_5$). Moreover, mobility is an important characteristic of healthcare applications, to help patients equipped with different sensors with their daily activities. However, except for Masip-Bruin *et al.* [66] and Gia *et al.* [69], other reviewed architectures do not meet this criterion ($C_4$).

*2) Architectures for Connected Vehicles*

Several works have used the fog in the context of connected vehicles. Hou *et al.* [71] propose an architecture called Vehicular Fog Computing (VFC) for vehicular applications. It uses vehicles as the infrastructure for communication and computation. The authors have investigated the communication and computational capability of the vehicles and conducted an empirical analysis to study the impact of the mobility of vehicular network on its connectivity and computational capacity. Their study shows a great enhancement in the communication and computation capacity that can be realized by VFC compared to Vehicular Cloud Computing (VCC). Specifically, using VFC gives a better connectivity, leading to more reliable communication with a higher capacity. In this work, the IoT/end-users stratum and the fog stratum are merged. Indeed, the authors use vehicles as the IoT devices and, at the same time, those vehicles act as the fog nodes. Here, two ways of communications are supported. The vehicles can either interact with each other directly (V2V) or via the infrastructure (V2I). For instance, they can communicate through roadside infrastructure wireless nodes, called Roadside Units (RSUs). In V2V, the main constraint is whether the distance between the two vehicles is smaller than the communication range, and in V2I, the main constraint is the energy consumption of the RSUs. Those vehicles are equipped with embedded computers. The fog stratum connects to the cloud stratum through RSUs.

In this paper, there is no discussion on the heterogeneity of the fog nodes. The authors have only considered a unique type of the fog node i.e., vehicles. Moreover, QoS is not discussed in terms of latency. So, the heterogeneity ($C_1$) and the QoS ($C_2$) criteria are not met. In addition, the authors do not mention if the proposed architecture is able to support the increasing number of vehicles and/or the fog domains. The scalability criterion ($C_3$) is then not met. However, this work meets the mobility criterion ($C_4$). Based on the conducted experiments, the authors show that VFC maintains the communication and the execution even when the fog nodes (i.e., the vehicles) move. However, it should be noted that they do not provide any details about the architectural module(s) responsible for that mobility. Moreover, the authors acknowledge that there is a need for mobility models to build an efficient VFC. In this work, the authors do not provide any description model to enable the federation between the cloud and the fog providers. The interactions are hard-coded and are provided as part of the proposed infrastructure. So, the federation criterion ($C_5$) is not met. Finally, the specification of the control and the data interfaces that such interactions are based on are not discussed. Consequently, this work does not meet the interoperability criterion ($C_6$).

In the same context, Datta *et al.* [72] propose an architecture for connected vehicles application that spans the cloud and the fog. The proposed architecture provides three consumer-centric services: M2M Data Analytics with Semantics, discovery, and the management of connected vehicles. The IoT/end-users stratum comprises of vehicular sensors. These sensors send data to the fog stratum in a uniform format by using the Sensor Markup Language. The fog stratum includes RSUs and M2M gateways acting as the fog nodes. Architectural modules that implement the three aforementioned services are included in this stratum. For instance, it includes a module that annotates the raw data generated by the sensors with semantic Web technologies to generate the inferred data. It also includes a component that tracks the mobile connected vehicle. The cloud stratum is responsible for data engineering. Components responsible for further analyzing and processing of the data based on the application needs are deployed in this stratum. The cloud and the fog strata communicate according to the REST principle.

Although the authors have considered that the components may be deployed either on the cloud or the fog, they did not consider the heterogeneity of the cloud and the fog nodes, affecting this deployment. Accordingly, the heterogeneity criterion ($C_1$) is not met. Moreover, managing QoS when executing some components on the fog and others on the cloud is not discussed. The scalability of the proposed architecture in terms of the supported number of sensors, the fog nodes, and/or the fog domains is not discussed, either. So, the QoS ($C_2$) and the scalability ($C_3$) criteria are not met. The mobility criterion ($C_4$) is met thanks to the component that keeps track of the mobility of vehicles. In this work, the authors do not discuss the federation between the cloud and the fog providers. The federation criterion ($C_5$) is then not met. The interoperability criterion ($C_6$) is met thanks to the REST interfaces between the cloud and the fog.

In the same context, Truong *et al.* [73] propose an architecture, called FSDN, for Vehicular Ad-hoc NETworks (VANETs). It leverages Fog and Software Defined Networking (SDN). The IoT/end-users stratum consists of SDN-based vehicles that act as end-users and/or the forwarding elements. In the fog stratum, application components responsible for storing local road system information and routing the required data to the cloud stratum are deployed. The fog stratum consists of an SDN Controller and several fog domains. Each fog domain includes specific types of the fog nodes such as SDN RSUs, cellular Base Station (BS), and SDN RSU Controller. The SDN Controller is responsible for coordinating the RSU Controllers and BSs. It includes several architectural modules. The *Resource Manager* orchestrates the execution of different components running on BSs and RSUCs. The *Fog Controller* is responsible for functionalities such as migrating VMs between the fog nodes. In the cloud stratum, the components responsible for storing the data for long terms are deployed.

The authors in [73] acknowledge that the cloud nodes and the fog nodes are highly heterogeneous. These nodes share their capabilities to provide control to vehicles. They are equipped with SDN capabilities and offer virtualization. Using this, the architecture provides a mechanism to model the capabilities of the nodes. So, the heterogeneity criterion ($C_1$) is met. Managing the QoS in terms of latency when the application components are distributed across the cloud and the fog strata is not discussed. Moreover, the scalability of the proposed architecture in terms of the supported number of connected vehicles, the fog nodes, and/or the fog domains is not discussed. Consequently, QoS ($C_2$) and the scalability ($C_3$) criteria are not met. The *Fog Controller* is responsible for migrating VMs, which can be applied to the vehicles' mobility case. Accordingly, the mobility criterion ($C_4$) is met. However, it supports migration across the fog nodes only and does not cover the cloud. The same applies to the *Resource Manager*. It only orchestrates the nodes belonging to fog domains and does not cover the cloud nodes. So, the federation ($C_5$) criterion is not met. The cloud and the fog communicate with each other through the SDN Controller. This SDN controller provides the control interfaces needed between the cloud and the fog. However, a specific model for the data interfaces is not provided. The interoperability criterion ($C_6$) is then not met.

None of the three works reviewed here meets the QoS and the scalability criteria. In such scenarios, the vehicles are mobile and their mobility affects the proposed solution. Accordingly, all the three reviewed works took this mobility into consideration.

*3) Architectures for Smart Living and Smart Cities*

In addition to the healthcare applications and vehicular network application, several works exploit the advantages of the fog in smart environments such as smart living and smart cities. In this subsection, we review the smart environment architecture proposed by Li *et al.* in [46], supporting smart living applications (e.g., smart healthcare and smart energy). The IoT/end-users stratum is comprised of smart objects such as sensors and laptops. Application components deployed in the fog stratum are responsible for filtering the collected data from the smart objects and ensuring real-time interactions. For instance, considering smart healthcare applications, monitoring and detecting heart problems can be provided in real time. The fog stratum consists of two types of fog nodes: The fog server and the fog edge nodes. The fog server includes modules that are responsible for management functionalities such as application deployment, network configuration, and billing. The fog edge node provides computing, storage, and communication capabilities to smart objects. The fog edge nodes include a module called foglet. It is responsible for functionalities such as orchestration and Service Level Agreement (SLA) management. Foglets are also responsible for the communication between the fog edge nodes and the fog servers. Meanwhile, the communications between the fog edge nodes and the cloud nodes are routed through the fog servers. The cloud stratum includes components that are responsible for functionalities such as storing the data received from the fog stratum as a backup. For instance, considering the same smart healthcare application, professionals can access these data to evaluate patients' health status.

In this paper, although the authors consider different types of fog nodes with different capabilities, there is no discussion on the heterogeneity of these nodes and the nodes in the cloud. Accordingly, the heterogeneity criterion ($C_1$) is not met. The authors perform a simulation and the results indicate, when the fog is employed, the latency drops by 73%. However, the related latency for such a processing fluctuates due to the absence of the QoS manager. So, the QoS ($C_2$) criterion is not met. The scalability of the proposed architecture in terms of the supported number of IoT devices, the fog nodes, and/or the fog domains is not discussed. Moreover, there is no architectural module that can handle the mobility of the fog nodes and/or the IoT devices. Consequently, the scalability ($C_3$) and the mobility ($C_4$) criteria are not met. In addition, there is no discussion of the federation that is needed to ensure a proper coordination of the interaction between the cloud and the fog providers. The federation ($C_5$) criterion is then not met. Finally, in this work, although the authors mention that the fog edge nodes communicate with the cloud nodes to send the data, they neither provide common data nor control the interfaces that could enable the interoperability between these involved nodes. So, the interoperability criterion ($C_6$) is not met.

In [74], Yan *et al.* propose an architecture for smart grid applications. The proposed architecture enables data storage and processing in order to improve the existing smart meters infrastructure. The IoT/end-users stratum consists of smart homes, the smart building, etc. In the fog stratum, the smart meters act as the fog nodes. Each smart meter acts as a datanode. A specific datanode is considered as the master node. The latter includes architectural modules that store the metadata of the file name and storage location. It also includes modules that duplicate and split the collected data and then distribute it to the datanodes. This is done at fixed time intervals. The cloud stratum stores the data received from the fog nodes as a backup.

In this work, the authors do not discuss the heterogeneity of nodes in the cloud and the fog. So, the heterogeneity criterion ($C_1$) is not met. The authors compare the average processing time by the centralized cloud and the one when the fog is integrated. The results demonstrate a better performance when the fog is integrated. However, managing QoS when the component placement between the cloud and the fog changes is not discussed. Accordingly, the QoS criterion ($C_2$) is not met. It should be noted that, according to the authors, the architecture is designed in such a way to easily add additional nodes to the architecture as the data repository grows, without the need to reconfigure the entire architecture. To that end, the scalability criterion ($C_3$) in terms of the number of the fog nodes is met. However, the authors consider the fixed fog nodes in the fog stratum and the fixed IoT devices in the IoT/end-users stratum. Accordingly, the mobility criterion ($C_4$) is not met. The federation between the different involved providers is not discussed, leaving by that the federation criterion ($C_5$) unmet. Finally, the common data interfaces between the several involved nodes are provided by using Hadoop MySQL-like language Hive. However, common control interfaces are missing. So, the interoperability criterion ($C_6$) is not met.

Considering the same context, i.e., smart environments, Brzoza-Woch *et al.* [75] propose an architecture for smart levee monitoring applications. They design a proper architecture for advanced telemetry systems that are able to support an automated flood risk assessment system. The proposed architecture is layer-based, consisting of three layers. It spans from the IoT/end-users stratum to the cloud stratum. In the IoT/end-users stratum, the authors propose the measuring layer. It includes sensors and their related networks. In the fog stratum, the authors propose the edge computing layer, consisting of many distributed telemetry stations. It also includes components that are responsible for collecting data from the measuring layer, processing it, and sending this data to the central part of the system. The cloud stratum includes the communication layer proposed by the authors. It provides the communication between the edge computing layer and the central part of the system. It includes components that are responsible for further data processing.

In this work, the heterogeneity of the fog nodes is not discussed. All the fog nodes are identical telemetry stations. The same applies to the considered cloud nodes. There is also no architectural module in the proposed architecture to manage QoS when the application components are distributed across the cloud and the fog. Accordingly, the QoS ($C_1$) and the scalability ($C_2$) criteria are not met. The authors acknowledge that there might be more than thousands of sensors in the measuring layer that sends the data, which is likely to need compression. So, the scalability criterion ($C_3$) in terms of the IoT devices is met. The mobility of the sensors and/or the fog nodes is not discussed. So, the work does not meet the mobility criterion ($C_4$). The federation of the different providers (i.e., the cloud and the fog providers) is not discussed, either. However, the authors developed mechanisms to transmit data from a telemetry station in the fog stratum into the central system by using the MQTT protocol. Therefore, they provided the data interfaces. Yet, authors do not provide application life-cycle management mechanisms, hence control interfaces are not provided. Accordingly, the federation ($C_5$) and interoperability criteria ($C_6$) are not met.

Tang *et al.* [76] propose an architecture for smart cities application. The proposed architecture is hierarchically distributed. Its goal is to support a big number of infrastructures and services in future smart cities. The authors propose a layer-based architecture comprising of four layers. These layers span from the IoT/end-users stratum to the cloud stratum. For the IoT/end-users stratum, the authors propose layer 4. It contains the sensing network including numerous sensory nodes. These sensors forward their raw data to the fog stratum. For the fog stratum, the authors propose two layers, layer 3 and layer 2. Layer 3 has many low-power and high-performance edge nodes. Each edge node is responsible for a local group of sensors. This layer includes components that are responsible for performing data analysis in a timely manner. Layer 2 consists of a number of intermediate computing nodes. Each node is connected to a group of edge nodes in layer 3. This layer includes components that make quick responses to control the infrastructure when hazardous events are detected. The data analysis results in these two layers (i.e., Layer 2 and 3) are reported to the cloud stratum. For the cloud stratum, the authors propose layer 1. It includes components that are responsible for very high-latency computing tasks such as long-term natural disaster detection and prediction.

In this paper, although the authors consider that the components span in both the cloud and the fog strata, they do not discuss the heterogeneity of the nodes that host these components. The heterogeneity criterion ($C_1$) is then not met. The authors present a prototype along with performance results demonstrating that the fog provides real-time interactions compared to the cloud. Using the fog, the data transmitted to the cloud is 0.02% of its total size, hence reducing the transmission bandwidth and the power consumption. However, the related latency for such processing may vary due to the absence of the QoS manager. So, the QoS ($C_2$) criterion is not met. Moreover, the authors do not discuss if their architecture is scalable in terms of the supported number of sensors, the fog nodes, and/or the fog domain. The mobility of the sensors and/or the fog nodes is also not taken into consideration, leaving by that the scalability ($C_3$) and the mobility ($C_4$) criteria unmet. In this work, the authors do not discuss the coordination of the execution flow between the components belonging to different providers. Moreover, they do not provide any common data or control interfaces between the cloud and the fog strata. Accordingly, the federation ($C_5$) and the interoperability ($C_6$) criteria are not met.

None of the works reviewed in this subsection meets most of the criteria, except for [74] and [75] which meet the scalability ($C_3$) criterion in terms of number of fog nodes and number of IoT devices respectively.

*4) Architectures for Other Applications*

In addition to the already reviewed applications, various other applications that range from Wireless Sensor Networks (WSNs), industrial IoT, data analytics, energy management, to emergency applications rely on the fog. Lee *et al.* [77] present an architecture for WSANs applications. In the proposed architecture, gateways act as the fog nodes. The IoT/end-users stratum comprises of WSANs. The proposed architecture for the fog stratum consists of two layers: The slave layer and the master layer. The slaves layer includes conventional gateways and microservers acting as the fog nodes. This layer includes modules responsible for flow management, virtual gateway, and resource management. It also includes modules that provide WSAN virtualization. This virtualization is event-driven. It creates virtual networks from the sensors and actuators and shares them with various applications. The master layer includes relatively more powerful and smarter gateways acting as the fog nodes. This layer includes the modules responsible for control functionalities. For the cloud stratum, the authors do not mention the components it includes or the modules it involves.

In this work, the authors deploy components on both conventional and smart gateways. However, the heterogeneous capabilities of this gateway are not taken into consideration when the application components are deployed. Moreover, they do not discuss the heterogeneity of the nodes on the cloud and on the fog. Furthermore, there is no discussion on the management of application's QoS when the components are distributed across the cloud and the fog. The scalability of the architecture is left for future research. So, the heterogeneity ($C_1$), QoS ($C_2$), and scalability ($C_3$) criteria are not met. The authors do not discuss the mobility aspect of the sensors, despite the fact that WSANs usually include mobile sensors and/or actuators. So, the mobility criterion ($C_4$) is not met. In this paper, the authors do not discuss the federation between the cloud and the fog providers. They do not provide or discuss any control or data interfaces between the cloud and the fog strata, either. Accordingly, the federation ($C_5$) and the interoperability ($C_6$) criteria are not met.

Gazis *et al.* [78] propose an architecture to support the IoT applications in industrial domains. The proposed architecture allows predictive maintenance for industrial equipment by taking into account the configuration of each particular infrastructure machine. It consists of several components that span from the cloud to the fog. The IoT/end-users stratum includes sensors that can monitor the operational behavior of a machine (e.g., temperature). For the fog and the cloud strata, the authors propose a layer-based architecture consisting of three layers: Fog infrastructure, Operational Support System (OSS), and adaptive operations platform (AOP). The fog stratum consists of components responsible for filtering the data received from the sensors, based on some specific rules received by AOP. It includes the fog infrastructure layer and the OSS layers. The fog infrastructure includes the fog nodes such as gateways and routers. OSS provides management functions through several modules such as *provisioning* and *maintenance*. AOP belongs to the cloud stratum. It includes several components, responsible for collecting data about equipment failure models, generating rules, and sending the rules to the fog nodes. An example of those rules is to send only particular values (i.e., anomalies) to the cloud.

Although the authors consider that the components are deployed on both the cloud and the fog, they have not taken into account the heterogeneity of the targeted nodes on the cloud and the fog, which affects this deployment. So, the heterogeneity criterion ($C_1$) is not met. Moreover, the QoS criterion ($C_2$) is met thanks to OSS. The scalability of the architecture in terms of the number of the supported sensors, the fog nodes, and/or the fog domains is not discussed, hence leaving the scalability criterion ($C_3$) unmet. Furthermore, the authors do not consider mobile equipment and do not provide any architectural module to handle the mobility aspect of the fog nodes. Consequently, the mobility criterion ($C_4$) is not met. Although OSS is necessary to provide federation and interoperability, it is not enough. The complementary BSS is needed in order to enable federation and interoperability. Accordingly, the federation ($C_5$) and the interoperability ($C_6$) criteria are not met.

Xu *et al.* [79] propose an architecture for data analytics. The proposed architecture is based on SDN. The IoT/end-users stratum includes the IoT devices acting as MQTT publishers. In the fog stratum, the nodes with MQTT broker functionalities act as the fog nodes. This stratum includes an Open vSwitch[6] (OvS) that is virtual switch licensed under the Apache license. It supports standard management interfaces and protocols. It also includes modules that support the QoS control. The authors added an *SDN Controller* to this switch with the *Analytics* module. The *Analytics* module performs the analytics by parsing the MQTT payload content and retrieving the data such as temperature. It also performs real-time analytics such as detecting temperature beyond the threshold. The cloud stratum involves modules responsible for storage and exhaustive deferred analytics.

In this work, the authors do not take into account the heterogeneous fog nodes. The heterogeneity criterion ($C_1$) is then not met. Moreover, using OvS, the QoS criterion ($C_2$) is met as a result of the OvS's feature that provides QoS management. The authors also demonstrate an improved delivery delay performance. The scalability of the proposed architecture in terms of the supported number of IoT devices, the fog nodes, and/or the fog domains is not taken into consideration. In addition, the mobility of the IoT devices and/or the fog nodes is not discussed, hence leaving the scalability ($C_3$) and the mobility ($C_4$) criteria unmet. Furthermore, the authors do not provide any description model to enable the federation between the cloud and the fog providers. Accordingly, the federation criterion ($C_5$) is not met. Finally, the fog nodes communicate with other nodes using OvS. So, the interoperability criterion ($C_6$) is met thanks to the interfaces that OvS supports.

Al-Faruque *et al.* [80] deal with another application domain. They propose an architecture for energy management applications. The proposed architecture is designed according to the SOA specifications. As an example for the energy management, they studied the case of Home Energy Management (HEM). Their proposed architecture for energy management is implemented over the fog. The IoT/end-users stratum consists of different sensors and actuators. In the fog stratum, the HEM control panels act as the fog nodes. This stratum involves application components responsible for gathering, storing, processing, and analyzing the data. Moreover, the modules in this stratum are responsible for monitoring and managing the power and energy consumption of each device at home, by controlling the devices efficiently. For the cloud stratum, the authors have made an assumption that they do not send data from the fog to the cloud and they do not deploy any components on the cloud.

In this work, SOA abstracts the heterogeneity of the fog nodes' capabilities. However, based on their assumptions, the authors do not discuss the heterogeneity of the fog and the cloud nodes. The heterogeneity criterion ($C_1$) is then not met. Although the authors do not deploy any components on the cloud, managing QoS is not discussed even when the application components are distributed across different fog domains. So, the QoS criterion ($C_2$) is not met. The proposed architecture has an open software architecture and hardware infrastructure, which provides the ability to scale the architecture. However, the scalability in terms of the IoT devices, the fog nodes, and/or the fog domains is not discussed. Moreover, the mobility of the fog nodes or the IoT devices is not considered since the authors consider a case study of HEM with the fixed sensors and the fog nodes. Accordingly, the scalability ($C_3$) and the mobility ($C_4$) criteria are not met. It should be noted that based on their assumptions, the authors do not discuss the federation between the cloud and the fog providers. The federation criterion ($C_5$) then is not met. In addition, they do not discuss the interoperability between the cloud and the fog strata. So, the interoperability criterion ($C_6$) is not met.

The fog can also have its benefits in an emergency situation. Aazam *et al.* [81] exploit the advantages of the fog in emergency cases. They present an architecture for emergency alerts, called Emergency Help Alert Mobile Cloud (E-HAMC). Here, an application is installed on the user's smartphone allowing the user to contact the emergency department in case of an accident. At the IoT/end-users stratum, smartphones are used to send alert to the appropriate emergency department and the family members of the victim. In case of an emergency, the user needs to choose the type of the accident by pressing a single button on the application. The fog stratum includes the application components responsible for maintaining the list of the user's contacts. These components inform the family members by sending messages to the already stored contact numbers. It also includes the components responsible of deciding the particular department to send the alert to. In addition, the components that filter the data and forward it to the cloud stratum are deployed. In this stratum, there is a module that is responsible for monitoring the location of the users. Once it detects a user's movement or relocation (e.g., moving to another city), it updates the contact of the emergency department. The cloud stratum includes the application

---

[6] http://www.openvswitch.org/

TABLE III

Main features and requirements for the works proposing architecture for Applications Specific Fog Systems. In the Evaluation column, P is prototype, S is simulation, and x means no evaluation is conducted. In the criteria columns, ✓ means the criterion is met and x means the criterion is not met.

| Scope | | Paper | Evaluation | Performance metrics | Major Contribution | Criteria | | | | | |
|---|---|---|---|---|---|---|---|---|---|---|---|
| | | | | | | $C_1$ | $C_2$ | $C_3$ | $C_4$ | $C_5$ | $C_6$ |
| Healthcare Applications | Health care at Large | Stantchev et al. [63] | X | n/a | Propose an architecture for elderly-care applications. Includes a process-oriented view of healthcare applications modeled using BPMN. | x | x | x | x | x | x |
| | Healthcare with Focus on COPD | Fratu et al. [64] | X | -n/a | Propose an architecture for applications offering support for people affected by COPD. | x | ✓ | x | x | x | x |
| | | Masip-Bruin et al. [66] | P | -Heartbeat and oxygen volume vs mobility -Incidence Rate Ratio (IRR) for Re-admission rate and Respiratory Mortality | Propose an architecture called F2C in order to support people affected by COPD. | x | x | x | ✓ | x | x |
| | Healthcare with Focus on Speech Disorders | Monteiro et al. [67] | P | -Speech signal and its corresponding loudness in dB -Spectral centroid -Zero-crossing rate -Short-time energy | Propose an architecture called FIT which processes and analyzes the clinical speech data of patients with Parkinson and speech disorders. | x | x | x | x | x | x |
| | | Dubey et al. [68] | P | -Percentage data reduction achieved for difference scenarios by performing processing on fog -Processing time on fog for different methods and algorithms | Propose a service-oriented architecture for telehealth applications. They study a case for speech disorders. | ✓ | x | x | x | x | x |
| | ECG and EEG Feature Extraction | Gia et al. [69] | P | -Data rate and latency comparison | Propose an IoT-based health monitoring architecture. They study a case for ECG feature extraction to help in diagnosing cardiac diseases. | ✓ | x | x | ✓ | x | ✓ |
| | | Zao et al. [70] | P | n/a | Propose an architecture that applies the cloud and the fog in physiological signal processing and data management. | x | x | x | x | x | x |
| Connected Vehicles | | Hou et al. [71] | X | n/a | Propose an architecture for vehicular applications called VFC. Vehicles are used as the infrastructure for communication and computation. | x | x | x | ✓ | x | x |
| | | Datta et al. [72] | X | n/a | Propose an architecture for connected vehicles, where the fog is co-located at RSUs and M2M gateways. | x | x | x | ✓ | x | ✓ |
| | | Truong et al. [73] | X | n/a | Propose a VANET architecture that leverages fog computing and SDN. | ✓ | x | x | ✓ | x | x |
| Smart Living and Smart Cities | | Li et al. [46] | S | -Latency | Propose an architecture for the fog that supports smart living application. | x | x | x | x | x | x |
| | | Yan et al. [74] | P | -Comparison of average data processing time considering fog and cloud processing | Propose an architecture for smart grid applications that enables data storage and processing. | x | x | ✓ | x | x | x |
| | | Brzoza-Woch et al. [75] | P | n/a | Propose an architecture for smart levee monitoring application to support automated flood risk assessment system. | x | x | ✓ | x | x | x |
| | | Tang et al. [76] | P | -Amount of data transmitted to cloud -Response time | Propose a hierarchical distributed architecture to support the integration of a big number of infrastructure components in smart cities. | x | x | x | x | x | x |
| Other Applications | | Lee et al. [77] | X | n/a | Propose an architecture for WSANs application. | x | x | x | x | x | x |
| | | Gazis et al. [78] | P | -Number of packets -CPU utilization | Propose a platform called (AOP), where the fog is co-located at the gateway/router. | x | ✓ | x | x | x | x |
| | | Xu et al. [79] | P | -Congestion window -Throughput | Propose an architecture for data analytics based on SDN. | x | ✓ | x | x | x | ✓ |
| | | Al-Faruque et al. [80] | P | n/a | Propose an architecture for energy management applications based on SOA specifications. | x | x | x | x | x | x |
| | | Aazam et al. [81] | P | -Data upload and synchronization delay to cloud and to fog | Propose an architecture for emergency alert and management through fog. | x | x | x | ✓ | x | x |

components responsible for further analyzing the data and creating an extended portfolio of services.

The authors do not discuss the heterogeneity of the cloud and the fog nodes, which needs to be taken into consideration when the application components across the cloud and the fog strata are deployed. Besides, they do not discuss how to maintain QoS in terms of latency when the application components are distributed across the cloud and the fog. So, the heterogeneity ($C_1$) and the QoS ($C_2$) criteria are not met. The scalability of the architecture in terms of the supported number of users' mobile devices, the fog nodes, and/or the fog domains is also not taken into consideration. The scalability ($C_3$) criterion is then not met. The monitoring module in the fog stratum allows meeting the mobility criterion ($C_4$). Meanwhile, the authors do not provide any unified models between the cloud and the fog providers, hence leaving the federation criterion ($C_5$) unmet. In order for the fog stratum to synchronize its contact list based on the available departments, it needs to contact the cloud stratum. It also needs to send emergency-related information to the cloud. However, the authors do not discuss any specifications for the data and control interfaces enabling this. Consequently, the interoperability criterion ($C_6$) is not met.

The works in reviewed this subsection deal with different application domains. However, none of them meets the heterogeneity ($C_1$), the scalability ($C_3$), and the federation ($C_5$) criteria. Moreover, the mobility criterion ($C_4$) is met only by Aazam *et al.* [81] and the interoperability criterion ($C_6$) is met only by Xu *et al.* [79].

## V. ALGORITHMS FOR THE FOG SYSTEM

Like for any large-scale computing system, application-agnostic and application-specific algorithms have been proposed for fog. Furthermore, the application-agnostic algorithms cover the computing, storage/distribution, and energy consumption, like in most large-scale distributed computing systems. This section discusses the application-agnostic algorithms and the application-specific algorithms for fog systems. The 3 first subsections are devoted to the application agnostic algorithms and cover computing, storage/distribution, and energy consumption. The last subsection is devoted to the application-specific algorithms.

It should be noted that not all algorithmic criteria are relevant for every reviewed paper. Let us assume, for instance, the case of an algorithm that optimizes vehicles routing to their destinations. It will not make sense to evaluate it with the heterogeneity criterion ($C_1$), as the algorithm runs over a compute node selected by a task scheduling algorithm. The vehicles routing algorithm would thus operate independently of the heterogeneity of nodes. When a criterion is not applicable to a particular work, we explicitly mention it in our discussion. Table IV, V, and VI provide a summary of the main features of the papers reviewed in this section. For each paper, we outline its scope, the approach that is followed, the evaluation methodology, its major contribution as well as the criteria it meets.

### A. Algorithms for Computing in The Fog Systems

Many algorithms have been proposed for computing in the fog systems. In the following, we review them. We first present compute resource sharing algorithms. We then cover task scheduling algorithms and present after that offloading and load redistribution algorithms.

#### 1) Resource Sharing

When it comes to computing in the fog systems, a first aspect that has been investigated is the compute resource sharing and cooperation among the nodes. These aspects have been tackled so far in the fog stratum, with the objective of executing compute demands. Abedin *et al.* [82], Oueis *et al.* [83], and Nishio *et al.* [84] cover these aspects.

Abedin *et al.* [82] introduce an algorithm that enables compute resource sharing among the fog nodes inside the same fog domain, in the fog stratum, in order to enable the execution of the users' compute demands. They define a utility metric for a couple of nodes that accounts for the communication cost and pricing benefits in case they share their resources. Using this metric, their algorithm first determines an ordered list of preference pairing nodes for each node. Then, each node in the fog domain sends requests to its preferred pairing nodes. On the reception of a pairing request, and depending on the preference level of the previously received requests, a target node decides whether to accept or reject the request. This operation leads to a one-to-one pairing. Accordingly, depending on the capabilities of a node, additional pairings are considered in a subsequent step, following the ordered list of rejected requests. The evaluation of this strategy shows that it outperforms a greedy approach when it comes to total utility. In this work, the limitations of relevant system resources inside the fog domain are modeled, which enables the support for heterogeneous resources. Therefore, the heterogeneity criterion ($C_1$) is met. The QoS criterion ($C_2$) is not met, as the QoS is not considered as part of the resource sharing decisions. Moreover, the evaluation is conducted over a small-scale and disregards the mobility of devices or the fog nodes. Accordingly, the elastic scalability ($C_3$) and support for mobility ($C_4$) criteria are not met. The federation criterion ($C_5$) is not relevant for this work since the proposed algorithm operates inside a single fog domain.

Oueis *et al.* [83] also address the problem of compute resource sharing among the fog nodes to execute compute demands, while they particularly focus on fog-enabled small cells in cellular networks. The authors aim at forming clusters of small cells, where each cluster represents a group of small cells that share resources for offloading mobile devices from their workload. Their objective is to do so at the lowest power consumption cost. To that end, they formulate their problem as an optimization problem. It aims at simultaneously forming clusters and allocating computational and communication resources there while respecting each user's latency constraint. By comparing their scheme with other clustering strategies, the authors show that their method can satisfy a higher percentage of user demands. However, this comes at the price of intermediate power consumption per user. In this work, the authors take the limitations in computational and network resources into account. They also consider the latency constraint for the users. They thus meet the heterogeneity ($C_1$) and the QoS ($C_2$) criteria. However, the elastic scalability ($C_3$) and the support for mobility ($C_4$) criteria are not met; the authors test their strategy over a small-scale scenario and do not consider the user's mobility. The federation criterion ($C_5$) is not met, as the authors do not consider the possibility of having several providers.

Nishio et al. [84] in turn tackle the same issue but consider the case of a mobile fog system. It includes a fog stratum formed by mobile devices and a cloud reachable through the cellular network. Accordingly, they target CPU optimization, bandwidth, and storage sharing to serve the compute demands. Due to the heterogeneity of these resources, they map them into time resources to allow quantifying them in the same unit. They separately study the optimization with the following two objectives: Maximizing the sum and maximizing the product of utility functions. They solve their problems by using convex optimization. The evaluation of their strategy over a set of three nodes with real-world measurements shows that it allows reducing service latency and leads to high-energy efficiency. In this work, the authors meet the heterogeneity ($C_1$) and the QoS ($C_2$) criteria by representing them in their model. Nevertheless, the evaluations are only conducted over a small-scale and mobility is not covered; thus, the elastic scalability ($C_3$) and the support for mobility ($C_4$) criteria are not met. Finally, the federation criterion ($C_5$) is not met, as the authors do not account for the existence of various providers.

Our review shows that the proposed compute resource sharing algorithms meet in general the heterogeneity criterion ($C_1$) and the QoS criterion ($C_2$). However, the elastic scalability criterion ($C_3$) and the support for mobility criterion ($C_4$) are not met in any work. None studies the federation among domains, which enables resource sharing among different domains and therefore the federation criterion ($C_5$) is not addressed.

*2) Task Scheduling*

As the fog systems provide additional computing capabilities at the edge of the network, a major question that they raise is how to manage task execution. More precisely, how to decide which tasks to execute in the IoT/end-users stratum, the fog stratum, and the cloud stratum? On a finer level, to which nodes a particular task should be assigned? What metrics to consider in deriving decisions? Several studies have addressed these questions, considering only the fog stratum, as in the case of Oueis et al. [85], Intharawijitr et al. [86], Aazam et al. [87]; considering the IoT/end-users and the fog strata simultaneously, as in the case of Zeng et al. [88], and considering the fog and cloud strata at the same time as in the case of Agarwal et al. [25] and Deng et al. [89]. These contributions are discussed in details, as follows:

Starting by works focusing on the fog stratum, Oueis et al. [85][7] study the task scheduling problem in a cellular network-based fog stratum, where small cells are enabled with computing capabilities and form the fog nodes. They propose a task scheduling strategy that operates according to two major steps. The first step allocates computational resources at the level of each individual small cell over an ordered list of users associated with it and based on a specific objective. At the end of this step, some requests may not have been processed due to the lack of available resources. Accordingly, in the second step, computation clusters are built for their processing. Again in this step, the requests are served based on a certain order and following a particular objective. The authors test three variants of their algorithm with various ordering metrics and clustering objectives. Their results show that all the strategies outperform static clustering and no clustering in terms of the user's satisfaction ratio. In addition to that, latency-oriented variants achieve lower latency in comparison to others, while the power-centric variant leads to low power consumption per user. Overall, in this work, the authors represent the heterogeneity of resources in their algorithm and study the QoS on the user's side. So, they meet the heterogeneity ($C_1$) and the QoS ($C_2$) criteria. However, their evaluation is conducted in a small-scale scenario and the user's mobility is not discussed. Accordingly, the elastic scalability criterion ($C_3$) and the support for mobility criterion ($C_4$) are not met. The need for the federation criterion ($C_5$) is not met, either as the authors do not account for the possibility of having several providers.

Intharawijitr et al. [86] also study the task scheduling problem in the fog stratum. However, in contrast to Oueis et al. [85], they consider that the fog nodes in the fog stratum represent compute servers. In this context, they aim at finding a mapping between tasks and servers that minimizes the task blocking probability, while respecting the latency constraint on the user's side. To solve the problem, they propose three different policies. The first one adopts a random approach: A fog node is randomly selected to execute a task upon its arrival. The second one is a lowest-latency policy selecting the fog node that implies the lowest total latency according to the system state, to execute a new arriving task. The third policy targets the capacity and attributes to an arriving task, the fog node with the maximum remaining resources in a candidate list. These policies are compared in a simulative environment. The results show that the blocking probability is the lowest in the case of the lowest latency policy. In this work, the authors do not account for differences in resource limitations. They thus do not meet the heterogeneity criterion ($C_1$). However, they meet the QoS criterion ($C_2$) as they impose a threshold on the maximum latency. The elastic scalability criterion ($C_3$) and the support for mobility criterion ($C_4$) are in turn not met. The authors evaluate their policies only in a small-scale scenario and do not study the impact of mobility on the algorithm's performance. Besides, the need for the federation criterion ($C_5$) is not met and the possibility of outsourcing tasks is not considered.

---

[7] This work is also discussed in Section V.C

In turn, Aazam *et al.* [87] and [90] study the problem of task scheduling in the fog stratum, by considering adaptive solutions with respect to the user's behavior. In [87], the authors propose a proactive resource allocation algorithm to reserve all types of resources for customers. Their scheme incorporates the users' historical data and attributes more resources to users that are loyal to the requested service and to the service provider in general, and thus offers them higher QoS. In [90], the authors extend their resource allocation strategy by accounting for differences in the type of devices. Additionally, they complement their resource allocation strategy with a pricing strategy that also accounts for the users' loyalty. The evaluation of their strategies over a set of 10 users validates that their strategy enables an adaptive allocation of resources, with respect to the various parameters. It also shows that it allows avoiding resource wastage. In these works, the authors model the heterogeneity in the available resources and consider QoS on the user's side. Accordingly, they satisfy the heterogeneity ($C_1$) and the QoS ($C_2$) criteria. However, their evaluations do not cover a large-scale scenario and the algorithms do not cover the case of moving devices. Therefore, the elastic scalability ($C_3$) and the support for mobility ($C_4$) criteria are not met. Finally, the authors do not account for the presence of different providers, hence leaving the federation criterion ($C_5$) unmet. By considering the fog and the devices strata, Zeng *et al.* [88] study task scheduling and task image placement simultaneously. More precisely, they consider that tasks can run either on computation server in the fog stratum or on the embedded devices and that task images can be saved on the storage servers. Their strategy aims at jointly optimizing task scheduling and image placement in order to minimize the maximum completion time. This implies balancing loads between client devices and computation servers, efficiently placing task images on the storage servers and balancing the I/O requests among storage servers. They formulate their problem as a mixed-integer nonlinear problem. To solve it, they propose an algorithm that operates over three steps and allows minimizing the three elements of completion time: Computation, I/O, and transmission time. In the first two steps, the algorithm minimizes the I/O and computation time independently, based on a mixed-integer linear program model. In the third step, the obtained results are combined to minimize the task completion time. The results show that the proposed algorithm outperforms greedy solutions, focusing on the server or the client, for different task arrival rates, client processing rates, computational servers processing rates, and disk reading rates. In both of their model and algorithm, the authors take into consideration the limitations of storage and computational resources. Moreover, they aim at minimizing the total task completion time. Thus, they meet the heterogeneity ($C_1$) and the QoS ($C_2$) criteria. However, they do not cover user's mobility, with a significant impact on the total completion time. Additionally, their evaluations are conducted over a small-scale scenario, with no clear motivation for their parameters, hence leaving the elastic scalability criterion ($C_3$) and the support for mobility criterion ($C_4$) unmet. Finally, they do not account for the presence of various operators, hence leaving the federation criterion ($C_5$) unmet.

While previous studies disregard the possibility of enabling executions in the cloud stratum, Agarwal *et al.* [25][8] account for its presence and propose an algorithm that allows efficiently distributing the workload over the fog and the cloud strata. They consider that in a fog domain in the fog stratum, a *fog server manager* receives the user's requests and is responsible for matching the fog resources and the user demands. Upon the receipt of a request, the *fog server manager* verifies whether enough computational resources are available in the fog domain. Depending on the available resources, either it executes all tasks, executes part of them and postpones the execution of others, or it transfers the demand to nodes in cloud stratum, in order to run tasks over the cloud nodes. Simulation results show that the proposed algorithm is more efficient than other existing strategies aiming at optimizing the response time or enabling load-balancing. It leads to a lower maximum response time and lower maximum processing time values, at lower costs. In this work, the authors model the heterogeneity in limitations of the computational resources in the fog and the cloud strata and they consider that enough network resources are available for the communication between the two. Moreover, they aim at meeting the latency constraint on the user's side. Accordingly, they meet the heterogeneity criterion ($C_1$) and the QoS criterion ($C_2$). Nevertheless, they conduct their evaluation in a small-scale simulative environment, with no clear motivation for the chosen parameters. They do not cover the user's mobility, either – hence, violating the elastic scalability criterion ($C_3$) and the support for mobility criterion ($C_4$). Finally, they do not consider the presence of various providers and consequently do not meet the federation criterion ($C_5$).

Deng *et al.* [89][9] also tackle the problem of task scheduling over the cloud and the fog nodes. In contrast to Agarwal *et al.* [25], they aim at doing so at the lowest system power consumption while taking additional system constraints into consideration. In particular, they account for the limitations in the fog and the cloud nodes in terms of the computational capabilities, communication bandwidth limitations between the fog and the cloud nodes, the delay constraints on the user's side, and the workload balancing between the fog and the cloud nodes. Due to its complexity, the authors divide the problem into three parts to solve it. In the first part, they focus on the optimization of power in the fog stratum for a certain input workload, using convex optimization methods. In the second part, they optimize the power in the cloud stratum for an input workload, based on a linear optimization heuristic. In the third step, they optimize the network communication between the fog and the cloud, using a combinatorial optimization algorithm. The authors evaluate their strategy based on a set of a few fog and cloud nodes. Their results show that as a higher workload is attributed to the fog nodes, the power consumption of the system grows while system delay decreases. This is because nodes in the cloud stratum are more powerful and energy-efficient than the fog nodes while imposing additional communication delays. In this work, the limitations in resources, as well as the delay constraint on the user's side, are covered. Accordingly, the heterogeneity criterion ($C_1$) and the QoS criterion ($C_2$) are met. Nevertheless, the elastic scalability ($C_3$) and the support for mobility ($C_4$) criteria are not met, as a result of considering small-scale evaluation scenario and disregarding devices' mobility. The need for the federation criterion ($C_5$) is not met, either; only one fog provider and one cloud provider are considered.

---

[8] This work is also discussed in Section IV.A.2

[9] This work is also discussed in Section V.C

Based on this analysis, we conclude that the introduced task scheduling algorithms have mainly targeted the fog stratum, with only a couple of contributions operating over different strata. Moreover, most works meet the heterogeneity criterion ($C_1$) and the QoS criterion ($C_2$) when making decisions. However, the elastic scalability criterion ($C_3$), the support for mobility criterion ($C_4$), and the federation criterion ($C_5$) are met by none.

*3) Offloading and Load Redistribution*

As discussed in the previous section, several task-scheduling algorithms have been proposed so far in the context of the fog systems. While they allow distributing compute tasks over compute nodes across the three strata of the system, they have not considered the possible unbalance among nodes in terms of workload. In fact, these algorithms focus on minimizing the task blocking probability, the latency in the system or the energy consumption, and they may even lead to such unbalanced loads among nodes. This stresses the need for the algorithms that perform offloading and load redistribution in the system. In this context, Hassan *et al.* [91] and Ye *et al.* [92] have focused on offloading devices in the IoT/end-users stratum to nodes in the fog stratum. Instead, Fricker *et al.* [93], Ningning *et al.* [94] have only focused on the fog stratum and addressed offloading and load redistribution. In turn, Li *et al.* [95] also focus on the fog stratum only and propose coding schemes that lead to redistributing tasks and/or injecting redundant ones in the system. Finally, Ottenwalder *et al.* [96] study migrations in the overall system, an orthogonal aspect to offloading and load redistribution. We study these contributions below.

Hassan *et al.* [91] tackle the problem from the mobile devices perspective and aim at offloading them from their workload. To study the problem, they model the interactions among functions through a graph structure. In this graph, functions are modeled with nodes and the interactions among them are mapped to edges. They aim at splitting the graph into two parts: One that runs on the mobile device and the other that is offloaded to the nodes in the fog stratum, at the lowest possible execution time. The experimental results show that offloading to the fog stratum outperforms offloading to the cloud stratum or running everything on the mobile device from both the response time and energy consumption perspectives. In this work, the authors account for the various limitations in available resources and aim at meeting the QoS constraints, hence satisfying the heterogeneity ($C_1$) and the QoS ($C_2$) criteria. However, the evaluation is done in a small-scale environment, and thus the elastic scalability criterion ($C_3$) is not met. The mobility is not covered either in the study. Besides, the federation is not considered, hence leaving the support for mobility criterion ($C_4$) and the need for the federation criterion ($C_5$) unmet.

Ye *et al.* [92] extend the scope of the work by Hassan *et al.* [91] and consider offloading simultaneously cloudlets and mobile devices to a fog domain in the fog stratum. More precisely, they consider that a cloudlet system is deployed over roadside units in an urban area and they propose to enable buses with the fog computing capabilities. In this context, the authors propose an offloading strategy that is based on a genetic algorithm. It aims at offloading devices at low transmission and energy costs, while ensuring a high utility and meeting the user's application deadline. The proposed method operates over a set of solutions, evaluates their fitness, and keeps those that imply the highest fitness values. It then relies on these solutions in order to form new possibilities in the crossover and mutation steps of genetic algorithms. The solutions with the highest fitness are again selected. The procedure keeps running until the number of iterations exceeds a set limit. The evaluation of the strategy shows that it allows offloading devices at a low cost, even in the presence of a low and high number of users. Also, the cost is shown to significantly decrease in presence of a high number of buses. In this work, the authors consider that the devices are homogeneous inside the fog domain. So, they do not meet the heterogeneity criterion ($C_1$). Instead, they set a limit on the maximum response time, hence satisfying the QoS criterion ($C_2$). Evaluations are conducted in a small area of 2x2 km$^2$. Accordingly, the elastic scalability criterion ($C_3$) is not met. As for the support for mobility criterion ($C_4$), it is satisfied since the authors lead their study while accounting for the movements of the buses. Finally, the authors focus on offloading to a single fog domain, and thus the federation criterion ($C_5$) is not relevant for their study.

Fricker *et al.* [93] only target offloading in the fog stratum and focus on a scenario with data centers deployed in the fog stratum. In this context, they study the problem of offloading small data centers to the big neighboring ones. More precisely, in case of blocked requests at a small data center, they forward them to a big backup neighboring server according to an offloading probability. The authors characterize the performance of the system analytically and complement it with numerical results. Based on their analysis, the authors show that the proposed strategy significantly reduces the requests blocking rate at the small data center, with minor impact on the blocking rate over big data centers. In this work, the authors consider the heterogeneity of data centers; meeting accordingly the heterogeneity criterion ($C_1$). As they also aim at reducing the requests blocking rate, they satisfy the QoS criterion ($C_2$). The elastic scalability criterion ($C_3$) is satisfied, as the authors consider the case of heavy high loads. Instead, the support for mobility criterion ($C_4$) is not relevant for this work, as the methods operate at the level of data centers. Finally, the federation criterion ($C_5$) is not met, as the authors do not cover the possibility of having data centers belonging to different providers.

From a similar perspective, Ningning *et al.* [94] propose a dynamic load balancing algorithm in the fog stratum. It allows coping with the dynamic arrival and exit of nodes inside a fog domain. The algorithm starts with an atomization step that maps physical resources into virtual resources. A graph representing the system is then built. In this graph, each virtual resource is represented by a node and has a certain capacity. An edge links each couple of virtual nodes and is weighted by the bandwidth of communication link between them. In an iterative process, the graph is partitioned by removing edges one after the other according to a minimum weight threshold that guarantees a compatible degree of distribution of tasks over the virtual machines. Upon the arrival of a new fog node, the algorithm redistributes loads in its neighborhood in order to balance the loads, while accounting for the task distribution degree and the links among the nodes. A reverse strategy is adopted in case of node removal. The algorithm is evaluated against a static one from the literature. The evaluation results show that the proposed strategy leads to a lower number of moves in the graph, implying a lower migration cost. Additionally, it requires less time to derive results with respect to the static strategy. In this work, the authors consider the fog nodes as homogeneous, leaving by that the heterogeneity criterion ($C_1$) unmet. However, they aim at satisfying the user demands in terms of tasks and accordingly meet the QoS criterion ($C_2$). However, the elastic scalability criterion ($C_3$) is not met as the evaluations are conducted over a scenario with only 10 fog nodes and 10 mobile devices. The work meets the support for the mobility criterion ($C_4$) as the authors account for the arrival and exit of the fog nodes as a result of their mobility. Finally, the need for the federation criterion ($C_5$) is not relevant for this work as the authors operate at the level of a single fog domain.

Li *et al.* [95] introduce a coding framework that allows to redistribute tasks and/or inject redundant ones in the fog stratum. The framework operates by considering the trade-off between communication load and computation latency. Depending on the system's characteristics and imposed requirements, one of two coding schemes is used. The first coding scheme aims at minimizing bandwidth usage. It runs more computations at each node that would require a lower exchange of information among nodes. This leads thus to a reduced communication load. The second coding scheme targets the minimization of latency. It operates by injecting redundant computations over nodes that allow minimizing the computation time in case some nodes are slower than others or blocked. The two coding schemes and the trade-off are illustrated in the paper. The coding schemes used in this paper are designed in such a way to handle the heterogeneity among nodes in terms of computing capabilities. By that, the work meets the heterogeneity criterion ($C_1$). The framework allows selecting the coding scheme with respect to the computation latency. Thus, the QoS management criterion ($C_2$) is met. The coding schemes enable operations as part of large-scale systems. However, the evaluation does not cover such a scenario. The scalability criterion ($C_3$) is therefore not met. The mobility criterion ($C_4$) is unmet, as the impact of mobility is not covered. Finally, the federation criterion ($C_5$) is not met in this work, targeting a single provider.

A complementary aspect to workload redistribution is covered by Ottenwalder *et al.* [96]. The authors cover migration in the fog system, with a particular attention to devices' mobility. Their objective is to enable migrations while minimizing the placement and migration costs, meeting the consumer latency restriction, and accounting for the user's mobility. To do so, they propose to build a migration plan for each operator, triggering migrations at discrete time steps. The migration plan is obtained from a time-graph structure, showing possible migrations for an operator over time according to the user's mobility. In this time-graph structure, the shortest path reflecting the lowest network utilization is chosen as the migration plan. In a following step, migration plans are coordinated between operators, in order to ensure there are enough resources for the migration. A similar algorithm is introduced for uncertain mobility patterns, with migration plans including additional targets linked with weighted transfer probabilities. The evaluation of this strategy shows that it saves 49% and 27% of the network utilization resources in case of the complete and uncertain mobility patterns respectively. In this study, the limitations in heterogeneous resources are taken into account, the user's QoS is considered and evaluations are conducted based on a sample of 1000 vehicles with realistic mobility patterns, indicating that the method can be employed at large scales in a real-world environment. Based on this, the heterogeneity ($C_1$), the QoS ($C_2$) and the elastic scalability ($C_3$) criteria are met. Mobility remains at the core of the proposed method. Therefore, the support for mobility criterion ($C_4$) is met. Moreover, various resource costs are considered, enabling outsourcing decisions inside a federation. As a result, the need for the federation criterion ($C_5$) is met.

Based on our review in this section, it is concluded that offloading and load redistribution algorithms have mainly focused so far on the IoT/end-users and the fog strata. All contributions meet the QoS criterion ($C_2$) while the remaining heterogeneity criterion ($C_1$), the elastic scalability criterion ($C_3$), the support for mobility criterion ($C_4$) and the federation

TABLE IV
The main features and criteria for works proposing algorithms for fog systems. In the approach column, A is analysis, E is exact algorithm, G is Graph-based algorithm, H is Heuristic, P is policy, S is Scheme and x means no approach is proposed. In the evaluation column, E is Experimental, S is Simulative and x means no evaluation is performed. In the criteria columns, ✓ means the requirement is met, x means the requirement is not met and - means the requirement is not applicable.

| Scope | | Paper | Approach | Evaluation | Major Contribution | Criteria | | | | |
|---|---|---|---|---|---|---|---|---|---|---|
| | | | | | | $C_1$ | $C_2$ | $C_3$ | $C_4$ | $C_5$ |
| Computing in The Fog Systems | Resource Sharing | Abedin et al. [82] | H | S | Propose an algorithm to enable resource sharing among fog nodes. | ✓ | x | x | x | - |
| | | Oueis et al. [83] | E | S | Propose an algorithm to cluster small cells to enable resource sharing among them. | ✓ | ✓ | x | x | x |
| | | Nishio et al. [84] | E | E | Present a strategy to optimize the sharing of resources with the objective of maximizing the corresponding utility. | ✓ | ✓ | x | x | x |
| | Task Scheduling | Oueis et al. [85] | H | S | Introduce an algorithm to manage the execution of tasks in small-cell fog stratum. | ✓ | ✓ | x | x | x |
| | | Intharawijitr et al. [86] | P | S | Present three policies to select fog nodes to execute tasks. | x | ✓ | x | x | x |
| | | Aazam et al. [87] | H | S | Propose a loyalty-based task scheduling algorithm. | ✓ | ✓ | x | x | x |
| | | Aazam et al. [90] | H | S | Propose a loyalty-based task scheduling algorithm by considering different types of devices. | ✓ | ✓ | x | x | x |
| | | Zeng et al. [88] | H | S | Introduce a task scheduling and image placement algorithm that aims at minimizing the overall completion time. | ✓ | ✓ | x | x | x |
| | | Agarwal et al. [25] | H | S | Introduce an algorithm to distribute workload to reduce response time and cost. | ✓ | ✓ | x | x | x |
| | | Deng et al. [89] | H | S | Introduce an algorithm to distribute workload in cloud/fog systems, at lowest power cost. | ✓ | ✓ | x | x | x |
| | Offloading and Load Redistribution | Hassan et al. [91] | G | E | Propose a strategy to offload applications on mobile devices. | ✓ | ✓ | x | x | x |
| | | Ye et al. [92] | H | S | Propose a strategy to offload cloudlets and mobile devices to fog-enabled buses. | x | ✓ | x | ✓ | - |
| | | Fricker et al. [93] | H | S | Evaluate the performance of a data center offloading strategy in the fog layer. | ✓ | ✓ | ✓ | - | x |
| | | Ningning et al. [94] | H | S | Introduce a dynamic load balancing algorithm in the fog layer that allows coping with the dynamic arrival and exit of fog nodes. | x | ✓ | x | ✓ | - |
| | | Li et al. [95] | S | S | Propose a coding framework to handle redundancy in tasks computation in fog computing. | ✓ | ✓ | x | x | x |
| | | Ottenwalder et al. [96] | G | S | Propose a strategy that allows migrating operators at minimal migration cost. | ✓ | ✓ | ✓ | ✓ | ✓ |

criterion ($C_5$) are not always met and are not relevant in some cases.

B. *Algorithms for Content Storage and Distribution in The Fog Systems*

Besides computing in the fog systems, an important aspect to tackle is content storage and distribution. Previous works have mainly considered this aspect in the specific context of Fog-Radio Access Networks (F-RAN), as in the case of Tandon et al. [97], Park et al. [98], Hung et al. [99], and Xiang et al. [100]. Other works have covered this aspect for the fog systems, operating on top of any underlying communication network, as in the case of Do et al. [101], Jingtao et al. [102], Malensek et al. [103], and Hassan et al. [91].

F-RAN is an extension of the Cloud-RAN (C-RAN) concept. More precisely, C-RAN enables the cloudification of the radio functionalities in cellular networks, offering a significant degree of flexibility in the management of the cellular radio access network. However, this comes at the cost of a high latency in the system. F-RAN complements it in this sense, by keeping part of radio functionalities close to the users and enabling content caching there as well to serve content to users with low latency [23]. Tandon *et al.* [97] analyze the performance of the F-RAN system with a particular focus on the trade-off that can be obtained among latency, caching, and fronthaul capacity metrics. They derive analytical results in the case of two users that are served by two edge nodes. Their analysis shows that depending on the fronthaul capacity, two different regimes can be identified. The first one is a low-fronthaul capacity regime that implies an optimal latency in case caches are of high capacity. The second one is a high-fronthaul capacity regime that requires both caching and using the cloud to reach optimal latency. In this study, the authors consider that all caches are of the same size, hence leaving the support for the heterogeneity criterion ($C_1$) unmet. However, they aim at optimizing the latency in the system, meeting by that the QoS criterion ($C_2$). Meanwhile, their evaluations are conducted only for a few number of users and fog nodes, leaving the elastic scalability criterion ($C_3$) unmet. Similarly, they do not address the users' mobility and accordingly leave the support for mobility criterion ($C_4$) unmet as well. Finally, as they consider operations in the context of one cellular network provider, the federation criterion ($C_5$) is not relevant for their work.

Park *et al.* [98] analyze the performance of the hard and soft transfer modes in the F-RAN system to serve content to users. In the hard transfer mode, they consider that baseband processing takes place only on the Baseband Unit (BBU) side. Instead, in the soft transfer mode, they consider that the centralized precoding is performed over the BBU and is complemented by a local precoding at the RRH side. In this context, they study the problem of minimizing the delivery latency and accordingly assess the performance of the two modes. Their results indicate that the soft-transfer mode leads to lower latency in case the fronthaul capacity is low or the Signal to Noise Ratio (SNR) is high. Instead, the hard transfer mode is more efficient in other regimes. In this work, the authors consider different capacity limitations of caches over RRHs. Consequently, the work meets the heterogeneity criterion ($C_1$). The QoS criterion ($C_2$) is also met since the authors aim at minimizing it. However, the elastic scalability ($C_3$) and the support for the mobility ($C_4$) criteria are not met, as the authors neither cover a large-scale scenario and nor assess the impact of the user's mobility. Finally, as they consider operations over a single cellular network provider, the federation criterion ($C_5$) is not relevant for their work.

Park *et al.* [104] extend the scope of their work in [98] and consider a hybrid delivery mode that combines both hard- and soft-transfer modes. They assess its performance as part of a delivery phase optimization problem in which they aim at maximizing the delivery rate while satisfying fronthaul capacity and power constraints. Their results indicate that the proposed hybrid-mode outperforms the performance of hard- or soft-transfer modes alone. In this work, the authors consider that different caches have different sizes. Accordingly, the heterogeneity criterion ($C_1$) is met. However, QoS is not covered although latency is relevant for the work. Hence, the QoS criterion ($C_2$) is not met. The authors do not cover the elastic scalability. Similarly, they do not consider the impact of the number of users on their study. Neither do the authors discuss the end-users' mobility. Therefore, the scalability ($C_3$) and the mobility ($C_4$) criteria are unmet. Finally, the need for the federation criterion ($C_5$) is not relevant for this study as it targets one cellular network provider.

Chen *et al.* [105][10] study instead the problem of radio resource sharing and cooperation among fog-enabled radio units in cellular networks to serve the content to users. In their system, they consider that the content can be either stored on local caches placed over radio units or over central processors. Moreover, users are served in a multicast fashion. In this context, the authors propose an algorithm to optimize the beamforming and clustering of radio units, in order to minimize the power consumption while still meeting a user's QoS and the capacity of backhaul links. The authors solve the problem by designing an algorithm that combines relaxation and approximation strategies. The evaluation of the algorithm shows that it can reach optimal solutions after several iterations. In this work, the authors do not model the heterogeneity among the fog nodes. So, they do not meet the heterogeneity criterion ($C_1$). However, they meet the QoS criterion ($C_2$), as they aim at satisfying QoS on the user's side, by setting a threshold on the Signal to Interference and Noise Ratio (SINR). The elastic scalability criterion ($C_3$) is met, as the authors consider the presence of a large number of users over several fog nodes. The support for the mobility ($C_4$) criterion is not met, as the authors do not take into account the users' mobility. Finally, the need for the federation criterion ($C_5$) is not met, since the authors do not consider that the radio units can belong to various providers.

Hung *et al.* [99] also focus on the analysis of an F-RAN system. They aim at identifying which content to cache in the cloud and which content to cache in the fog stratum, according to the content features. Their objective is to do so at the lowest communication cost while respecting communication link capacities and cache sizes. The results indicate that it is better to save high-rank Internet contents with big size over the cloud. Instead, small-size files are better saved in the small proximity of users. In this work, the authors do not account for differences in the capabilities of nodes. So, they do not satisfy the heterogeneity criterion ($C_1$). Instead, they consider the delay as part of their analysis, meeting by that the QoS criterion ($C_2$). The elastic scalability criterion ($C_3$) and the support for mobility criterion ($C_4$) remain unmet, as the authors conduct evaluations in case of a single fog node with 100 users and do not analyze what happens in case the devices are moving. Finally, the need for the federation criterion ($C_5$) is not relevant for their study, as they target one cellular network provider in particular.

Xiang *et al.* [100][11] go beyond caching in the cloud and the fog strata of the F-RAN system and consider that the content can be cached over the user equipment. In this scenario, the authors aim at maximizing the energy efficiency of the system, by optimizing the mode selection and resource allocation when serving content to users. In particular, they consider that three different communication modes exist: Device to device model, single serving antenna, and coordination among antennas. They propose an algorithm that relies on particle swarm optimization and show that it leads to a tradeoff between average energy efficiency and average delay. In their work, the authors do not account for the heterogeneity among devices, leaving the heterogeneity criterion ($C_1$) unmet. However, they account for the latency on the user's side and try to satisfy it, hence meeting the QoS criterion ($C_2$). However, the authors do not cover the elastic scalability and support for mobility ($C_3$ and $C_4$), leaving them unmet. Finally, the need for the federation criterion ($C_5$) is not relevant for their work, as the study targets one cellular network provider.

---

[10] This work is also discussed in Section V.C

[11] This work is also discussed in Section V.C

Do *et al.* [101] consider the management of content in a fog-based content distribution network. They aim at optimizing the content distribution from the data centers to the fog nodes, in a way that maximizes the corresponding utility, while minimizing the carbon footprint. Given the large number of fog nodes that content distribution networks cover, the authors propose to solve the problem via a distributed proximal algorithm that divides it into many subproblems and solve them in a small number of iterations. Their numerical results confirm the expected performance. Based on simulations over a set of 100 users, the algorithm is shown to converge to near optimum in a few iterations. In this work, the authors take into consideration the limitation of the workload capacity and assume that enough bandwidth is available between the data centers and the fog nodes to enable content distribution. Accordingly, they fulfill the heterogeneity criterion ($C_1$). Despite its relevance, the QoS on the user's side is not considered as part of the content distribution phase, thus the QoS criterion ($C_2$) is unmet. While the algorithm is designed to cope with large-scale fog systems, its evaluation is conducted in a small-scale scenario with arbitrary parameters. Therefore, the elastic scalability criterion ($C_3$) is not met. Also, no discussion concerning the user's mobility is covered, leading as well to the violation of the support for the mobility criterion ($C_4$). Finally, the need for the federation criterion ($C_5$) is not relevant for the study, as the authors operate at the level of a single content delivery network provider.

Jingtao *et al.* [102] tackle the problem of connecting the caching nodes at the lowest connectivity cost in the fog systems. In particular, they assume the presence of static fog clusters, where each cluster is formed by a movie, web, file, and game servers, with each web server including its own cache. They model their problem by using a graph structure, where each node represents a generic server and each edge links a couple of servers with a connection. Each edge is weighted based on the connection cost. To solve the problem, the authors consider a Steiner tree scheme. It aims at interconnecting a set of nodes of interest by a network of shortest length while allowing of extra vertices and edges to the network. The proposed algorithm operates in the following steps: First, it constructs a new graph structure over the set of nodes of interest, linked through edges whose weights represent the shortest paths between them. Then, the algorithm generates the corresponding spanning tree. In the next step, the tree is expanded to include the hidden nodes. After that, the minimum spanning of the expanded graph is derived. The evaluation of the strategy over a set of four clusters shows that it outperforms the shortest-path scheme. In this study, the authors consider the limitations in terms of network resources and thus meet the heterogeneity criterion ($C_1$). However, they do not take into consideration the QoS on the devices side, leaving the QoS criterion ($C_2$) unmet. Evaluations are also conducted over a small-scale scenario, with arbitrary parameters. So, the elastic scalability criterion ($C_3$) is not met. Besides, devices mobility is not covered and, as a result, the support for mobility criterion ($C_4$) is unmet. The need for the federation criterion ($C_5$) is in turn not met, as the authors do not discuss the possibility of operating across several fog providers.

Malensek *et al.* [103] consider sampling techniques over data streams that help enable a better usage of storage resources in the fog system. They propose a hierarchical data structure called spillway to handle sampling efficiently. On top of it, they employ an extension of the reservoir sampling technique. The reservoir sampling technique is a random sampling technique that operates over an array of a fixed size. There, entries are inserted in the array, as long as it is not full. Once full, new entries replace existing ones with a certain probability. The extended technique employs the same concept over the spillway structure, which is a group of reservoirs organized in a hierarchical structure. By that, the spillway structure allows for both short-term and long-term analyses. The technique is evaluated with experiments in a real-world environment. The results show that the employed technique leads to an error that remains below 0.25%. The heterogeneity criterion ($C_1$) is met in this work as heterogeneous nodes in terms of storage technology and capabilities are handled. The QoS management criterion ($C_2$) is not met, as the authors do not study the impact of sampling on QoS. Evaluations are conducted in a real-world environment with a few devices only. Therefore, the scalability criterion ($C_3$) is not met. The mobility is not addressed; as a result, the mobility criterion ($C_4$) is not met. Finally, the federation criterion ($C_5$) is met as the authors employ specific mechanisms to enable federated storage and retrieval of information among multiple domains.

The works discussed in this section focus on problems at the system level. Hassan *et al.* [91] focus instead on the individual user level and propose to distribute mobile phone data over a personal storage space, formed by the user's personal devices, e.g., personal computers, as part of the fog stratum. The authors propose to place content on this personal storage space, to minimize the communication overhead, taking into account the disk space limitation. Experimental results show that higher throughputs can be obtained when the fog is used with respect to local or Dropbox storages. In this work, the authors consider distinct limitations in available resources and aim at meeting the QoS constraints. Accordingly, they meet the heterogeneity ($C_1$) and the QoS ($C_2$) criteria. Nevertheless, the evaluation is conducted in a small-scale scenario, leaving the elastic scalability criterion ($C_3$) unmet. The support for mobility criterion ($C_4$) is met, as the authors propose to keep metadata in the cloud to enable downloads independently of the user's location. Finally, the need for federation criterion ($C_5$) is not relevant for this study, as it targets personal storage.

Based on our analysis, it can be concluded that previous research on the content-related aspects in the fog system has mainly focused on the case of F-RAN systems. The heterogeneity criterion ($C_1$) and the QoS criterion ($C_2$) are met by the majority of contributions, while the elastic scalability criterion ($C_3$) and the support for mobility criterion ($C_4$) are not met in general. Instead, the federation criterion ($C_5$) is not relevant for almost all contributions.

*C. Algorithms and Energy Consumption in The Fog Systems*

Environmental concerns, alongside growing fuel prices, are channeling research efforts to energy consumption in the computing systems and in particular in the fog systems. Sarkar *et al.* [106] [107], Jalali *et al.* [108], Oueis *et al.* [83], Nishio *et al.* [84], and Cao *et al.* [109] have analyzed and assessed energy consumption in the overall system. Oueis *et al.* [85], Chen *et al.* [105], Xiang *et al.* [100], Deng *et al.* [89], and Ye *et al.* [92] have considered the design of strategies aiming at reducing energy consumption in the system.

Sarkar *et al.* [106] [107] and Jalali *et al.* [108] focus on the analysis of energy consumption in the fog systems. Sarkar *et al.* [106] [107] consider in particular the performance of the fog systems in the context of IoT. They compare several metrics including power consumption, service latency, and $CO_2$ emission in a fog system to the case of the cloud system. By simulating real-time IoT services in 100 cities served with 8 data centers, the authors conclude that fog computing is more efficient than cloud computing. From a latency perspective, they observe that with 25% of applications requesting real-time services, the service latency decreases by 30%. In turn, the power consumption decreases by 42.2% and $CO_2$ emissions decrease by more than 50%, translating into significant reductions in cost. Better results are even obtained for a higher portion of real-time services. In this work, the authors disregard the heterogeneity among devices, leaving the heterogeneity criterion ($C_1$) unmet. However, they evaluate the delay in the system and thus meet the QoS criterion ($C_2$). In turn, they meet the elastic scalability criterion ($C_3$) as they conduct their analysis with hundreds of thousands of nodes over 100 cities. The support for the mobility criterion ($C_4$) is not met as the authors do not consider the movements of the nodes in the system. Finally, the need for federation ($C_5$) is not relevant for this study, as it does not affect the conducted analysis.

Jalali *et al.* [108] also focus on the analysis of energy consumption in the fog systems with respect to the cloud computing system. They consider a fog stratum where nano data centers form the fog nodes. The results of their analysis accord with those by Sarkar *et al.* [106] [107]. They show that nano data centers are more energy-efficient than the data centers in the cloud, by pushing content close to end-users and decreasing the energy consumption in the network. However, in case the connecting network is inefficient from an energy perspective, or the active time of the nano data center is high, an increase in the energy consumption can be obtained. In this study, the heterogeneity among the fog nodes and the cloud nodes is taken into account. The heterogeneity criterion ($C_1$) is therefore met. The authors also aim at satisfying the arriving requests and, as a result, the QoS criterion ($C_2$) is met. However, the elastic scalability criterion ($C_3$) is not met, as the evaluation is conducted over a small-scale scenario. The support for mobility criterion ($C_4$) is not met, either as the movements of the nodes in the system are not taken into account. Finally, the need for federation ($C_5$) is not relevant for this work, as it does not affect the analysis.

Other works that propose algorithms to operate in the fog system have evaluated the impact of their strategies on the energy consumption. Oueis *et al.* [83] propose a clustering strategy that allows small cells to share their resources in order to offload mobile devices from their workload. In their evaluation, they show that their strategy allows satisfying a higher percentage of the user demands with respect to other clustering strategies, at the price of intermediate power consumption per user[12]. Nishio *et al.* [84] also focus on resource sharing in the fog systems in the context of cellular networks and they optimize the utilization of CPU, bandwidth, and content. In their evaluation, they show that the proposed algorithm allows reducing latency and leads to high energy efficiency[12]. Instead, Cao *et al.* [109] design fall detection and analysis strategies as part of strokes mitigating application. The components of their application are split between the fog and the cloud strata. The evaluation of their fall detection algorithm shows that it leads to a low missing rate and a low false alarm rate. Moreover, its response time and energy consumption are close to the existing strategies[13].

Several works have also aimed at designing algorithms that aim at minimizing the energy consumption in the system. Oueis *et al.* [85] introduce a method to cluster small cells into computational clusters to process the users' requests that could not be served by individual small cells. They consider three variations of the algorithm to show that the power-centric one leads to low power consumption per user and, as a result, to a low energy consumption per user[12]. Chen *et al.* [105] consider the problem of radio resource sharing among radio units in a fog-RAN system in order to cooperatively serve content to users. They design an algorithm that allows optimizing

---

[12] See section V.A for more details about this contribution

[13] See section V.D for more details about this contribution

TABLE V

The main features and criteria for works proposing algorithms for fog systems. In the approach column, A is analysis, E is exact algorithm, G is Graph-based algorithm, H is Heuristic, P is policy, S is Scheme and x means no approach is proposed. In the evaluation column, E is Experimental, S is Simulative and x means no evaluation is performed. In the requirements columns, ✓ means the requirement is met, x means the requirement is not met and - means the requirement is not applicable.

| Scope | Paper | Approach | Evaluation | Major Contribution | $C_1$ | $C_2$ | $C_3$ | $C_4$ | $C_5$ |
|---|---|---|---|---|---|---|---|---|---|
| Content Storage and Distribution in The Fog Systems | Tandon et al. [97] | A | - | Analyze trade-offs in the F-RAN system among latency, fronthaul capacity and caching storage capacity. | x | ✓ | x | x | - |
| | Park et al. [98] | A | S | Analyze the performance of the hard and soft transfer modes in the F-RAN system. | ✓ | ✓ | x | x | - |
| | Park et al. [104] | A | S | Analyze the performance of a hybrid transfer mode in the F-RAN system. | ✓ | x | x | x | - |
| | Chen et al. [105] | H | S | Introduce a method to optimize sharing of radio resources over radio units in order to serve content to users. | x | ✓ | ✓ | x | x |
| | Hung et al. [99] | A | S | Analyze the performance of a caching system in an F-RAN system. | x | ✓ | x | x | - |
| | Xiang et al. [100] | H | S | Introduce a mode selection algorithm maximizing the energy efficiency in F-RAN systems while serving content from caches to users. | x | ✓ | x | x | - |
| | Do et al. [101] | H | S | Propose an algorithm to optimize the distribution of video contents over fog nodes. | ✓ | x | x | x | - |
| | Jingtao et al. [102] | G | S | Propose a strategy to place content on caches at the lowest communication cost. | ✓ | x | x | x | x |
| | Malensek et al. [103] | H | E | Propose a data structure and technique to handle sampling of data streams in fog computing. | ✓ | x | x | x | ✓ |
| | Hassan et al. [91] | E | E | Propose a strategy to place mobile data over expanded storage. | ✓ | ✓ | x | ✓ | - |
| Energy Consumption | Sarkar et al. [106] [107] | E/A | S | Analyze the performance of fog computing in the context of IoT, in terms of power consumption, service latency and $CO_2$ emission. | x | ✓ | ✓ | x | - |
| | Jalali et al. [108] | A | S | Analyze energy consumption in fog systems with respect to cloud computing systems. | ✓ | ✓ | x | x | - |
| | Oueis et al. [83] | E | S | Propose a small cell clustering algorithm, efficient at serving requests, at the cost of intermediate power consumption. | ✓ | ✓ | x | x | x |
| | Nishio et al. [84] | E | E | Present a strategy to optimize the sharing of resources, leading to high energy efficiency. | ✓ | ✓ | x | x | x |
| | Cao et al. [109] | S | E | Present a fall detection algorithm leading to energy consumption values that are close to other existing strategies. | - | ✓ | - | - | - |
| | Oueis et al. [85] | H | S | Introduce a power-centric method to form computational clusters of small cells to process computational requests. | ✓ | ✓ | x | x | x |
| | Chen et al. [105] | H | S | Optimize sharing of radio resources over radio units in order to serve content to users at lowest power consumption. | x | ✓ | ✓ | x | x |
| | Xiang et al. [100] | H | S | Introduce a mode selection algorithm maximizing the energy efficiency in F-RAN systems while serving content from caches to users. | x | ✓ | x | x | - |
| | Deng et al. [89] | H | S | Introduce an algorithm to distribute workload in cloud/fog systems, at lowest power cost. | ✓ | ✓ | x | x | x |
| | Ye et al. [92] | H | S | Propose a strategy to offload cloudlets and mobile devices to fog-enabled buses at low transmission and energy costs. | x | ✓ | x | ✓ | - |

beamforming and clustering at the minimum power consumption. The evaluation of their algorithm shows that it can reach an optimal solution in terms of power consumption in a few iterations[14]. Xiang et al. [100] aim at maximizing the energy efficiency of the fog-RAN system, by optimizing the mode selection and resource allocation when serving content to users. In particular, they consider that three different communication modes exist: Device to device model, single serving antenna, and coordination among antennas. They propose an algorithm that relies on particle swarm optimization and show that it leads to a tradeoff between average energy efficiency and average delay[13].

---
[14] See section V.B for more details about this contribution

Deng *et al.* [89] distribute workloads among the cloud nodes and the fog nodes at the lowest system power consumption. Their results show that as a higher workload is attributed to the fog nodes, the system power consumption grows while the system delay decreases[15]. This is because the cloud nodes are more powerful and energy-efficient than the fog nodes while imposing additional communication delays. Ye *et al.* [92] cover the problem of the cloudlet and the devices offloading to a fog stratum. They introduce an algorithm that manages the offloading strategy at low transmission and energy costs. The evaluation of this strategy shows that it allows offloading the devices according to the set objective[15].

Based on our discussion, it is concluded that significant savings in terms of energy can be obtained in the fog systems. When it comes to the evaluation criteria, all studies meet the QoS criterion ($C_2$). The heterogeneity criterion ($C_1$) is met by some, while the elastic scalability criterion ($C_3$) and the support for mobility criterion ($C_4$) are not met. Instead, the federation criterion ($C_5$) remains not relevant for most of the contributions.

*D. Specific End-User Application Algorithms for The Fog Systems*

While the majority of previous algorithmic efforts target applications at large, a few works introduce algorithms that target specific applications. In this section, the algorithms that target healthcare applications and then those that target other applications are discussed.

*1) Algorithms for Healthcare Applications*

Healthcare applications have been covered by Gu *et al.* [110], Cao *et al.* [109], and Craciunescu *et al.* [111]. Gu *et al.* [110] aim at optimizing resource utilization in the fog stratum for medical cyber-physical systems in order to minimize the communication cost as well as virtual machines deployment cost. In their system, they consider that the cellular network-base stations constitute the fog nodes. Accordingly, they consider in their model base station-user associations, task distribution, and virtual machine placement with QoS constraints. They formulate their problem as a mixed-integer non-linear program. Due to its complexity, they propose an algorithm that leads to a near-optimal solution. The algorithm relies on two phases: The first phase allows of minimizing the user's uplink communication cost while the second phase allows of minimizing the inter-BS communication cost and the VM deployment cost. The simulation results show that the algorithm outperforms a greedy algorithm when varying a list of parameters including the number of BSs, subcarriers, and users and the request arrival rate. In this work, the authors take into consideration the limitations in heterogeneous resources as well as QoS constraints, hence meeting the heterogeneity ($C_1$) and the QoS ($C_2$) criteria. While their evaluation is conducted over a larger scale than in other studies, it is still limited to 100 users and 50 BSs, with no realistic system parameters. Moreover, they do not consider the user's mobility. Thus, the elastic scalability criterion ($C_3$) and the support for mobility criterion ($C_4$) remain unmet. Finally, the need for the federation criterion ($C_5$) is not relevant for this study, as it targets a single network operator.

A fall detection and an analysis application for mitigating strokes is introduced by Cao *et al.* [109]. They consider that the application components are split between the fog and the cloud strata. In the fog stratum, a simple fall detection algorithm analyzes the acceleration magnitude values measured by the mobile device. It compares recorded values to two thresholds in order to detect the three consecutive stages of a fall: Free fall, hitting, and inactivity. To filter the detections linked to usual daily activities, e.g., jumping into a surface, a filtering process is also run in the fog stratum. The filtering process relies on the comparisons of the magnitude of acceleration and free fall interval duration recorded to those of typical daily activities. On the cloud, a more complex procedure is conducted to identify false events. It relies on the analysis of time series evolution of measurements. The evaluation of the strategy shows that it is characterized by a low missing rate and a low false alarm rate. Moreover, its response time and energy consumption are close to the existing strategies. In this work, the authors consider that enough resources are made available in the system, by the resource allocation entity, to allow of running their application. Based on that, the heterogeneity criterion ($C_1$) is not relevant. Instead, the authors aim at providing a low response time and thus meeting the QoS criterion ($C_2$). The elastic scalability criterion ($C_3$), the support for mobility criterion ($C_4$) and the need for the federation criterion ($C_5$) are in turn not relevant for this work, as the proposed algorithm targets an individual instance of an application and the mobility support should be offered by another algorithm.

Craciunescu *et al.* [111] also study the feasibility of the fog computing paradigm in the context of e-health applications. To this end, they consider a home-based fog system, used to process sensitive real-time data and a cloud system, where historical data is stored. In the home-based fog system, a fall detection algorithm is introduced for patients, which aims at detecting cases of a patient's falls. The fall is detected according to the peaks in acceleration variations. The evaluation of the proposed strategy shows that, in more than 90% of the cases, falls are detected. Moreover, the encountered delay is in the order of 1sec. Instead, in the case of computations in a cloud environment, this value can go up to 5 sec, which is not tolerable in case of health emergencies. The heterogeneity criterion ($C_1$) is not relevant for this work, as the proposed algorithm operates by assuming that the resource allocation algorithms provide enough resources. The QoS criterion ($C_2$) is met, as the latency is covered in the evaluation results. The elastic scalability criterion ($C_3$), the support for mobility criterion ($C_4$) and the need for the federation criterion ($C_5$) are not relevant for this work, as the proposed algorithm targets an individual instance of an application and another algorithm is needed for the mobility support.

---

[15] See section V.A for more details about this contribution

TABLE VI
The main features and criteria for works proposing algorithms for fog systems. In the approach column, A is Analysis, E is Exact algorithm, G is Graph-based algorithm, H is Heuristic, P is Policy, S is Scheme and x means no approach is proposed. In the evaluation column, E is Experimental, S is Simulative and x means no evaluation is performed. In the criteria columns, ✓ means the criterion is met, x means the criterion is not met and - means the criterion is not relevant.

| Scope | | Paper | Approach | Evaluation | Major Contribution | Criteria | | | | |
|---|---|---|---|---|---|---|---|---|---|---|
| | | | | | | $C_1$ | $C_2$ | $C_3$ | $C_4$ | $C_5$ |
| Specific End-user Application Algorithms | Healthcare Applications | Gu et al. [110] | H | S | Introduce an algorithm to manage resources in medical cyber-physical systems. | ✓ | ✓ | x | x | - |
| | | Cao et al. [109] | S | E | Present an algorithm to detect individual falls. | - | ✓ | - | - | - |
| | | Craciunescu et al. [111] | S | E | Present an algorithm to detect individual falls. | - | ✓ | - | - | - |
| | Other Applications | He et al. [112] | H | S | Propose a framework for scheduling transcoding tasks in fog computing. | ✓ | ✓ | x | x | - |
| | | Tang et al. [113] | H | S | Propose a download cooperation model among users for video streaming. | ✓ | ✓ | x | x | - |
| | | Mei et al. [114] | S | E | Propose an algorithm to measure UV radiations. | - | x | x | - | - |
| | | Zhu et al. [115] | x | x | Propose to optimize website performance at fog nodes. | - | x | x | - | - |
| | | Kim et al. [116] | S | S | Introduce an algorithm to match vehicles and parking lots. | - | x | x | ✓ | - |
| | | Li et al. [117] | H | S | Propose algorithms for gaming applications in fog systems. | ✓ | ✓ | ✓ | x | - |

Our review of algorithms for healthcare applications shows that the number of such contributions is limited. When it comes to the evaluation criteria, all works meet the QoS criterion ($C_2$). Other criteria are either not applicable or not met.

*2) Algorithms for Other Applications*

Besides healthcare, other applications with different targets have been covered in the literature: video streaming by He et al. [112] and Tang et al. [113], UV radiation by Mei et al. [114], website performance by Zhu et al. [115], smart parking associations by Kim et al. [116], and gaming by Li et al. [117].

He et al. [112] focus on crowdsourced livecast service platforms, where users broadcast live streams to other viewers. In this context, real-time video transcoding is needed to transcode a source stream into different quality versions needed, according to the capabilities of receivers and network connectivity characteristics. The authors introduce a transcoding framework that allows to do so using fog computing. It assumes the presence of multiple regional data centers, each responsible for managing a specific region with fog computing nodes. Transcoding tasks are offloaded to fog computing nodes. A scheduling algorithm is introduced to select the most adequate node for executing a transcoding task. It aims at minimizing transcoding reassignments and cross-regional assignments in the system. It relies on a tree structure organization of the candidate pool by order of preference. In addition to that, a distributed rate adaptation mechanism is proposed. It enables rate allocations by considering each viewer's requirements and the streaming server's capacity. The efficiency of the two proposed algorithms is shown through a prototype. In this work, the heterogeneity criterion ($C_1$) is met as the capabilities of each node are considered. The QoS management criterion ($C_2$) is also met as the authors aim at meeting the QoS of livecast streaming. The authors organize the nodes in a tree-like structure that aims at handling scalability issues. However, evaluation remains limited to a small-scale scenario. The scalability criterion ($C_3$) is thus unmet. The authors do not cover the case of moving users and nodes. Therefore, the mobility criterion ($C_4$) is left unmet. As for the federation criterion ($C_5$), it is not relevant for the work targeting a specific application.

In turn, Tang *et al.* [113] target video streaming applications. They study cooperation among devices to enhance users QoE while streaming videos in cellular networks. They consider in particular a download cooperation model among users named crowdsourced cooperation model. The model aims at maximizing the social welfare represented by the gap between users QoE and cost. The model pools users download capacities in order to handle channel variations and enable an efficient utilization of network resources. An offline scheduling scheme, assuming complete knowledge of future variations in the network and an online scheduling scheme which operates in real time. The efficiency of the proposed solutions is outlined based on the derivation of theoretical bounds and simulations. In this study, the heterogeneity criterion ($C_1$) is met as the designed solutions and simulations cover heterogeneous users in terms of cellular network link capacity. The QoS management criterion ($C_2$) is met as it is covered as part of the objective of the proposed solutions. The scalability criterion ($C_3$) is not met since the evaluations are only conducted over a set of 50 users. The mobility of users is not addressed and as a result, the mobility criterion ($C_4$) is unmet. Finally, the federation criterion ($C_5$) is not relevant for the paper targeting a specific application.

Mei *et al.* [114] aim at determining the level of UV radiation by combining measurements from closely located mobile phone cameras. They propose an algorithm that allows of operating in the fog stratum. The algorithm gathers measurements at regular intervals of 10 minutes from devices. Then, if the number of collected samples is less than a threshold, i.e., too few samples are collected, the fog node does not return any UV measure to the mobile device. Otherwise, the algorithm eliminates the outliers and computes the UV radiation level as the average of all the remaining samples and returns it to the mobile user. By performing real-world experiments and comparing them to the official reports by an environment protection agency, the authors show that the results provided by their methodology are very close to those of the official reports, with deviations in the order of 3%. As an algorithm targeting the operations of a particular application, the heterogeneity criterion ($C_1$), the mobility support mobility criterion ($C_4$) and the need for federation criterion ($C_5$) are not relevant for this work. The algorithm operates by assuming that enough resources exist to run the algorithm and the mobility support and the federation support are provided by other algorithms in the system. However, the QoS criterion ($C_2$) and the elastic scalability criterion ($C_3$) remain relevant for the actual implementation of the proposed algorithm, but the authors do not address them. Hence, they are not met.

Zhu *et al.* [115] in turn focus on the optimization of websites' performance, through fog computing. As the fog nodes are located at the edge of the network, they are able to capture a close view of the performance on the user's side. The authors use this information to perform the website optimization procedures, implying file modifications at the level of a fog node, prior to sending the content to the users. While the authors expect such a scheme to improve the user's performance, they do not provide a proper strategy for the corresponding evaluation. Accordingly, they leave the relevant QoS criterion ($C_2$) and the elastic scalability criterion ($C_3$) unmet. As an algorithm targeting the operations of a particular application, the heterogeneity criterion ($C_1$), the support for mobility criterion ($C_4$), and the need for the federation criterion ($C_5$) are not relevant for this work.

Kim *et al.* [116] propose an algorithm to optimize the associations between vehicles and parking lots in a smart parking application, operating in the fog system. Their algorithm runs in the fog stratum at the level of roadside units. The latter are coordinated by a roadside cloud. The algorithm iteratively runs over a list of parking slots, controlled by the roadside unit, while it takes into consideration the vehicles' parking preferences. It associates to each parking slot the vehicle with the highest revenue, e.g., occupying the parking slot for the longest duration. The comparison of the proposed algorithm to two other state-of-the-art strategies shows that it leads to a balance of costs for the users and to benefits for the parking owners. The heterogeneity criterion ($C_1$) is not relevant for this work, as the proposed algorithm operates by assuming the resource allocation strategies that ensure the sufficiency of resources. The relevant QoS criterion ($C_2$) is not met despite the criticality of the response time in dynamic vehicular environments. The relevant elastic scalability criterion ($C_3$) is met. In fact, the authors conduct evaluations over a set of 1000 vehicles, indicating the capability of operating in a large-scale environment. As an important element here, the authors take the mobility support into consideration through the cooperation between roadside units and by using the roadside cloud, hence meeting the mobility support criterion ($C_4$). Finally, the need for the federation criterion ($C_5$) is not relevant for this study, as it targets a specific application.

Li *et al.* [117] focus on a cloud gaming system called CloudFog. It uses fog computing in order to improve the user's QoE. In particular, the cloud performs the intensive computation such as the generation of the new game state of the virtual world (e.g., the inclusion of new shapes or change of objects position) and shares it with the fog nodes. In turn, the fog nodes referred to as supernodes, render the game video and stream it to the user. To maintain playback continuity, the authors propose a rate adaptation strategy that is driven by the receiver. In particular, it adapts the encoding rate of the video to the segment size in the player's buffer, by considering the game's delay and loss rates. To further improve the QoE, the authors also introduce a buffer scheduling strategy. The latter delays and drops packets according to video game loss and delay tolerance degree. CloudFog was compared with cloud gaming and EdgeCloud. EdgeCloud is composed of powerful servers to perform all tasks of the cloud. The results demonstrate that CloudFog enables a reduced latency with respect to cloud gaming and EdgeCloud. In this work, the authors account for the limitations in resources of nodes. They thus meet the heterogeneity criterion ($C_1$). The QoS is covered, as the authors propose strategies that target the improvement of the user's QoE. Therefore, the QoS criterion ($C_2$) is met. The elastic scalability criterion ($C_3$) is also met, as CloudFog is evaluated over large-scale scenarios with thousands of players. The mobility support criterion ($C_4$) is met, as the authors allow the user to use the closest fog node to him. Finally, the federation criterion ($C_5$) is not relevant for this study, as it targets a specific application.

It is concluded that the reviewed literature covers a set of specific applications besides healthcare applications. Most of the discussed works do not meet the QoS criterion ($C_2$) and the elastic scalability criterion ($C_3$). Other criteria including the heterogeneity criterion ($C_1$), the support for mobility criterion ($C_4$) and the federation criterion ($C_5$) are generally not relevant for these contributions.

## VI. CHALLENGES AND RESEARCH DIRECTIONS

As shown in Table II, Table III, Table IV, Table V, and Table VI, none of the reviewed architectures and algorithms meets all the identified criteria. This section discusses the remaining challenges, i.e., not yet tackled in the reviewed literature, and the related research directions. First, the architectural challenges and their research directions and then, the algorithmic challenges and their research directions are addressed. In each case, the remaining challenges and research directions for each of the evaluation criteria are discussed.

### A. Architectural Challenges and Research Directions

This section discusses the most important architectural challenges and the research directions that will be invaluable as fog computing matures. Table VII provides a summary of these challenges and research directions, the relevant works from the literature, and potential solutions and starting points.

#### 1) Heterogeneity

Although several of the reviewed works (e.g., resources description APIs [52], label-based nodes classification [53]) provide potential solutions for the heterogeneity criterion, none has proposed semantic-based approach. In fact, most of them put the burden of handling heterogeneity on application developers (e.g., [25], [62], [68], [69]). Only one work proposes a dynamic matching procedure between the application requirements and the resource capabilities (i.e., [54]) but the proposed approach does not rely on semantic-based matching.

In general, semantic ontologies are explicit formal specifications of the terms used in a given domain and the relations among them [118]. They can be assimilated as formal representations and naming of the properties, types, and relationships of the entities that set up a particular universe. One of the main goals of the use of these ontologies is sharing a common understanding of the structure of information among entities [119][120]. This common understanding can be done by using well-defined taxonomies and vocabularies.

In the specific fog universe, defining appropriate ontologies that could cover the strong variety and specificities of the involved nodes from the cloud, the fog, and IoT would contribute to the homogenization and simplification of the way applications are provisioned over these nodes. Indeed, the several providers, as part of the fog system, might rely on heterogeneous description models, schemes, naming, and/or vocabularies. For instance, IoT devices in the IoT/end-users stratum can be described by models such as the one proposed by the IoT-A initiative [121] while the cloud nodes in the cloud stratum can be described by models such as the OASIS TOSCA [122]. Obviously, the heterogeneity of the models is unsuitable for collaborative environments such as the fog system where providers from all strata and domains need a common understanding of resources when provisioning applications.

A research direction here is the design of appropriate and exhaustive ontologies that could support this heterogeneity. Based on these ontologies and semantic Web technologies, one could enable the unification of the representation of these resources, described by different models.

The work in the context of multi- and heterogeneous cloud providers (see e.g., [123][124] for multi-IaaS and [125] for multi-PaaS) can serve as a starting point in order to define a relevant ontology for the fog system. Although most of the cloud-based ontologies are object-oriented and extensible [123], they are not flexible enough to cater to the fog and/or IoT strata. Examples of the required extensions may include information on the power autonomy of the hosting fog nodes since smartphones and laptops are among the most used fog nodes.

#### 2) QoS Management

Service Level Agreements (SLA) management is a potential research direction related to QoS management. Appropriate SLA management techniques are critical to maintaining acceptable QoS in highly dynamic environments like in the fog system. Apart from [53], none of the reviewed papers proposes fog-enabled SLA. However, the introduced SLAs in [53] represent a simple extension of a cloud-based schema that aims at describing the applications' latency. In other terms, the associated model does not cover all the providers and domains as part of a fog system. In fact, based on the definition in Section II, a fog system may consist of several providers and domains in the cloud stratum, as well as, several fog providers and domains in the fog stratum. Each one of these providers may have its own business model (e.g., different metering and billing methods, different scalability and elasticity procedures). Consequently, SLA schema and management techniques and solutions may differ from one provider to another.

A potential solution can be based on cloud-management SLA approaches. However, cloud-based SLA management solutions only cover resources that belong to the same provider (e.g., [126], [127]). More sophisticated solutions support the aggregation of SLA management in the case of multi-cloud provisioning. For instance, in [128], the authors introduce a system that enables the SLA management of distributed data centers. However, all these solutions are still based on the same and unique business model that makes them not suitable for operating in the case of a fog system. Novel SLAs definition and management techniques need to be designed in order to support the several involved business models. In addition, the new SLAs should cover the specificities of the fog system compared to a pure cloud system such as the energy-limitation, mobility-sensibility, and resource constraint of the fog nodes.

The work presented in [129] can serve as a starting point. It introduces an approach for multi-provider service negotiation and SLA management in the context of Network Virtualization Environment (VNE). The associated business model may include several infrastructure providers. Firstly, SLA should be extended in order to cover the previously mentioned specificities of the fog providers' resources. Secondly, the operational framework (called V-Mart) and its business workflow should be adapted accordingly to integrate the fog providers.

*3) Scalability*

A few papers in the reviewed literature have addressed the scalability criterion (i.e., [52], [53], [55], [5], [74], [75], [61]). However, the presented solutions are not general. They only tackle one part of the whole system. For instance, the proposed architectures in [53] and [74] only support the scalability of the fog nodes while the one presented in [75] only supports the scalability of IoT devices. To address scalability in a fog system, besides scaling the resources from the cloud, it is required to support scaling the resources from the several involved domains in the fog stratum and the used devices in the IoT/end-users stratum. Distributed and hybrid cloud solutions, such as the one discussed in [130], are unsuitable since they do not provide such global view in the fog and IoT/end-users strata. A potential solution to make them fog-enabled is to extend and adapt them. An alternative is to design upstream modules with a global view of the system (e.g., CloudScale [131]) so that they can check the components' health and address their prospective scalability problems. This requires the design of novel mechanisms that enable discovering, monitoring, and acting on the nodes in order to allow the system to (i) be aware of the current status of the available resources from all providers and strata and (ii) to execute the appropriate scalability procedures on them.

Cloud brokers, such as CompatibleOne Broker [132] or mOSAIC [133], can be used as a starting point to design such mechanisms. By definition, cloud brokers list the several available resources and act as intermediaries to assist when discovering the suitable resources [132]. These brokers and their related SLAs can be extended in order to integrate and index the additional resources from the fog and IoT.

*4) Mobility*

Several works in the reviewed literature aim at supporting the mobility. However, they propose no general solution. For instance, the work on vehicular applications (e.g., [71], [72], [73]) focuses on the fog nodes' mobility when other works, such as healthcare-oriented ones (e.g., [69], [81]), exclusively focused on IoT devices' mobility.

Supporting mobile entities in all fog system strata is a complex and challenging research direction. In order to address it, one alternative could be the extension of the existing work in order to support all the mobile entities in a fog system. In [73], for instance, the VANET architecture leveraging the fog stratum can be extended in order to support the mobility of end-users and IoT devices. The addressing and the routing of the packets can be then handled by using SDN as it is the case between the cloud and the fog strata. In such architectures, the orchestration and network management modules and the SDN controller are centralized nodes placed in the cloud stratum to benefit from the overall view of the system. A challenge here is to distribute them over the several mobile nodes instead and to enable their cooperation (e.g., direct forwarding or hop relaying messaging systems as the ones in [71]) in order to reduce the latency fluctuation when the network is managed during node movements.

Another alternative for this research direction is Mobile Cloud Computing (MCC). MCC aims at providing cloud computing services in a mobile environment and overcomes obstacles for the hosting nodes (e.g., heterogeneity, availability) and performance (e.g., battery autonomy, limited computation capabilities) [134]. Based on the fog system definition and characteristics introduced in Section II, these obstacles are also very relevant in the fog context. Furthermore, the networking between the cloud and the mobile entities in MCC is done in the same way as in the fog systems. The bindings are wireless and performed via access points such as satellites and/or Base Transceiver Station (BTS). Several realistic MCC use cases (e.g., mobile gaming, mobile learning) and architectures (e.g., Mobile Service Clouds, VOLAIRE) are already discussed in the relevant literature [135]. However, one major difference between MCC and fog systems should be taken into account. Only end-users are considered as mobile in MCC while some hosting nodes could be mobile as well in the fog case. MCC networking and routing mechanisms can be then reused while the placement and management techniques should be adapted to the new context.

*5) Federation*

Reference [62] is the only work in the reviewed literature that discusses a fog-enabled solution for addressing the federation criterion as the main objective. Actually, the proposed solution is designed for hybrid clouds federation and is extended to also cover the fog. The author claims that enabling federation in this environment necessarily implies local hardware awareness of the cloud platforms. However, he provides neither architecture nor feasibility prototype for it. In particular, the execution of applications with components provisioned over the several strata and domains as part of a fog system is not discussed. Such a system requires the implementation of appropriate mechanisms that compose the distributed applications' components while executing them.

As a research direction for the federation of several domains as part of the fog system, appropriate composition mechanisms for the applications' components are designed. Such a composition should be performed in a well-defined order with respect to the business functionality of the application.

In general, there are two existing composition techniques: Orchestration and choreography [136]. The former allows a central entity to control an application's components and their interactions. On the contrary, the latter allows an application's components to collaborate in a decentralized way. It should be noted that some existing works already acknowledge the need for orchestration between the cloud and the fog [16]. In addition, [5] proposes an early architecture for orchestrator in the fog systems. The proposed approaches in [16] and [5] are orchestration-based. However, these solutions are not efficient because the orchestrator has to be deployed as part of the cloud stratum with the overall view. Such an architecture is not scalable and may lead to important overhead and delays when dealing with the remote fog stratum.

A better approach would be the design of a distributed composition engine made up of several local engines that communicate and cooperate when executing an application. Similar distributed composition solutions are already proposed in the cloud environments. These solutions are various and can be easily adapted in the fog systems. For instance, for the choreography-based approach, as it is the case in the cloud, there are two choreography modeling styles that involve the distributed components and can be used in the fog: Interaction modeling and interconnected interfaces modeling [137]. The interaction modeling designs the choreography as a workflow in which the activities implement the message exchange between the application's components. Examples of such modeling in the cloud that can be used in the fog are Let's Dance [138] and Web Service Choreography Description Language (WS-CDL) [139]. The interconnected interfaces modeling designs the choreography as a set of participants grouped by their roles in that choreography. In other terms, the participants are grouped by their expected messaging behavior. The roles are connected by properties such as message flows and communication channels. An example of such modeling in the cloud that can be reused in fog is BPEL4Chor [140].

*6) Interoperability*

A research direction for enabling interoperability in the fog system is the design of signaling, control, and data interfaces between the several domains (the cloud and the fog) that are part of the system. To that end, two critical requirements should be fulfilled: (i) Inter-domain agreement and (ii) operational interfaces standards implementing such an agreement.

A few of the reviewed works address the federation issues in the fog systems (i.e., [52], [53], [25], [62], [69], [72], [61]). However, none meets the two requirements discussed below.

The first is the requirement for the inter-domain agreements as the common contract between all the involved domains, such as common policies and common model descriptions. It can be met by designing a common description model that allows describing and consequently hiding the specificities of the nodes and data in the cloud and the fog. Similar cloud-based models are already proposed in the literature in order to enable interoperability in multi-clouds systems. Cloud Infrastructure Management Interface[16] (CIMI) and Open Cloud Computing Interface[17] (OCCI) recommendation for a standard are among the examples. Both models can be used as starting point and can be extended in order to cover the fog domains. CIMI tries to standardize the interactions between several cloud IaaSs in order to achieve interoperable management between them. It provides an open standard API specification, also extendable to include the fog domains. OCCI is a set of specifications that define a meta-model for abstract cloud resources and an HTTP rendering for their management. It provides a flexible specification with a strong focus on interoperability while still offering a high degree of extensibility. The main specification is OCCI core, defining a meta-model for the cloud resources at large [141]. Some extensions of the core meta-model are already defined for specific cloud resources (e.g., see [142] for IaaS resources, see [143] for PaaS resources). A novel OCCI extension can be designed and considered as a unified model for the resources as part of the fog system. For the data part, standards such as Cloud Data Management Interface[18] (CDMI) can be adapted. An example of the adaptations is the support of real-time database systems that are critical in several fog-driven IoT use cases.

Concerning the second requirement, the operational interfaces standards comprise unified operations and procedures for the inter-boundary control, signaling, and data exchanges between the fog system domains. These interfaces will implement the newly defined resources and data models. Existing CIMI implementations such as Apache DeltaCloud can be adapted to implement control and signaling interfaces [144]; OCCI-compliant implementations such as COAPS API [145][146] can also serve this purpose. For the data interfaces, a CDMI-based solution such as ODBAPI can be used as a starting point for the data interface implementation [147]. A potential solution is to extend and adapt ODBAPI in order to integrate the support of the fog stratum as part of its negotiation and discovery capabilities for data management and exchange. The most common exchange would be the fog sending data to the cloud since the former domain is closer to the devices that generate the data (e.g., sensors).

*B. Algorithmic Challenges and Research Directions*

In this section, the most important algorithmic challenges and the research directions that will be invaluable as fog computing matures are discussed. Table VIII provides a summary of these challenges and research directions, the relevant works from the literature, and potential solutions and starting points.

---

[16] dmtf.org/standards/cloud
[17] occi-wg.org
[18] snia.org/cdmi

TABLE VII
Summary of Architectural Challenges and Research Directions

| Reference Evaluation Criteria | Research Direction | Relevant Work from the Reviewed Literature | Noticed Limitations | Potential Solutions and Starting Points |
|---|---|---|---|---|
| **Heterogeneity $C_1$** | The design of exhaustive and flexible semantic ontologies. | [52] [53] [25] [62] [56] [68] [69] [61] | Put the burden of heterogeneity handling on application developers. | • Extend object-oriented ontologies of cloud system (e.g., see [124], [125]) when defining common taxonomies and vocabularies for the heterogeneous fog system resources. |
| **QoS Management $C_2$** | The design of appropriate SLA management techniques. | [53] [56] | • Represent a simple extension of a cloud-based schema that aims at describing the applications latency.<br>• Does not cover all the providers and domains part of a fog system. | • Reuse and extend pure cloud SLA (e.g., [129]) in order to include missing elements such as the energy-limitation, mobility-sensibility, and resource constraint of the fog nodes. |
| **Scalability $C_3$** | The design of mechanisms that in addition to scale the resources from cloud, can support scaling the resources of involved domains in fog stratum and devices in IoT/end-users stratum. | [52] [53] [55] [5] [74] [75] [61] | • Do not propose a general solution: tackle one part of the whole system.<br>• Supports either the scaling of fog nodes or IoT devices. | • Design and integrate upstream modules with global view to the fog system for heath checking (e.g., CloudScale [131]).<br>• Extend existing cloud brokers for fog resources publication and discovery (e.g., CompatibleOne [132], mOSAIC [133]). |
| **Mobility $C_4$** | The design of mechanisms that considers the mobility of IoT and fog nodes in addition to the cloud nodes. | [71] [72] [73] [53] [69] [81] | Do not propose a general solution: Focus either on fog nodes mobility or IoT devices mobility. | • Extend the existing work on fog in order to support all the mobile entities in a fog system (e.g., The VANET system in [73]).<br>• Reuse MCC networking and routage mechanisms and adapt placement and management techniques to fog system specificities (e.g., [134]). |
| **Federation $C_5$** | The design of appropriate composition mechanisms for the applications' components. | [62] [61][56] | No architecture or feasibility prototype. | • Design of a distributed composition engine made up of several local engines that communicates and cooperates with each other when executing an application.<br>  ▪ Adaptation of the cloud choreography-based solutions (e.g., [138]).<br>  ▪ Adaptation of the cloud orchestration-based solutions (e.g., [139] [140]). |
| **Interoperability $C_6$** | The design of signaling, control, and data interface between the several domains part of the fog system. | [52] [53] [25] [62] [69] [72] [61] | • No Inter-domain agreement.<br>• No operational interfaces standards implementing such an agreement. | • Extend cloud description model to describe and hide the specificities of the nodes and data in fog system (e.g., OCCI model [141]).<br>• Define control, signaling and data interfaces implementing such a model (e.g., [142] for control interfaces, [143] for data interfaces). |

*1) Heterogeneity*

Many of the algorithmic contributions fail to meet the heterogeneity criterion (e.g., [92], [106]) in terms of node computing and storage capabilities in the fog system. Only some meet it (e.g., [84], [88]) by accounting for resource limitations as part of their study, allowing to model the differences between the nodes with respect to their capabilities. However, even these works do not have the same understanding of the degree of heterogeneity among the nodes of the system.

By the heterogeneity degree, we mean the level up to which the fog system nodes differ in their computing and storage capabilities. In fact, evaluations of the proposed algorithms, conducted in simulative and experimental fog system environments, have relied on different assumptions concerning the degree of heterogeneity. In reality, this degree of heterogeneity in the system can have a significant impact on the performance of algorithms and today the relevant literature still lacks an agreement on it. In this respect, two complementary challenges are to be addressed.

First challenge concerns existing computing and storage devices around us, e.g., end-user mobile devices that can serve as fog nodes. The question is to what extent these devices impose the degree of heterogeneity in the system in terms of computing and storage capabilities. More precisely, how much of their resources can they spare to serve as fog nodes, capable of processing other devices requests and storing their content? A research direction there consists in deriving the actual usages of the computing and storage resources of the existing devices in order to draw corresponding actual usage patterns. When conducted over large scales and covering a large variety of potential devices, the analysis allows assessing the capabilities that these devices can offer to the fog system. As a result, this allows determining the degree of heterogeneity that these devices impose on the system in terms of computing and storage capabilities. Adequate prediction models can then be derived accordingly to enable decision-making over the future. Such models can be inspired from those proposed in the context of volunteer computing systems [148].

Second, besides existing devices around us that can serve as fog nodes, additional computing and storage nodes can be added in the system. These devices participate in turn to defining the degree of heterogeneity in the system. Here, the question is how to decide their dimensioning and placement. Studying the corresponding optimization problem is a research direction. It aims to fulfill the demand of IoT/end-user devices in the fog system considered as an input. It would then optimize the dimensioning of nodes considering a number of aspects as part of the objective or constraints. These can include power consumption and resource costs. The problem can be mapped to a facility location problem [149] that has been extensively studied in the operations research community. Exact, approximation and heuristic algorithms have been employed to solve it. The problem is also similar to the problem of cloudlet design optimization, where cloudlet facilities are to be placed. Therefore, corresponding solutions can be considered as a starting point[150].

*2) QoS Management*

The QoS criterion is met by most of the reviewed contributions. This is generally achieved by incorporating a latency constraint as part of the addressed problem. Although the latency is an important metric for the system, other important performance metrics or relevant costs such as uplink and downlink bandwidth or resource usage costs on the user's side are generally not taken into account – which has to be addressed as another challenge. An exception is [87], where resource usage cost is considered. However, this cost is only accounted for as part of a pricing strategy. Other algorithms, targeting resource utilization or particular applications, need in turn to account for different performance metrics and relevant costs. Moreover, these metrics should be integrated as part of an algorithm's objective, instead of forming constraints only.

This can improve QoS, rather than imposing a limit on a certain QoS metric. In this context, [85] and [88] have integrated latency and completion time as individual objectives for their proposed algorithms. Other metrics, such as uplink and downlink bandwidth or resource usage, remain open for consideration.

Besides covering QoS metrics as individual objectives, considering them together with other objectives has not received significant attention so far and thus remains a challenge. In reality, several objectives that can even be contradictory may need to be covered at the same time. One example is to minimize completion time and minimize power consumption costs in the fog system when managing compute resources. In this case, minimizing completion time implies the need for more resources in the system, in contrast to minimizing power consumption. There is thus a trade-off to consider here. Due to the complexity of such problems, so far, only [101] has addressed such contradictive objectives. However, the study remains limited to a video streaming application in content delivery networks and only targets the distribution of content from the data centers to the fog nodes. Accordingly, today, we lack algorithms that manage the system resources as well as algorithms that manage other particular applications, considering several simultaneous optimization objectives. To study these problems, one research direction is to consider multi-objective optimization strategies [151], in order to derive optimal solutions in the presence of trade-offs between various objectives.

*3) Scalability*

An algorithm that runs in the fog system should be operational over a large scale. Validating an algorithm in a small-scale environment, with a few devices and nodes does not guarantee it performs well over a large scale, in terms of quality of obtained solution as well as execution time. Despite the importance of this criterion, most of the proposed algorithms have been evaluated over small-scale scenarios. Exceptions are [105], [93], [96], and [106], with algorithms for the fog system at large, and [11] for a smart parking application in the fog system. All other algorithms for the fog systems have been validated at a small scale, not guaranteeing that they perform well at a large scale, thus leaving it as a challenge to address.

Besides the considered scale, proposed algorithms need to operate in real-world conditions. For instance, a task scheduling algorithm needs to be operational for real-world traffic patterns. Except for a few experimental evaluations as in [91] and [85], proposed algorithms have been mainly validated in simplistic simulation environments, with no clear motivation for system parameters. As a result, in a real-world environment, these algorithms may diverge from the expected performance, posing by that a second challenge. To handle these challenges, a possible research direction is to run real-world experimental evaluations over a large scale. However, the latter is infeasible due to the high costs they imply. Large-scale realistic simulative evaluations remain instead a tractable option enabling the comparison of different algorithms. They should thus be taken into account. However, to do so, two major aspects are to be highlighted. First, still today, we lack a clear knowledge of real-world system deployments, as discussed in Section VI.B.1. Second, we do not yet acquire a clear understanding of the real-world evolution of application and services consumption on the side of IoT/end-user devices. Such a characterization can be achieved through the analysis of corresponding real-world traces, through machine learning techniques [152] [159]. When acquired, proper models need to be derived based on it to enable proper evaluations through realistic simulations.

TABLE VIII
Summary of Algorithmic Challenges and Research Directions

| Reference Evaluation Criteria | Research Direction | Relevant Work from the Reviewed Literature | Noticed Limitations | Potential Solutions and Starting Points |
|---|---|---|---|---|
| **Heterogeneity $C_1$** | Acquire a clear vision on the degree of heterogeneity in terms of computing and storage capabilities. | [92] [106] [84] [88] | • No agreement on the degree of heterogeneity among nodes and devices in terms of computing and storage capabilities. | • Analyze and predict computing and storage resource usage of existing nodes (e.g., extend the work in [148]).<br>• Plan the dimensioning and placement of additional nodes in the system (e.g., extend the solutions of cloudlet design optimization [150]). |
| **QoS Management $C_2$** | Consider various QoS metrics. | [85] [88] [101] | • Consider mainly the latency. | • Integrate different QoS metrics in the constraints and objectives of problems and consider multi-objective optimization strategies [151] in order to solve them. |
| **Scalability $C_3$** | Validate algorithms over large scale in real-world environment. | [105] [93] [96] [106] [116] [91] [84] | • Validate over small scale in an unrealistic environment. | • Use machine learning techniques [152] in order to acquire a clear understanding of the real-world evolution of application and service consumption on the side of IoT/end-user devices. |
| **Mobility $C_4$** | Ensure the continuity of offered services despite the movement of IoT/end-user devices and/or fog nodes. | [96] [91] [94] | • No realistic mobility models.<br>• No consideration of the fog nodes mobility. | • Consider real-world mobility traces [153] [154].<br>• Design mobility prediction methods [155].<br>• Manage fog domains in the light of fog nodes mobility (e.g., Extend clustering solutions from WSN [156]). |
| **Federation $C_5$** | Design algorithms for federation in fog systems. | [88] [89] | • No consideration of the possibility of operating inside federated systems. | • Consider dynamic strategies [157] and game-theoretic approaches [158] to design adequate pricing and insourcing/outsourcing algorithms. |

*4) Mobility*

Whether it concerns IoT/end-user devices or fog nodes, mobility poses significant challenges in fog systems. Consider the mobility of an IoT/end-user device, the continuity of offered services needs to be ensured, despite the movement of a device across various fog domains. This stresses the need for proper strategies that allow handling the mobility. Efforts in this direction remain limited. In [96], authors propose a migration strategy that allows moving components across fog domains according to the movement of devices. In [91], authors propose instead to keep the content metadata in the cloud to enable downloads independently of the user's location. However, neither work builds upon realistic mobility models that can affect the accuracy of the evaluation results. A research direction here is to derive realistic mobility models. To do so, real-world mobility traces need to be integrated into the analysis, as done in the case of VM migration in mobile cloud systems in [153] and [154]. Moreover, accurate mobility prediction methods [155] are needed to complement algorithms operating in real-time. Depending on the context, these schemes would either target individual or group mobility and can be built based on collected traces of IoT/end-user devices.

Similarly, to manage the mobility of a fog node, we also need to ensure the offered services are not interrupted. In fact, the fog nodes' mobility is even more complex to handle, as it involves the serving resources availability. For instance, as a fog node leaves a fog domain, it implies the need to offload tasks assigned to it to other fog nodes in the system. Instead, as a fog node joins/creates a fog domain, additional/novel resources would be available for IoT/end-user devices that can be connected to it. So far, only one work studies this problem (i.e., [94]) with the objective of enabling dynamic load balancing among fog nodes in a fog domain upon the arrival and exit of a fog node. However, the authors do not account for the possibility of having completely new fog domains created in the system. Such a problem is in fact similar to traditional clustering problems in Wireless Sensor Networks (WSNs) [156], targeting, in this case, the formation, management, and ending of a fog domain. As in case of IoT/end-user device, mobility prediction and modeling schemes representing the movements of the fog nodes are also crucial and need to be incorporated in the clustering problem. For example, a fog domain can be mapped into a cluster, with a cluster coordinator representing the fog node that moves the least, in order to ensure the stability of the cluster coordinator.

*5) Federation*

The review of algorithmic contributions in the fog systems has shown that the federation among operators is a challenge that has been largely disregarded. First, resource sharing algorithms, e.g., [83] and [105], remain limited to the case of a single operator. They thus do not consider the possibility of sharing resources from various operators. Second, compute and storage resources management algorithms, e.g., [88] and [89], do not take into account the possibility of operating inside federated systems. Indeed, the federation among providers extends the capabilities of the system and has not received so far any attention from an algorithmic perspective.

Two main questions are to be answered there. First, inside a single federation, how does an operator set the prices for leasing its resources? Second, when does a provider derive insourcing or outsourcing decisions with respect to other providers inside the same federation? The same questions are covered in case of cloud computing federations, where cooperation among several cloud providers inside a single federation is managed. Proposed strategies can be used here as a basis to design algorithms for federation in the fog systems as a research direction. For instance, dynamic strategies [157] and game-theoretic approaches [158] can be considered to design adequate pricing and insourcing/outsourcing algorithms.

## VII. LESSONS LEARNED AND PROSPECTS

### A. Lessons Learned

Our literature review allows us to derive several lessons relevant to fog systems. First, fog systems do indeed enable reduced latency with respect to traditional cloud systems. Both experimental and simulative measurements confirm that significant reductions in terms of latency can be obtained. This is important for real-time applications such as the IoT ones. Several references demonstrated this reduced latency such as Krishnan *et al.* [59], Gia *et al.* [69], Tang *et al.* [76], Li *et al.* [46], and Sarkar *et al.* [106] [107]. For instance, Gia *et al.* [69] and Li *et al.* [46] showed a reduction in the latency by 48% and 73% respectively when employing fog system. However, it should be noted that the latency reduction does not come automatically and depends on where the application components are placed. As shown in reference [5], there may be sometimes worse response time by locating some application components in the fog and others in the cloud, compared to locating all of them in the cloud. Besides the fog system's inherent capabilities, many algorithms have been proposed to further aid in latency reduction, by managing the resource utilization accordingly, with notable reductions. There, several issues have been addressed. Resource sharing at minimum service latency is covered by Nishio *et al.* [84]. The problem of task scheduling with a minimum latency is studied by Oueis *et al.* [85], Zeng *et al.* [88], and Agarwal *et al.* [25]. Devices offloading at lowest latency is considered by Hassan *et al.* [91]. Content caching with latency considerations is covered by Xiang *et al.* [100].

It should be noted that the reduced latency advantage is not critical for every IoT scenario. For instance, Industrial IoT solutions [160] usually require low-latency ingestion but immediate processing of data [161]. The high delays of data transmissions to the cloud and their remote processing cannot always be afforded in such solutions. It is also the case with most of the healthcare applications as well. This brings us to the second lesson learned related to the traffic reduction over communication links towards the cloud. Local processing offered by fog nodes saves the bandwidth and enables faster processing by impeding (or eventually avoiding) the turnaround with the cloud stratum. It may prevent inappropriate or irrelative data to be sent to the cloud (for processing and/or storage purposes). This results in a reduction of the volume of data transmitted and considerably reduces the traffic over the several strata of the system. Several references conducted experiments indicating that when using fog, a smaller traffic is sent to the cloud as in Tang *et al.* [76], Krishnan *et al.* [59], and Gia *et al.* [69]. For instance, Tang *et al.* demonstrated that the data sent to the cloud is 0.02% of the total size and Gia *et al.* showed a reduction in the data size up to 93% when using fog system.

Third, fog systems are confirmed to be energy-efficient. In addition to offering low latencies and reducing network traffic to the cloud, fog systems are observed to be efficient when it comes to energy consumption. Assessment of the overall energy consumption in the system shows that fog systems lead to lower energy consumption, with respect to cloud computing systems, implying in turn low costs, as concluded by Sarkar *et al.* [106] [107] and Jalali *et al.* [108]. Nevertheless, Jalali *et al.* [108] underline the fact that it is not the case when the connecting network is not energy-efficient. Furthermore, several algorithms have been proposed in the literature with the objective of reducing energy consumption. There, resource sharing has been tackled by Nishio *et al.* [84] and Chen *et al.* [105]. Task scheduling has been covered by Oueis *et al.* [85] and Deng *et al.* [89]. Offloading strategies are introduced by Hassan *et al.* [91] and Ye *et al.* [92].

Fourth, despite the large interest in fog systems, there is still a lack of related initiatives and consortiums. To the best of our knowledge, there is only OpenFog Consortium[19]. The consortium just published (early 2017) an OpenFog Reference Architecture [12] which represents the baseline to developing an open fog-enabled architecture environment. The main objective announced by the OpenFog consortium is to accelerate the adoption of IoT in the enterprise. Relevant consortiums definitely need to produce reference architectures, developer guides, samples, and SDKs in order to articulate the value of fog to developers and IT companies. In terms of standards, there is none in the area of fog systems. However, there are standards in related areas such as MEC, which provides terminology [162], requirements [163], and framework [42] specifications.

In addition to the lessons learned from the reviewed paper, we have also learned a few lessons from the remaining challenges and research directions. From architectural perspective, the first lesson learned is the need for semantic Web technologies to handle the strong heterogeneity between cloud, fog and IoT nodes. The second lesson learned concerns the absence of appropriate monitoring and reconfiguration mechanisms in fog systems. These mechanisms have not yet been considered by the community despite the fact that they are critical when it comes to tackling the QoS management, mobility, and scalability challenges. Finally, the third lesson learned is the suitability of choreography-based solution rather than orchestration based when addressing the federation criterion. Indeed, we have shown that fog-enabled orchestrators such as the ones proposed in [5] and [16] are not appropriate because of the centralized approach they adopt. Choreography-based solutions would be more appropriate.

When it comes to the remaining algorithmic challenges and research directions, the first lesson is that there is a need for algorithms that enable decisions on the design of fog systems deployments, an aspect that has not received any attention so far. The second is that algorithms managing the QoS in fog systems need to be extended to cover various QoS metrics. The third is that large-scale and realistic evaluation scenarios need to be designed to enable accurate evaluation of algorithms. The fourth and last lesson is that realistic mobility models need to be derived. Finally, algorithms for federation in fog systems need to be designed.

### B. Prospects

We expect fog computing to play a decisive role in emerging technologies such as Tactile Internet. Tactile Internet (References [164], [165], [166], [167]) is expected to enable skill sets delivery over networks in addition to the content delivery (e.g., text, voice, video) over networks enabled by the current Internet. The skill-set delivery will be done via haptic communications, meaning the remote real-time control of physical tactile experiences. Some examples of potential applications are telesurgery, telerehabilitation, vehicle platoons, and augmented reality.

The functional architecture of Tactile Internet comprises a possibly distributed master domain with operators acting through tactile-human interfaces; a network domain with core and edges (with the edges hosting intelligent Tactile Support Engines); and a controlled domain which may include for instance remotely controlled robots as part of a tactile edge [165]. The fundamental pre-requisite is an ultra-responsive and ultra-reliable connectivity. An end-to-end latency of 1 ms or less and a maximum of a second of outage a year are required. 5G is expected to be a key enabler of Tactile Internet by providing the ultra-responsive and ultra-reliable connectivity.

In addition to 5G, cloud and edge computing are often mentioned as enablers of Tactile Internet ([164], [165]). This makes fog computing an ideal enabler due to its holistic approach which integrates end-users and/or IoT devices (IoT/end-users stratum), edges (fog stratum), clouds (cloud stratum) and the related interactions. As shown, by the Venn diagram of Fig. 5, no other mobile edge concept follows this holistic approach in which the interactions between clouds and edges are fully integrated.

The Tactile Internet functional architecture can actually be mapped quite naturally onto fog systems. Controlled domain (e.g., remotely controlled robots) and master domains (i.e., operators with tactile human-systems interface) are naturally part of the fog IoT/end-user stratum with the former belonging to the end-user devices domain of the stratum and the latter to the IoT devices domain. On the other hand, the edges (with the intelligent Tactile Support Engines) are naturally parts of the fog stratum. The cloud itself could be used whenever powerful processing and storage are required. The prospect of fog system based-tactile Internet brings architectural and algorithmic challenges and research directions that go far beyond the challenges and research directions previously discussed in this paper.

From architectural perspective, an example of a challenge is the design of the ultra-responsive and ultra-reliable higher layer APIs and protocols for fog system inter-strata and intra-strata communications. The protocols are the transport and application protocols that will run on top of the ultra-responsive and ultra-reliable physical and MAC layer protocols expected from 5G. Recent efforts to design novel ultra-high data rates (e.g., [168]) will need to be taken into account.

Yet another example of a challenge is the functionality split between cloud stratum and fog stratum. While intelligent Tactile Support Engines will reside in the fog stratum, they may be fed from algorithm repositories residing in the cloud stratum. One may even envision these engines as having some of their components located in fogs and the others located in cloud to harness both proximity (fog stratum) and powerful processing/storage (cloud stratum).

In turn, algorithms operating in the system need to ensure ultra-responsiveness and ultra-reliability are guaranteed for tactile applications. A fog system brings the possibility of running the different tasks related to Tactile Internet applications in fogs and/or cloud strata. Clearly, the end-to-end delay may far exceed the 1 ms threshold, depending on where these tasks are executed, the network traffic conditions, the load on the computing nodes and other factors. There is, therefore, a need for novel task scheduling algorithms. The goal is to ensure that tasks are executed, in a way that the overall threshold of 1 ms is not exceeded.

---

[19] http://www.openfogconsortium.org/

Besides tasks scheduling, novel machine learning and artificial intelligence algorithms are needed. They may run solely in the fog stratum or even in a distributed manner across cloud and/or fog strata, to predict actions and reactions. Actually, when they run in the fog stratum, the overall network load will be reduced and this will aid in meeting the 1ms latency requirement. A variety of techniques can be considered ranging from simple regression models to complex neural network-based techniques [169].

## VIII. Conclusion

This manuscript surveys the literature on fog computing. Based on a detailed description of fog systems, related concepts are identified and differences among them are presented. Illustrative use cases covering two different application domains (i.e., IoT and CDN) are introduced and used as a basis to derive a set of evaluation criteria for fog systems. These criteria are used to critically review the architectures and algorithms proposed so far in the area of fog computing. A set of lessons learned are derived based on the literature review. The remaining challenges and corresponding research directions are discussed. In addition, we have discussed the prospects of fog computing with a focus on the role it may play in emerging technologies such as Tactile Internet. The discussions include examples of challenges and research directions.

## Acknowledgment

This work is partially supported by CISCO systems through grant CG-589630 and by the Canadian National Sciences and Engineering Research Council (NSERC) through a Discovery Grant.

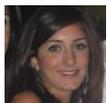
**Carla Mouradian** is currently pursuing her Ph.D. degree in Information System Engineering at Concordia University, Montreal, Canada. She obtained her Master's degree in Electrical and Computer Engineering from Concordia University (2014) and received her Bachelor's degree in Telecommunication Engineering from University of Aleppo, Syria (2009). Her research interests include cloud computing, fog computing, Internet of Things, Wireless Sensor Networks, and Network Function Virtualization. She is a member of the IEEE Communications Society. She currently serves as a Reviewer for many international journals.

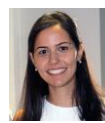
**Diala Naboulsi** is a postdoctoral researcher in the TSE research lab of CIISE at Concordia University, Canada. She obtained the Ph.D. degree in Computer

Science from INSA Lyon, France, in 2015. She received the M.S. degree in Computer Science from INSA Lyon, France and the M.Eng. degree in Telecommunications from the Lebanese University, Lebanon, in 2012. She was a visiting Ph.D. student at the Politecnico di Torino, Italy, in 2013. Her research interests include Mobile Networks, Network Function Virtualization, Internet of Things, and Content Delivery Networks.

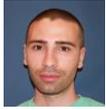

**Sami Yangui** is an Associate Professor with Institut National des Sciences Appliquées (INSA), *Toulouse*, France. He is member of the CNRS LAAS research team. His research interests include distributed systems and architectures, service-oriented computing, and Internet of Things. He is working on different aspects related to these topics, such as cloud/fog computing, network functions virtualization, and content delivery networks. He is involved in different European and International projects, as well as, standardization efforts. He published several scientific papers in high ranked conferences and journals in his field of research. He also served on many program and organization committees of International conferences and workshops. He holds a Ph.D. in computer science (2014) from Telecom SudParis, Institut Mines-Telecom, France and a M.Sc in computer science (2010) from the University of Tunis El-Manar, Tunisia.

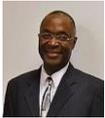

**Roch Glitho** holds a Ph.D. (Tekn. Dr.) in tele-informatics (Royal Institute of Technology, Stockholm, Sweden), and M.Sc. degrees in business economics (University of Grenoble, France), pure mathematics (University of Geneva, Switzerland), and computer science (University of Geneva). He is an associate professor and Canada Research Chair at Concordia University. He is also an adjunct professor at several other universities including Telecom Sud Paris, France, and the University of Western Cape, South Africa. In the past, he has worked in industry and has held several senior technical positions (e.g., senior specialist, principal engineer, expert) at Ericsson in Sweden and Canada. His industrial experience includes research, international standards setting, product management, project management, systems engineering, and software/firmware design. He has also served as an IEEE Distinguished Lecturer, Editor-In-Chief of IEEE Communications Magazine, and Editor-In-Chief of IEEE Communications Surveys & Tutorials Journal.

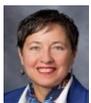

**Monique Morrow** held the title of CTO Cisco Services. Ms. Morrow's focus is in developing strategic technology and business architectures for Cisco customers and partners. With over 13 years at Cisco, Monique has made significant contributions in a wide range of roles, from Customer Advocacy to Corporate Consulting Engineering. With particular emphasis on the Service Provider segment, her experience includes roles in the field (Asia-Pacific) where she undertook the goal of building a strong technology team, as well as identifying and grooming a successor to assure a smooth transition and continued excellence. Monique has consistently shown her talent for forward thinking and risk taking in exploring market opportunities for Cisco. She was an early visionary in the realm of MPLS as a technology service enabler, and she was one of the leaders in developing new business opportunities for Cisco in the Service Provider segment, SP NGN. Monique holds 3 patents, and has an additional nine patent submissions filed with US Patent Office. Ms. Morrow is the co-author of several books, and has authored numerous articles. She also maintains several technology blogs, and is a major contributor to Cisco's Technology Radar, having achieved Gold Medalist Hall of Fame status for her contributions. Monique is also very active in industry associations. She is a new member of the Strategic Advisory Board for the School of Computer Science at North Carolina State University. Monique is particularly passionate about Girls in ICT and has been active at the ITU on this topic - presenting at the EU Parliament in April of 2013 as an advocate for Cisco. Within the Office of the CTO, first as an individual contributor, and now as CTO, she has built a strong leadership team, and she continues to drive Cisco's globalization and country strategies.

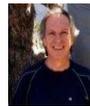

**Paul Polakos** was a Cisco Fellow and member of the Mobility CTO team at Cisco Systems focusing on emerging technologies for future Mobility systems. Prior to joining Cisco, Paul was Senior Director of Wireless Networking Research at Bell Labs, Alcatel-Lucent in Murray Hill, NJ and Paris, France. During his 28 years at Bell Labs he worked on a broad variety of topics in Physics and in Wireless Networking Research including the flat-IP cellular network architecture, the Base Station Router, femtocells, intelligent antennas and MIMO, radio and modem algorithms and ASICSs, autonomic networks and dynamic network optimization. Prior to joining Bell Labs, he was a member of the research staff at the Max-Planck Institute for Physics and Astrophysics (Munich) and visiting scientist at CERN and Fermilab. He holds BS, MS, and Ph.D. degrees in Physics from Rensselaer Polytechnic Institute and the University of Arizona, and author of more than 50 publications and 30 patents.